\documentstyle{ruthesis}

\begin{document}


    \phd
    \title{\bf Representation theory and
               tensor product theory for vertex operator algebras}

    \author{ Haisheng Li}
    \campus{New Brunswick}
    \program{Mathematics}
    \director{James Lepowsky and Robert L. Wilson}
    \approvals{4}
    \submissionmonth{May}
    \submissionyear{1994}
\beforepreface


\acknowledgements{
I'd like to express my deep gratitude to my Ph.D. advisers Professors
James Lepowsky and Robert Wilson for their insightful
advice, constant encouragement, and interesting
lectures.  Being one of their students, I have learned from them
not only mathematics but also English writing.
I am grateful to Professor Chongying Dong for many
stimulating discussions and for helpful advice. I would also like to thank
Professors Ruqi Zhou and Zhihuang Zhang for their many years
encouragement.}

\abstract{
We first formulate a definition of tensor product for two modules for
a vertex operator algebra in terms of a certain universal property and
then we give a construction of tensor products. We prove the unital property of
the adjoint module and the commutativity
of tensor products, up to module isomorphism. We relate this tensor
product construction with Frenkel and Zhu's $A(M)$-theory. We give a
proof of a formula of Frenkel and Zhu for fusion rules.

We also give the analogue of the ``Hom''-functor of
classical Lie algebra theory for vertex operator algebra theory by
introducing a notion of ``generalized intertwining operator.'' We
prove that the space of generalized
intertwining operators from one module to another for a vertex
operator algebra is a generalized
module. From this result we derive a general form of
Tsuchiya and Kanie's ``nuclear democracy
theorem'' for any rational vertex operator algebra. This
proves that the fusion rules obtained from our construction of tensor
products are the same as the fusion rules obtained by using Tsuchiya
and Kanie's method, for both WZW models and minimal models.
We prove that if $V$ satisfies certain ``finiteness'' and
``semisimplicity'' conditions, then
there exists a unique maximal submodule inside the generalized module.
Furthermore, we prove that this maximal submodule is isomorphic to
the contragredient module of a
certain tensor product module. This gives another construction of
tensor product modules and this
result turns out to be closely related to Huang and Lepowsky's construction.}

\afterpreface 

\makeatletter
\@addtoreset{equation}{chapter}
\def\theequation{\thechapter.\arabic{equation}}
\makeatother

\chapter{Introduction}

Vertex (operator) algebras ([B], [FLM2], [FHL]) are ``algebras''
equipped with an infinite sequence of multiplications satisfying
the so-called ``Jacobi identity.''
The notion of vertex (operator) algebra is the algebraic counterpart of
the notion of what is now usually called  ``chiral
algebra'' in conformal field theory. In fact, much of conformal field
theory, basic algebraic features of whose axiomatic structure were
formulated in such works as [BPZ] and later, [MS], can be understood
in terms of the theory of vertex operator algebras and their
representations.

Vertex (operator) algebras can be viewed as ``complex
analogues'' of classical Lie algebras and also of commutative
associative algebras.
In particular, any commutative associative algebra with identity is a
vertex (operator) algebra [B].
In [L2], motivated by a classical analogue we found a
new proof that ``commutativity'' implies ``associativity'' for vertex
(operator) algebras, so
that the commutativity property implies the Jacobi identity for vertex
(operator) algebras (see  [FLM2], [FHL], [DL], [L2]) and we
developed an analogue of the notion of endomorphism ring of a vector
space for vertex operator algebra theory [L2].

The main theme of this thesis is to develop analogues of the tensor
product functor ``$\otimes$'' and the ``Hom''-functor
for vertex operator algebra theory
to deepen the analogy between vertex operator algebras and classical
commutative associative algebras. We also rebuild Zhu's 1-1
correspondence [Zhu] by using the notion of ``generalized Verma
module'' for ${\bf Z}$-graded Lie algebras (cf. [Lep]) and we
define the notion of
``generalized Verma (weak) module'' for vertex operator algebras as an
analogue.

Vertex operator algebras are not classical algebras, but there are
classical algebras associated to each vertex operator algebra. In
his thesis [Zhu], as one of his remarkable results, Zhu has
constructed an associative algebra $A(V)$ for any
vertex operator algebra $V$ and established
a 1-1 correspondence between the set of equivalence classes of
irreducible lower-truncated ${\bf Z}$-graded weak $V$-modules and the set of
equivalence classes of irreducible $A(V)$-modules. Roughly speaking,
Zhu's algebra $A(V)$ is the associative algebra generated by all
weight-zero components of vertex operators modulo certain relations
arising from the $L(-1)$-derivative property and the Jacobi identity.
For an $A(V)$-module $U$, to construct a lowest-degree (weak)
$V$-module with $U$ as the lowest-degree subspace, Zhu first constructs
a free space $F(U)$ generated by symbols $a(n)$ (linear in $a\in V$) from
$U$ and then he defines a linear function on $U^{*}\otimes_{{\bf C}}
F(U)$ inductively. Using induction he
proves the rationality, the commutativity and the associativity for
the desired correlation functions, so that the quotient space
$\bar{F}(U)$ of $F(U)$ modulo a certain subspace is a (weak) $V$-module.
In the last step, he proves that $\bar{F}(U)$ is a lowest-degree (weak)
$V$-module with $U$ as its lowest-degree subspace.

To a certain
extent, Zhu's $A(V)$-theory is analogous to lowest-weight module
theory for an affine Lie algebra $\tilde{{\bf g}}$, where $A(V)$
roughly plays the
role of $U({\bf g})$, the universal enveloping algebra of the
finite-dimensional simple Lie algebra ${\bf g}$.
The Jacobi
identity for vertex operator algebras is equivalent to the
combination of Borcherds' commutator formula (2.1.3)  and the (associative)
iterate formula (2.1.4). With the help of the
notion of affine vertex algebras (recalled in Section 2.1.3) we  prove
that for any vertex algebra $V$, the quotient space $g(V)$ of
${\bf C}[t,t^{-1}]\otimes V$ modulo the subspace $({d\over dt}\otimes
1+1\otimes L(-1))V$ is a Lie algebra with the commutator
formula as its Lie bracket (cf. [B]). This Lie algebra
has also been studied by Feingold, Frenkel and Ries [FFR].
As a consequence, $g(V)$ is the
Lie algebra with underlying space $\hat{V}={\bf
C}[t,t^{-1}]\otimes V$ and with the commutator formula providing its
defining relations.
For any ${\bf Z}$-graded Lie algebra $g=\oplus_{n\in {\bf Z}}g_{n}$,
we have a triangular decomposition $g=g_{+}\oplus g_{0}\oplus g_{-}$,
where $g_{\pm}=\oplus_{n=1}^{\infty}g_{\pm n}$. Generalizing the
notion of ``generalized Verma module'' we can define the notion
of ``generalized Verma module'' for any ${\bf Z}$-graded Lie algebra
$g$. It follows from the PBW theorem that there is a 1-1 correspondence
between the set of equivalence classes
of irreducible $g_{0}$-modules and the set of equivalence classes of
irreducible lowest-degree $g$-modules.
Since $g(V)$ is a ${\bf Z}$-graded Lie algebra (graded by weights),
the lowest-degree module theory applies to $g(V)$.

Although the lowest-degree $g(V)$-module theory is not equivalent to
the lowest-weight (weak) $V$-module theory, this theory is really useful for
studying Zhu's $A(V)$-theory.
It follows from [Zhu] that the natural projection map from $g(V)_{0}$
to $A(V)$ is a Lie algebra homomorphism where $A(V)$ is viewed as a
Lie algebra $A(V)_{Lie}$, so that any $A(V)$-module $U$ is a natural
$g(V)_{0}$-module. Let $M(U)$ be the generalized Verma $g(V)$-module
with the lowest-degree subspace $U$. Let $L(U)$ be the quotient
$g(V)$-module of $M(U)$ divided by the (unique) maximal graded submodule
 intersecting $U$ trivially. Then we prove that associativity holds
on $L(U)$ so that the Jacobi identity follows (since the
commutativity, the $L(-1)$-derivative property and the vacuum property
have already been built to the $g(V)$-module structure on $L(U)$).
Thus $L(U)$ is a (weak)
$V$-module with $U$ as its lowest-degree subspace.
It follows from an abstract argument that the universal lowest-degree
$V$-module $\bar{M}(U)$ with lowest-degree subspace $U$ exists. That is,
$\bar{M}(U)$ is the Verma $V$-module with lowest-degree subspace $U$.
In this way, the
analogy between Zhu's $A(V)$ theory and the lowest-degree module theory
is manifest.

The study of tensor product theory for representations for a chiral
algebra or a vertex operator algebra was initiated by physicists due
to the need to describe the coupling of vertices (see for example [MS]).
The first mathematically rigorous notion of tensor product was
given by Kazhdan and Lusztig [KL0]-[KL4] for
modules of certain levels for an affine Lie algebra.
Then Huang
and Lepowsky [HL0]-[HL2] give an approach to tensor product theory for modules
for a vertex operator algebra. Huang and Lepowsky's approach is
analytic in nature, so that it is not quite analogous to the
classical tensor product construction. (Professor James Lepowsky
informed us that some years ago, Borcherds
also began considering a notion of tensor product of modules for a
vertex algebra.)

In the classical Lie algebra theory, if $U^{1}$ and $U^{2}$ are
modules for a Lie algebra ${\bf g}$, the tensor product vector space
$U^{1}\otimes U^{2}$ has a natural ${\bf g}$-module structure with the
following action:
\begin{eqnarray}
a(u^{1}\otimes u^{2})=au^{1}\otimes u^{2}+u^{1}\otimes au^{2}
\;\;\;\mbox{for }a\in {\bf g},u^{1}\in U^{1},u^{2}\in U^{2}.
\end{eqnarray}
Of course, this is due to the Hopf algebra structure of the universal
enveloping algebra $U({\bf g})$. But it is important to notice that the
formula (1.1) is symbolically the classical Jacobi
identity.
If $U^{3}$ is another module, an intertwining operator of type
$\left(\begin{array}{c}U^{3}\\U^{1},U^{2}\end{array}\right)$ is
defined to be a ${\bf g}$-homomorphism $I$ from the tensor product module
$U^{1}\otimes U^{2}$ to $U^{3}$. Equivalently, it can also be defined
(at least superficially) without using the notion of tensor product
module as a linear map $I$ from $U^{1}$ to ${\rm Hom}_{{\bf
C}}(U^{2},U^{3})$ satisfying the following condition:
\begin{eqnarray}
aI(u^{1})u^{2}=I(u^{1})au^{2}+I(au^{1})u^{2}\;\;\;\mbox{for }a\in {\bf
g},u^{i}\in U^{i}.
\end{eqnarray}
Denote by $F$ the standard bilinear map from $U^{1}\times U^{2}$ to
$U^{1}\otimes U^{2}$. Then $F$ is an intertwining operator of type
$\left(\begin{array}{c}U^{1}\otimes
U^{2}\\U^{1},U^{2}\end{array}\right)$ and the pair $(U^{1}\otimes
U^{2},F)$ satisfies the
following universal property: For any ${\bf g}$-module $U^{3}$ and any
intertwining operator $I$ of type
$\left(\begin{array}{c}U^{3}\\U^{1},U^{2}\end{array}\right)$, there
exists a unique ${\bf g}$-homomorphism $\psi$ from $U^{1}\otimes U^{2}$ to
$U^{3}$ such that $I=\psi\circ F$. On the other hand, the
tensor product module $U^{1}\otimes U^{2}$ is nothing but the vector space
spanned by ``formal symbols'' $I(u^{1})u^{2}$ (linear in $u^{1}$ and
$u^{2}$) with the ${\bf g}$-action defined by (1.2).

In vertex operator algebra theory,
the notion of intertwining operator was defined [FHL] analogously to
the second classical definition, but
we initially don't have the notion of tensor product.
Let $V$ be a vertex operator algebra
and let $M^{i}$ $(i=1,2,3)$ be three $V$-modules. An intertwining
operator of type
$\left(\begin{array}{c}M^{3}\\M^{1},M^{2}\end{array}\right)$ is
defined  to be a
linear map $I(\cdot,x)$ from $M^{1}$ to $\left({\rm
Hom}(M^{2},M^{3})\right)\{x\}$
satisfying the truncation condition, the $L(-1)$-derivative property
and the following Jacobi identity:
\begin{eqnarray}
& &x_{0}^{-1}\delta\left(\frac{x_{1}-x_{2}}{x_{0}}\right)Y(a,x_{1})
I(u^{1},x_{2})- x_{0}^{-1}\delta\left(\frac{x_{2}-x_{1}}{-x_{0}}\right)
I(u^{1},x_{2})Y(a,x_{1})\nonumber\\
&=&x_{2}^{-1}\delta\left(\frac{x_{1}-x_{0}}{x_{2}}\right)
I(Y(a,x_{0})u^{1},x_{2}).
\end{eqnarray}
Motivated by the classical tensor product theory, we
formulate the following definition: Let $M^{1}$ and $M^{2}$ be two
$V$-modules. A
{\it tensor product} for the ordered pair $(M^{1}, M^{2})$ is a
pair $(M,F(\cdot,x))$ consisting of a
$V$-module $M$ and an intertwining operator $F(\cdot,x)$ of type
$\left(\begin{array}{c}M\\M^{1}, M^{2}\end{array}\right)$ such that the
following universal
property holds: For any $V$-module $W$ and any intertwining operator
$I(\cdot,x)$ of type
$\left(\begin{array}{c}W\\M^{1},M^{2}\end{array}\right)$, there
exists a unique $V$-homomorphism $\psi$ from $M$ to $W$ such that
$I(\cdot,x)= \psi\circ F(\cdot,x)$. (Here $\psi$ extends canonically
to a linear map from $M\{x\}$ to $W\{x\}$.)

An intertwining operator can be viewed as either a formal series
of operators or an operator-valued functional.
In [HL1], Huang and Lepowsky have also formulated certain notions of tensor
product involving some geometry, where an intertwining operator is
viewed as an operator-valued functional.

Next is our construction of tensor products. Roughly speaking,
our tensor product module is constructed as the quotient space of the
tensor product vector space ${\bf
C}[t,t^{-1}]\otimes M^{1}\otimes M^{2}$ (symbolically the linear span
of all coefficients of $Y(u^{1},x)u^{2}$ for $u^{i}\in M^{i}$)
divided by all the axioms for an intertwining operator of a certain type.

First we form the
formal vector space $F_{0}(M^{1},M^{2})={\bf C}[t,t^{-1}]\otimes
M^{1}\otimes M^{2}$. Taking
${\rm Res}_{x_{0}}$ of (1.3), we obtain the commutator formula:
\begin{eqnarray}
Y(a,x_{1})I(u^{1},x_{2})=I(u^{1},x_{2})Y(a,x_{1})
+{\rm Res}_{x_{0}}x_{2}^{-1}\delta\left(\frac{x_{1}-x_{0}}{x_{2}}\right)
I(Y(a,x_{0})u^{1},x_{2}).
\end{eqnarray}
Analogously to the classical case, using this formula we define a
tensor product action of $\hat{V}={\bf C}[t,t^{-1}]\otimes V$ on
$F_{0}(M^{1},M^{2})$ as follows:
\begin{eqnarray}
& &Y(a,x_{1})(Y_{t}(u^{1},x_{2})\otimes u^{2})\nonumber\\
&=&Y_{t}(u^{1},x_{2})\otimes Y(a,x_{1})u^{2}
+{\rm Res}_{x_{0}}x_{2}^{-1}\delta\left(\frac{x_{1}-x_{0}}{x_{2}}\right)
Y_{t}(Y(a,x_{0})u^{1},x_{2})\otimes u^{2},
\end{eqnarray}
where $Y_{t}(u^{1},x)=\sum_{n\in {\bf Z}}(t^{n}\otimes u^{1})x^{-n-1}$.
Then we prove that this defined action on
$F_{0}(M^{1},M^{2})$ satisfies the commutator formula,
or equivalently this defines a
$g(V)$-module structure on $F_{0}(M^{1},M^{2})$ (Proposition 5.2.1). Let
$F_{1}(M^{1},M^{2})$ be the quotient
$g(V)$-module of $F_{0}(M^{1},M^{2})$ divided by the truncation
condition. Then we define $F_{2}(M^{1},M^{2})$ to be the quotient
$g(V)$-module of $F_{1}(M^{1},M^{2})$ divided by the Jacobi identity
relation for an intertwining
operator. Then we prove that $F_{2}(M^{1},M^{2})$ is a lower-truncated
${\bf Z}$-graded weak $V$-module (Theorem 5.2.9). Furthermore, let
$T(M^{1},M^{2})$ be the
quotient module of $F_{2}(M^{1},M^{2})$ divided by its (unique) maximal graded
submodule with degree-zero subspace being zero. Then we prove that if $V$ is
rational in the sense of [Zhu],
$T(M^{1},M^{2})$ is a tensor
product for the ordered pair $(M^{1},M^{2})$ (Theorem 5.2.22). (The last step
corresponds to the $L(-1)$-derivative property for an intertwining operator.)

To summarize, it is very natural that the tensor product vector space ${\bf
C}[t,t^{-1}]\otimes M^{1}\otimes M^{2}$ divided by all the axioms for an
intertwining operator of a certain type is a (weak)
$V$-module. (In classical Lie algebra theory, the commutator
formula is just the Jacobi identity, so that we don't have to pass to
a quotient space.) Furthermore, we also
prove the unital property of the adjoint module $V$
(Proposition 6.1.2) and the commutativity of tensor products, up to
$V$-isomorphism (Proposition 6.2.3).

Compared with Huang and Lepowsky's (analytic) approach, this is a
formal variable approach. But it follows from the universal properties
that the tensor product modules from both constructions are isomorphic.

In [FZ], generalizing Zhu's $A(V)$-theory, Frenkel and Zhu have
developed a theory to calculate the fusion rules. Obviously,
their theory is closely related to tensor product theory in certain ways.
To any $V$-module $M$, Frenkel and Zhu [FZ] have
associated an $A(V)$-bimodule $A(M)$. Then one of their main theorem
says that for any three irreducible
$V$-modules $M^{i}$
(i=1,2,3) there is a natural linear isomorphism between ${\rm
Hom}_{A(V)}(A(M^{1})\otimes _{A(V)}M^{2}(0),M^{3}(0))$ and
$I\left(\begin{array}{c}M^{3}\\M^{1},M^{2}\end{array}\right)$, the
space of intertwining operators of the indicated type. This gives one
way to compute fusion rules.
Attempting to give a complete proof, we find that
this is not true in general.
(In Appendix $A$ we give a counterexample.)
However, we prove that a sufficient condition is that both $M^{2}$ and
$(M^{3})'$ are universal (Theorem 4.2.4).
Especially, it is true if $V$ is  rational. In the proof, we can see that
Frenkel and Zhu's theorem  is deeply related to our construction of
tensor products. We also show that the
degree-zero subspace of $T(M^{1},M^{2})$ is a quotient $A(V)$-module
of the $A(V)$-module $A(M^{1})\otimes_{A(V)}M^{2}(0)$. Then it follows
from Frenkel and Zhu's theorem and the universal property of
$T(M^{1},M^{2})$ that  $A(M^{1})\otimes_{A(V)}M^{2}(0)$ is
isomorphic to the degree-zero subspace of $T(M^{1},M^{2})$ as an
$A(V)$-module.

In Lie algebra theory, for any two modules $U^{1}$ and $U^{2}$ for a
Lie algebra ${\bf g}$, the space ${\rm Hom}_{{\bf C}}(U^{1},U^{2})$
has a natural ${\bf g}$-module structure with the action defined by:
\begin{eqnarray}
(af)(u^{1})=af(u^{1})-f(au^{1})\;\;\;\mbox{ for any
}a\in {\bf g}, f\in {\rm Hom}_{{\bf C}}(U^{1},U^{2}), u^{1}\in U^{1}.
\end{eqnarray}
Notice that the formula (1.6) is also symbolically the
classical Jacobi identity.
Furthermore, we have the following natural inclusion relations:
\begin{eqnarray}
(U^{1})^{*}\otimes U^{2}\longrightarrow {\rm Hom}_{{\bf C}}(U^{1},U^{2})
\longrightarrow (U^{1}\otimes (U^{2})^{*})^{*}.
\end{eqnarray}
If both $U^{1}$ and
$U^{2}$ are finite-dimensional, the arrows are isomorphism so that
the space of linear homomorphisms gives a construction of tensor
product modules.

As a complex analogue of ${\rm Hom}_{{\bf
C}}(M^{1},M^{2})$, the vector space $\left({\rm Hom}_{{\bf
C}}(M^{1},M^{2})\right)\{x\}$ is the natural object to study. Taking ${\rm
Res}_{x_{1}}$ of the Jacobi identity we obtain the iterate formula:
\begin{eqnarray}
& &Y(Y(a,x_{0})b,x_{2})\nonumber\\
&=&{\rm Res}_{x_{1}}\left(x_{0}^{-1}\delta\left(\frac{x_{1}-x_{2}}{x_{0}}
\right)Y(a,x_{1})
Y(b,x_{2})- x_{0}^{-1}\delta\left(\frac{x_{2}-x_{1}}{-x_{0}}\right)
Y(b,x_{2})Y(a,x_{1})\right).\nonumber\\
& &\mbox{}
\end{eqnarray}
Analogously to the classical case we define an action of $\hat{V}$ as follows:
\begin{eqnarray}
& &Y(a,x_{0})\circ A(x_{2})\nonumber\\
&=&{\rm Res}_{x_{1}}\left(x_{0}^{-1}\delta\left(\frac{x_{1}-x_{2}}{x_{0}}
\right)Y(a,x_{1})
A(x_{2})- x_{0}^{-1}\delta\left(\frac{x_{2}-x_{1}}{-x_{0}}\right)
A(x_{2})Y(a,x_{1})\right).
\end{eqnarray}
But the right hand side may not exist. Just as in the
definition of vertex operator algebras we need a certain truncation
condition for $A(x)$ so that the Jacobi identity really makes
sense. On the other hand, just as in the classical case, since we treat $A(x)$
as $I(u,x)$ for some intertwining operator $I(\cdot,x)$ and for some
$u\in M^{1}$, $A(x)$ should satisfy certain properties that $I(u,x)$ satisfies.
Based on this philosophy, we
introduce the notion of generalized intertwining operators.

A {\it generalized
intertwining operator} from a $V$-module $M^{1}$ to another
$V$-module $M^{2}$ is an element
$\phi(x)$ of $\left({\rm Hom}_{{\bf C}}(M^{1},M^{2})\right)\{x\}$
satisfying the truncation condition,
the $L(-1)$-derivative property: $[L(-1),\phi(x)]=\displaystyle{{d\over dx}}
\phi(x)$ and
the commutativity without involving in ``matrix-coefficient,'' i.e.,
for $a\in V$, there is a positive integer $k$ such that
\begin{eqnarray}
(x_{1}-x_{2})^{k}Y(a,x_{1})\phi(x_{2})=(x_{1}-x_{2})^{k}\phi(x_{2})Y(a,x_{1})
\end{eqnarray}
(cf. [DL, formula (9.37)]). These are the main features of an intertwining
operator of a certain type evaluated at a vector.
We prove that $G(M^{1},M^{2})$, the space of all generalized
intertwining operators from $M^{1}$ to $M^{2}$, is a (weak)
$V$-module under the action (1.9) (Theorem 7.1.6). We prove that for
any $V$-module $M$, there is a natural linear isomorphism from ${\rm
Hom}_{V}(M, G(M^{1},M^{2}))$ onto
$I\left(\begin{array}{c}M^{2}\\M,M^{1}\end{array}\right)$, the space
of intertwining operators of the indicated type.
For the special case
when $M^{1}=V$, we prove that $G(V,M^{2})$ is isomorphic
to $M$ as a $V$-module (Proposition 7.2.3). In general, we prove that
if $V$ satisfies certain ``finiteness'' and ``semisimplicity'' conditions,
then there exists a unique
maximal submodule $\Delta (M^{1},M^{2})$ inside the generalized module
$G(M^{1},M^{2})$ (Proposition 7.2.4). Under these same conditions,
we prove
that $\Delta (M^{1},M^{2})$ is a tensor product module for the ordered
pair $(M^{1},(M^{2})')$ (Theorem 7.2.6).

Let $V$ be the vertex operator algebra associated to a standard module
of level $\ell$ for an affine Lie algebra $\tilde{{\bf g}}$. Then it
has been proved ([DL], [FZ], [L2]) that $V$ is rational and that the
set of equivalence classes of standard $\tilde{{\bf g}}$-modules of
level $\ell$
are exactly the set of equivalence classes of irreducible $V$-modules. These
are rigorous results for the physical WZW model. In [TK] and many physics
papers, an intertwining operator $I(\cdot,x)$ is defined to be a
linear map from $M^{1}(0)$ to $\left({\rm Hom}_{{\bf
C}}(M^{2},M^{3})\right)\{x\}$ satisfying the $L(-1)$-bracket formula
and the commutator formula (1.4) for $a\in {\bf g}$. It has been
proved [TK] that such defined an intertwining operator can be uniquely
extended to an intertwining operator on the whole space $M^{1}$ (in
the sense of [FHL]). This is the well-known Tsuchiya and
Kanie's nuclear democracy theorem [TK].

Suppose $V$ is any rational vertex operator algebra. Let $M^{i}$
$(i=1,2,3)$ be three irreducible $V$-modules and let $I_{0}(\cdot,x)$
be a linear map from $M^{1}(0)$ to  $\left({\rm Hom}_{{\bf
C}}(M^{2},M^{3})\right)\{x\}$ satisfying the truncation condition, the
$L(-1)$-bracket formula and the following condition:
\begin{eqnarray}
& &(x_{1}-x_{2})^{{\rm wt}a-1}Y(a,x_{1})I_{0}(u,x_{2})
-(-x_{2}+x_{1})^{{\rm wt}a-1}I_{0}(u,x_{2})Y(a,x_{1})\nonumber\\
&=&x_{2}^{-1}\delta\left({x_{1}\over x_{2}}\right)I_{0}(\bar{a}u,x_{2})
\end{eqnarray}
for any $a\in V$, where $\bar{a}$ is viewed as an
element of $A(V)$ or $g(V)_{0}$ so that $\bar{a}$ acts on $M^{1}(0)$.
Then we prove that there is a unique intertwining operator
$I(\cdot,x)$ in the sense of [FHL] extending $I_{0}(\cdot,x)$
(Proposition 7.3.1 and Corollary 7.3.2). Therefore, we obtain
a generalized version of  Tsuchiya and Kanie's nuclear democracy theorem.
Furthermore, for WZW-models,
TK's nuclear democracy theorem [TK] implies that the
fusion rules in the sense of [FHL] and the fusion rules in the sense
of [TK] or [MS] are equal. Therefore, it follows from our universal property
of a tensor product module (in terms of intertwining operators in the
sense of [FHL]) that the fusion rules from our construction of a
tensor product module are equal to those obtained in [TK] and [MS].
Even for minimal models, since the corresponding VOA's are rational [W], TK's
nuclear democracy theorem holds.

This thesis is organized as follows: In Chapter 2 we
study Lie algebras
and their modules associated to a vertex operator algebra and present
some elementary results. In Chapter 3 we define the notion of generalized
Verma (weak) $V$-module and give a different approach to Zhu's
$A(V)$-theory. In Chapter 4 we give a complete proof
for a formula of Frenkel and Zhu for fusion rules.
In Chapter 5 we formulate a definition of tensor product and construct
a tensor product
module for two modules. In Chapter 6 we prove the unital property of
the adjoint module $V$
and the commutativity of tensor products. We also
relate our approach to Frenkel and Zhu's theory. In Chapter 7 we give
an analogue of the classical ``Hom''-functor and
derive a generalized version of Tsuchiya and Kanie's nuclear democracy
theorem for any rational vertex operator algebra. In Appendix A we
give an example to show that Frenkel and Zhu's theorem is not true in
general. In Appendix B we give an example to show that  a tensor product module
of two nonzero modules might be zero.

\newpage

\makeatletter
\@addtoreset{equation}{section}
\def\theequation{\thesection.\arabic{equation}}
\makeatother

\chapter{Lie algebras associated to a vertex algebra}
\def\inch#1{$#1''$}  

A vertex algebra is not a classical algebra although it is analogous
to a Lie algebra and an associative algebra. However there are Lie
algebras and associative algebras closely related to a vertex algebra.
These classical algebras might be useful for the study of vertex
algebras and representations.
For a vertex (operator) algebra $V$, since
components of vertex operators are operators on $V$, which satisfy
certain relations from the axioms for vertex (operator) algebra,
it follows from abstract nonsense that one can always get a Lie algebra
or an associative algebra in terms some generators and defining relations.
For instance if we consider Borcherds' commutator formula (2.2.6),
we get a Lie algebra (cf. [FFR]); if we consider the whole Jacobi identity
we have Frenkel and Zhu's universal enveloping algebra $U(V)$ [FZ].
(Some nonassociative commutative algebras such as Griess
algebra ([Gri], [FLM2]) and Chevalley algebra [FFR] are also related to
vertex algebras.) In this chapter we shall study certain Lie algebras
and their modules related to a vertex (operator) algebra and its modules.
Although most of the results presented in this chapter are
pretty well-known, it is very interesting to apply these to some deep
theories.

\section{Vertex algebras and modules}
In this section we first review some necessary definitions from [B], [FLM2] and
[FHL]. Then we shall present some elementary results we need later.
Throughout this section we
will typically use the letters {\it a, b, c},... to represent
elements of a given vertex (operator) algebra $V$ and the letters
{\it u, v, w},... to represent vectors of $V$-modules. We use standard
notations and definitions of [FLM2], [FHL] and [FZ].

The following definition of ``vertex algebra'' is a combination of
Borcherds' definition of vertex algebra [B] and FLM's definition of
vertex operator algebra [FLM2].

{\bf Definition 2.1.1.} A vertex algebra is a quadruple $(V, Y, {\bf
1}, d)$ where $V$ is a vector space, $Y(\cdot,x)$ is a linear map from
$V$ to $({\rm End} V)[[x,x^{-1}]]$, ${\bf 1}$ is a specified element
of $V$ and $d$ is an endomorphism of $V$ such that the following
conditions hold:

(V1)\hspace{0.25cm} $Y({\bf 1},x)={\rm id}_{V}$;

(V2)\hspace{0.25cm} $Y(a,x){\bf 1}\in ({\rm End} V)[[x]]$ and
$\displaystyle{\lim_{x\mapsto
0}Y(a,x){\bf 1}=a}$ for any $a\in V$;

(V3)\hspace{0.25cm} $Y(da,x)=\displaystyle{{d\over dx}}Y(a,x)$ for any $a\in
V$;

(V4)\hspace{0.25cm} $Y(a,x)b\in V((x))$ for any $a, b\in V$;

(V5)\hspace{0.25cm} The following {\it Jacobi identity} holds:
\begin{eqnarray}
& &x_{0}^{-1}\delta\left(\frac{x_{1}-x_{2}}{x_{0}}\right)Y(a,x_{1})Y(b,x_{2})c
-x_{0}^{-1}\delta\left(\frac{-x_{2}+x_{1}}{x_{0}}\right)Y(b,x_{2})Y(a,x_{1})c
\nonumber\\
&=&x_{2}^{-1}\delta\left(\frac{x_{1}-x_{0}}{x_{2}}\right)Y(Y(a,x_{0})b,x_{2})c
\end{eqnarray}
for any $a, b, c\in V$, where all variables are formal and commuting,
and $(x_{i}-x_{j})^{n}$ is understood to be
the formal Laurent series in nonnegative powers of the second variable $x_{j}$
for any
integer $n$, so that the three leading terms are formal Laurent series
in formal
variables $x_{0}, x_{1}$ and $x_{2}$. For $a\in V$, set
$Y(a,x)=\sum_{n\in {\bf Z}}a_{n}x^{-n-1}$, which is called the {\it
vertex operator} associated to $a$.

This completes the definition of vertex algebra.
The following are consequences:
\begin{eqnarray}
& &Y(a,x)b=e^{xd}Y(b,-x)a\;\;\;\;\mbox{({\it skew symmetry})};\\
& &[Y(a,x_{1}),Y(b,x_{2})]={\rm
Res}_{x_{0}}x_{2}^{-1}\delta\left(\frac{x_{1}-x_{0}}{x_{2}}\right)
Y(Y(a,x_{0})b,x_{2});\\
& &\mbox{ ({\it commutator formula})}\nonumber\\
& &Y(Y(a,x_{0})b,x_{2})=Y(a,x_{0}+x_{2})Y(b,x_{2})-Y(b,x_{2})
(Y(a,x_{0}+x_{2})-Y(a,x_{2}+x_{0}))\nonumber\\
& &\mbox{ ({\it associator formula})}
\end{eqnarray}
for any $a,b\in V$.

{\bf Definition 2.1.2.} A {\it module} for a vertex algebra $V$ is a pair
$(M, Y_{M})$ where $M$ is a vector space and $Y_{M}(\cdot,x)$ is
a linear map from $V$ to $({\rm End} M)[[x,x^{-1}]]$
satisfying the following conditions:

(M1)\hspace{0.25cm} $Y_{M}({\bf 1},x)={\rm id}_{M}$;

(M2)\hspace{0.25cm} $Y_{M}(da,x)=\displaystyle{{d\over dx}}Y_{M}(a,x)$ for any
$a\in V$;

(M3)\hspace{0.25cm} $Y_{M}(a,x)u\in M((x))$ for any $a\in V, u\in M$;

(M4)\hspace{0.25cm} For any $a,b\in V, u\in M$, the following
corresponding Jacobi identity holds:
\begin{eqnarray}
& &x_{0}^{-1}\delta \left( \frac
{x_{1}-x_{2}}{x_{0}}\right)Y_{M}(a,x_{1})Y_{M}(b,x_{2})u-z^{-1}_{0}\delta
\left(\frac{-x_{2}+x_{1}}{x_{0}}\right)Y_{M}(b,x_{2})Y_{M}(a,x_{1})u\nonumber\\
&=&x^{-1}_{2}\delta
\left(\frac{x_{1}-x_{0}}{x_{2}}\right)Y_{M}(Y(a,x_{0})b,x_{2})u.
\end{eqnarray}

As consequences of the Jacobi identity, we have the following
commutativity and associativity: For any $a,b\in V$, there is a
nonnegative integer $k$ such that for any $u\in M$
\begin{eqnarray}
(x_{1}-x_{2})^{k}Y_{M}(a,x_{1})Y_{M}(b,x_{2})u
=(x_{1}-x_{2})^{k}Y_{M}(b,x_{2})Y_{M}(a,x_{1})u;
\end{eqnarray}
For any $a\in V, u\in M$, there is a nonnegative integer $r$ such that
for any $b\in V$
\begin{eqnarray}
(x_{0}+x_{2})^{r}Y_{M}(a,x_{0}+x_{2})Y_{M}(b,x_{2})u
=(x_{0}+x_{2})^{r}Y_{M}(Y(a,x_{0})b,x_{2})u.
\end{eqnarray}

For a vertex operator algebra $V$ (we will recall the definition
later), these are equivalent to the well-known rationality,
commutativity and  associativity in terms of
``matrix-coefficient'' ([FHL], [FLM2]). For the purpose of reference
we call (2.1.6) and (2.1.7) the commutativity and the associativity ``without
involving matrix-coefficients,'' respectively.

{\bf Proposition 2.1.3 ([DL], [FHL], [L2]).} {\it The Jacobi identity
for a module for a vertex algebra is
equivalent to the commutativity and the associativity without involving
matrix-coefficients.}

{\bf Definition 2.1.4.} A ${\bf Z}$-{\it graded} vertex algebra
is a vertex algebra $V$ together with a ${\bf Z}$-graded decomposition
$V=\oplus_{n\in {\bf Z}}V_{(n)}$ satisfying the following conditions:
\begin{eqnarray}
dV_{(n)}\subseteq V_{(n+1)},\; a_{m}V_{(n)}\subseteq V_{(r+n-m-1)}
\;\;\;\mbox{ for any }m,n,r\in {\bf Z}, a\in V_{(r)}.
\end{eqnarray}

{\bf Examples 2.1.5 [B].} Let $A$ be a commutative
associative algebra with identity together with a derivation $d$.
Define
\begin{eqnarray}
Y(a,x)b=\left(e^{xd}a\right)b\;\;\;\;\mbox{for any }a,b\in A.
\end{eqnarray}
Since $d(1)=0$, $Y(1,x)={\rm id}_{A}$. For any $a,b\in A$ we have:
\begin{eqnarray}
{d\over dx}Y(a,x)b={d\over dx}\left(e^{xd}a\right)b=\left(e^{xd}da\right)b
=Y(da,x)b.
\end{eqnarray}
By definition, we have:
\begin{eqnarray}
Y(a,x_{1})Y(b,x_{2})=\left(e^{x_{1}d}a\right)\left(e^{x_{2}d}b\right)
=\left(e^{x_{2}d}b\right)\left(e^{x_{1}d}a\right)
=Y(b,x_{2})Y(a,x_{1})
\end{eqnarray}
and
\begin{eqnarray}
Y(a,x_{0}+z_{2})Y(b,x_{2})&=&\left(e^{(x_{0}+x_{2})d}a\right)
\left(e^{x_{2}d}b\right)\nonumber\\
&=&e^{x_{2}d}\left(\left(e^{x_{0}d}a\right)b\right)\nonumber\\
&=&Y(Y(a,x_{0})b,x_{2}).
\end{eqnarray}
Then it follows from Proposition 2.1.3 that $(A,Y,1,d)$ is a vertex algebra.
It is clear that a subspace $B$ of $A$ is a vertex subalgebra if and
only if $B$ is a subalgebra (containing the identity) of the
associative algebra $A$ such that the derivation $d$ preserves $B$.
Furthermore, let $M$ be a module for $A$ as an associative algebra.
Similarly, we define $Y_{M}(a,x)u=\left(e^{xd}a\right)u$ for
$a\in V,u\in M$. Then $(M,Y_{M})$ is a module for the vertex algebra
$(A,Y,1,d)$.

Let $A$ and $B$ be commutative associative algebras (with identity)
with derivations $d_{A}$ and $d_{B}$, respectively. Let $\psi$ be an
algebra homomorphism from $A$ to $B$ such that $\psi d_{A}=d_{B}\psi$.
Then $\psi$ is a vertex algebra homomorphism from $A$ to $B$.

In particular, let $A={\bf C}((t))$ and $\displaystyle{d={d\over dt}}$. Then
 $({\bf C}((t)),Y,1,{d\over dt})$ is a vertex algebra.
By definition, we have:
\begin{eqnarray}
Y(f(t),x)=e^{x{d\over dt}}f(t)=f(t+x)\;\;\;\mbox{ for }f(t)\in {\bf C}((t)).
\end{eqnarray}
It is clear that the Laurent polynomial ring  ${\bf C}[t,t^{-1}]$ is a
vertex subalgebra. For any $m\in {\bf Z}$, we have:
\begin{eqnarray}
Y(t^{m},x)=(t+x)^{m}
\;\left(=\sum_{k=0}^{\infty}
\left(\begin{array}{c}m\\k\end{array}\right)t^{m-k}x^{k}\right).
\end{eqnarray}
Let $\displaystyle{d(0)=-t{d\over dt}}$. Then
\begin{eqnarray}
[d(0),d]=d,\;\;\;d(0)t^{n}=-nt^{n}\;\;\;\;\mbox{for }n\in {\bf Z},
\end{eqnarray}
and
\begin{eqnarray}
[d(0),Y(t^{m},x)]&=&[-t{d\over dt},(t+x)^{m}]\nonumber\\
&=&-t{d\over dt}(t+x)^{m}\nonumber\\
&=&-mt(t+x)^{m-1}\nonumber\\
&=&-m(t+x)^{m}+mx(t+x)^{m-1}\nonumber\\
&=&(-m)Y(t^{m},x)+x{d\over dx}Y(t^{m},x).
\end{eqnarray}
Therefore, ${\bf C}[t,t^{-1}]$ is a ${\bf Z}$-graded vertex algebra.

We have seen that a vertex algebra constructed from a commutative
associative algebra with identity together with a derivation satisfies
the commutativity (2.1.6) with $k=0$ for any two elements $a$ and $b$.
Conversely, if a vertex algebra $V$
satisfies the commutativity (2.1.6) with $k=0$ for any $a,b\in
V$, then $V$ is  a
vertex algebra constructed from a commutative associative algebra $A$
with identity together with a derivation. Even more than this, we have
the following proposition, which gives an answer to one of
the questions raised by Professor Charles  Weibel.

{\bf Proposition 2.1.6.} {\it Let $(V,Y,1,d)$ be a vertex algebra such
that there is a fixed nonnegative integer $k$ such that the
commutativity (2.1.6) holds for any $a,b\in V$. Then $V$ is isomorphic to a
vertex algebra constructed from a commutative associative algebra $A$
with identity together with a derivation.}

{\bf Proof.} We first prove that under the assumption of this
proposition, the commutativity (2.1.6) holds with $k=0$ for any
$a,b\in V$. Let
$r$ be the smallest nonnegative integer such that (2.1.6) holds with
$k=r$ for any $a,b\in V$. If $r>0$, differentiating (2.1.6) with $k=r$
with respect to $x_{1}$ we obtain
\begin{eqnarray}
& &r(x_{1}-x_{2})^{r-1}Y(a,x_{1})Y(b,x_{2})+(x_{1}-x_{2})^{r}
Y(da,x_{1})Y(b,x_{2}) \nonumber\\
&=&r(x_{1}-x_{2})^{r-1}Y(b,x_{2})Y(a,x_{1})+(x_{1}-x_{2})^{r}
Y(b,x_{2})Y(da,x_{1}).
\end{eqnarray}
Then for any $a,b\in V$  we have:
\begin{eqnarray}
(x_{1}-x_{2})^{r-1}Y(a,x_{1})Y(b,x_{2})
=(x_{1}-x_{2})^{r-1}Y(b,x_{2})Y(a,x_{1}).
\end{eqnarray}
This is against the choice of $r$. Therefore the commutativity (2.1.6)
holds with $k=0$ for any $a,b\in V$. In other words,
\begin{eqnarray}
a_{m}b_{n}=b_{n}a_{m}\;\;\;\mbox{ for any }a,b\in V, m,n\in {\bf Z}.
\end{eqnarray}
Next we define a bilinear product $''\cdot''$ on $V$ as follows:
\begin{eqnarray}
a\cdot b=a_{-1}b\;\;\;\mbox{ for any }a,b\in V.
\end{eqnarray}
For any $a,b\in V$, by definition we have:
\begin{eqnarray}
& &a\cdot {\bf 1}=a_{-1}{\bf 1}=a={\bf 1}_{-1}a={\bf 1}\cdot
a;\\
& &a\cdot b=a_{-1}b=a_{-1}b_{-1}{\bf 1}=b_{-1}a_{-1}{\bf 1}=b_{-1}a=b\cdot
a;\\
& &d(a\cdot b)=d(a_{-1}b)=a_{-1}(db)+(da)_{-1}b=a\cdot d(b)+a\cdot d(b).
\end{eqnarray}
Then $(V,\cdot)$ is a commutative algebra with ${\bf 1}$ as its
identity and with $d$ as a derivation. It follows from the definition
of vertex algebra that $Y(a,x){\bf
1}=e^{xd}a$ for any $a\in V$. Then for any $a,b\in V$ we have:
\begin{eqnarray}
Y(a,x)b=Y(a,x)b_{-1}{\bf 1}=b_{-1}Y(a,x){\bf
1}=b_{-1}e^{xd}a=\left(e^{xd}a\right)\cdot b.
\end{eqnarray}
Thus $Y(a,x)$ involves in only nonnegative powers of $x$. This implies
that $Y(a,x_{0}+x_{2})=Y(a,x_{2}+x_{0})$ for any $a\in V$, so that
(from the associator formula (2.1.4))
\begin{eqnarray}
Y(a,x_{0}+x_{2})Y(b,x_{2})c=Y(Y(a,x_{0})b,x_{2})c\;\;\;\mbox{ for any
}a,b,c\in V.
\end{eqnarray}
By considering the constant term of (2.1.25) we obtain the associativity
$a\cdot (b\cdot c)=(a\cdot b)\cdot c$. Therefore $V$ is isomorphic to
the vertex algebra constructed from a commutative associative algebra
with identity together with a derivation.$\;\;\;\;\Box$

{\bf Tensor product vertex algebras 2.1.7.}
Let $(V^{1},{\bf 1},d_{1},Y_{1})$ and $(V^{2},{\bf 1},d_{2},Y_{2})$ be two
vertex algebras. Then it is well known ([B], [FHL]) that $V^{1}\otimes
V^{2}$ has a vertex algebra
structure where $Y(u^{1}\otimes u^{2},x)=Y(u^{1},x)\otimes Y(u^{2},x)$
for any $u^{i}\in V^{i}$, and ${\bf 1}={\bf 1}_{V^{1}}\otimes {\bf
1}_{V^{2}}, d=d_{V^{1}}\otimes 1+1\otimes d_{V^{2}}$. (For vertex
operator algebras, this was proved in [FHL] by proving the
rationality, the
commutativity and the
associativity in terms of ``matrix-coefficient.'' For vertex algebras,
this can be easily proved by proving the commutativity and the
associativity without involving matrix-coefficients.)

Similarly,
if $M^{i}$ is a $V^{i}$-module for $i=1,2$, then the tensor product
space $M=M^{1}\otimes M^{2}$ is a module for the tensor product vertex
algebra $V^{1}\otimes V^{2}$ where $Y_{M}(a^{1}\otimes
a^{2},x)=Y_{M^{1}}(a^{1},x)\otimes Y_{M^{2}}(a^{2},x)$ for
 $a^{i}\in V^{i}$ ([B], [FHL]).

\section{Lie algebras associated to a vertex algebra}
In this section, we shall associate  Lie algebras $g_{0}(V)$ and
$g(V)$ to a vertex algebra $V$ and study  certain of their modules.

{\bf Lemma 2.2.1 [B].} {\it Let $(V,Y,{\bf 1},d)$ be any vertex
algebra. Then the quotient space ${\bf g}_{0}(V):=V/dV$ is a Lie
algebra with the bilinear product:
$[\bar{a},\bar{b}]=\overline{a_{0}b}$ for any $a,b\in
V$. Furthermore, any $V$-module $M$ is a ${\bf g}_{0}(V)$-module with
the action given by: $au=a_{0}u$ for any $a\in V,u\in M$. }

{\bf Affinization of vertex algebras 2.2.2.}
Let $V$ be any vertex algebra. Then [B] we have a tensor product vertex algebra
$\hat{V}={\bf C}[t,t^{-1}]\otimes V$, which is called the affinization
of the vertex algebra $V$. (The
affinization of a vertex operator algebra has also been used in [HL].) Set
$\hat{d}={d\over dt}\otimes 1+1\otimes d$. Then from Lemma 2.2.1 ${\bf
g}_{0}(\hat{V})=\hat{V}/\hat{d}\hat{V}$ is a Lie algebra. For any
$m,n\in {\bf Z}, a \in V$, by definition we have:
\begin{eqnarray}
(t^{m}\otimes a)_{n}&=&{\rm Res}_{x}x^{n}Y(t^{m}\otimes a,x)\nonumber\\
&=&{\rm Res}_{x}x^{n}(t+x)^{n}\otimes Y(a,x)\nonumber\\
&=&\sum_{i=0}^{\infty}\left(\begin{array}{c}m\\i\end{array}\right)
t^{m+n-i}\otimes a_{i}.
\end{eqnarray}
Thus
\begin{eqnarray}
[\overline{(t^{m}\otimes a)}, \overline{(t^{n}\otimes b)}]
=\overline{(t^{m}\otimes a)_{0}(t^{n}\otimes b)}
=\sum_{i=0}^{\infty}\left(\begin{array}{c}m\\i\end{array}\right)
\overline{t^{m+n-i}\otimes a_{i}b}
\end{eqnarray}
for any $a,b\in V, m,n\in {\bf Z}$, where ``bar'' denotes the
natural quotient map from $\hat{V}$ to $g_{0}(\hat{V})$. Therefore, we have:

{\bf Proposition 2.2.3.} {\it Let $V$ be any vertex algebra. Then the
quotient space ${\bf g}(V):={\bf g}_{0}(\hat{V})$ is a Lie algebra with
the bilinear operation:}
\begin{eqnarray}
[\overline{t^{m}\otimes a},\overline{t^{n}\otimes b}]
=\sum_{i=0}^{\infty}\left(\begin{array}{c}m\\i
\end{array}\right)\overline{t^{m+n-i}\otimes a_{i}b}.
\end{eqnarray}

{\bf Remark 2.2.4.} Proposition 2.2.3 can
also be directly proved by checking the skew-symmetry and the Jacobi
identity for a Lie algebra. The current
approach is suggested by Professor James Lepowsky. In [FFR], it was
showed that there was a Lie algebra structure on $\hat{V}/\hat{d}\hat{V}$. But
it seems that the existence of a Lie algebra structure on
$\hat{V}/\hat{d}\hat{V}$ does not obviously follow from their argument.

{\bf Remark 2.2.5.} It follows from Proposition 2.2.3 that $g(V)$ is
exactly the Lie algebra with generators $a(n)$ (linear in $a$) for
$a\in V,n\in {\bf Z}$ and with defining relations:
\begin{eqnarray}
& &{\bf 1}(n)=0\;\;\;\;\mbox{if }n\ne -1,\\
& &(da)(n)=-na(n-1),\\
& &[a(m),b(n)]=\sum_{i=0}^{\infty}\left(\begin{array}{c}m\\i\end{array}\right)
(a_{i}b)(m+n-i).
\end{eqnarray}

It is clear that ${\bf 1}(-1)$ is a central
element of $g(V)$. If ${\bf 1}(-1)$ acts as a scalar $k$ on a
$g(V)$-module $M$, we call $M$ a $g(V)$-module of {\it level $k$}.
(This corresponds to level for affine Lie algebras.) It is clear that
any $V$-module $M$ is a $g(V)$-module of level one, where $a(m)$ is
represented by $a_{n}$. It follows from (2.2.5) that
$\overline{t^{-n-1}\otimes a}=\frac{1}{n!}\overline{t^{-1}\otimes
d^{n}a}$ for any $n\in {\bf N}$. It also follows from  (2.2.5) that
$\overline{t^{n}\otimes a}=0$ for any $n\in {\bf N}$ if $da=0$.
Let $U$ be a subspace of $V$ such that $V=\ker
d\oplus U$. Since
$\overline{t^{n}\otimes u}=-\frac{1}{n+1}\overline{t^{n+1}\otimes du}$  for
any $n\in {\bf N}, u\in U$, we may consider
$\overline{t^{n}\otimes U}$ as a subspace of $\overline{t^{n+1}\otimes
U}$. Let $\bar{U}$ be the union of all $\overline{t^{n}\otimes U}$
for $n\ge 0$. Then it is easy to see that $g(V)$ is linearly
isomorphic to $\bar{U}\oplus \overline{t^{-1}\otimes V}$.

To summarize, for any vertex algebra $V$ we have two Lie algebras
$g_{0}(V)$ and $g(V)$ which are related by the following inclusion
relations:
\begin{eqnarray}
g_{0}(V)\subseteq g(V)\simeq g_{0}(\hat{V})\subseteq g(\hat{V})\subseteq\cdots.
\end{eqnarray}

For any $V$-module $M$, since
$\hat{M}={\bf C}[t,t^{-1}]\otimes M$ is a $\hat{V}$-module,
it follows from Lemma 2.2.1 that $\hat{M}$ is a
$g_{0}(\hat{V})$-module. By definition $\hat{M}$ is
a ${\bf g}(V)$-module of level zero with the action given by:
\begin{eqnarray}
a(m)(t^{n}\otimes u)
=\sum_{i=0}^{\infty}\left(\begin{array}{c}m\\i\end{array}\right)
t^{m+n-i}\otimes a_{i}u.
\end{eqnarray}
Even more, we have:

{\bf Lemma 2.2.6.} {\it Let $V$ be a vertex algebra, $M$ be a $V$-module
and let $z$ be any nonzero complex number. For any $a\in V, u\in M,
m,n\in {\bf Z}$, we define}
\begin{eqnarray}
a(m)(t^{n}\otimes u)=\sum_{i=0}^{\infty}\left(\begin{array}{c}m\\i\end{array}
\right)z^{m-i}(t^{m-i}\otimes a_{i}u).
\end{eqnarray}
{\it Then this defines a $g(V)$-module (of level zero) structure on
$\hat{M}={\bf C}[t,t^{-1}]\otimes M$.}

{\bf Proof.} Let $\psi$ be the automorphism of the associative algebra
${\bf C}[t,t^{-1}]$ defined by: $\psi(f(t))=f(zt)$ for $f(t)\in {\bf
C}[t,t^{-1}]$. Given any module $U$ for the associative algebra ${\bf
C}[t,t^{-1}]$, we
denote by $U^{\psi}$ the $\psi$-twisted module for the associative
algebra ${\bf C}[t,t^{-1}]$. By Examples 2.1.5 $U^{\psi}$
is a module for the vertex algebra ${\bf C}[t,t^{-1}]$  with the
action given by:
\begin{eqnarray}
Y(f(t),x)u=\psi\left(e^{x{d\over dt}}f(t)\right)u=f(zt+x)u
\end{eqnarray}
for $f(t)\in {\bf C}[t,t^{-1}], u\in U$. In particular, ${\bf
C}[t,t^{-1}]^{\psi}$ is a module for the vertex
algebra ${\bf C}[t,t^{-1}]$. Then ${\bf C}[t,t^{-1}]^{\psi}\otimes M$
is a $\hat{V}$-module, so that it is a $g(V)$ ($=g_{0}(\hat{V})$)-module
(of level zero). Then the lemma follows immediately.$\;\;\;\;\Box$

Let $V$ be a graded vertex algebra. For each homogeneous element $a\in
V$ and for any $n\in {\bf Z}$, we define
\begin{eqnarray}
{\rm deg}\:a(n)=\deg\: (t^{n}\otimes a)={\rm wt a}-n-1.
\end{eqnarray}
Then $g(V)$ becomes
a ${\bf Z}$-graded Lie algebra. We write the homogeneous decomposition
as $g(V)=\oplus_{n\in {\bf Z}}g(V)_{n}$. Set
$\displaystyle{g(V)_{\pm}=\sum_{n=1}^{\infty}g(V)_{\pm n}}$.
For any $g(V)$-module $M$, we define
\begin{eqnarray}
\Omega (M)=\{u\in M|g(V)_{-}u=0\}.
\end{eqnarray}
Then it is clear that $\Omega (M)$ is stable under the action of
the ``parabolic subalgebra'' $P=g(V)_{-}\oplus g(V)_{0}$. Then  $\Omega
(M)$ is a $g(V)_{0}$-submodule of $M$.
{}From the classical Lie algebra theory, there is a one-one
correspondence between the set of equivalence
classes of irreducible lower-truncated ${\bf Z}$-graded
$g(V)$-modules and the set of equivalence classes of irreducible
$g(V)_{0}$-modules.

For homogeneous elements $a,b\in V$, we define
\begin{eqnarray}[a,b]=\sum_{i=0}^{\infty}\left(
\begin{array}{c}{\rm wt}a-1\\i\end{array}\right)a_{i}b.
\end{eqnarray}
Extending $[,]$ bilinearly to $V\times V$, we obtain a bilinear product
on $V$.

{\bf Proposition 2.2.7.} {\it The quotient space $V/(d+d(0))V$ is a Lie
algebra with the bilinear operation $[,]$ induced by the defined
bilinear operation on $V\times V$, which is isomorphic to the Lie
algebra $g(V)_{0}$.}

{\bf Remark 2.2.8.} The Lie algebra $g(V)_{0}$ is exactly the Lie
algebra with generating space $V$ and with defining relations (2.2.13) and
\begin{eqnarray}
[\omega, a]=0\;\;\;\mbox{or }(L(-1)+L(0))a=0\;\;\;\mbox{for any }a\in V.
\end{eqnarray}

{\bf Remark 2.2.9.} It follows from Lemma 2.2.1 that
$g_{0}(V)=V/L(-1)V$ has a Lie algebra
structure for any vertex operator algebra $V$, which is a subalgebra
of $g(V)$. It is interesting to notice that $g(V)_{0}$ is isomorphic to
$g_{0}(V)$. Following [Zhu], we define
\begin{eqnarray}
Y[a,x]=Y(a,e^{x}-1)e^{x{\rm wt} a}\;\;\;\mbox{for each homogeneous
element }a\in V.
\end{eqnarray}
Then Zhu proved that $(V,{\bf 1},\omega-{c\over 24},Y[\cdot,x])$ is a
vertex operator algebra isomorphic to $V$. Under this new V.O.A
structure, $\tilde{L}(-1)=L(-1)+L(0)$ and
\begin{eqnarray}
 a[0]b&=&{\rm Res}_{}(1+x)^{{\rm wt}a-1}Y(a,x)b\nonumber\\
&=&\sum_{i=0}^{\infty}\left(\begin{array}{c}{\rm wt}a-1\\i\end{array}\right)
a_{i}b.
\end{eqnarray}

Let $U$ be any $g(V)_{0}$-module. Considering $U$ as a $P$-module, we
have the induced module $U({\bf g})\otimes _{U(P)}U$. Following
Lepowsky [Lep], we call this module the
generalized Verma $g(V)$-module with lowest-degree subspace $U$ and
denote this module by $M(U)$. Let $M$ be any nonzero lower-truncated
${\bf Z}$-graded $g(V)$-module with lowest-degree $k$. Let
$J(M)$ be the maximal graded submodule of $M$ such that
$M(k)\cap J=0$. Set $S(M)=M/J(M)$. There is another way [Zhu] to describe
$J(M)$ as follows: Let $M(k)^{*}$ be the dual vector space and let $p$ be
the projection of $M$ on $M(k)$ with respect to the ${\bf Z}$-grading
decomposition. Define a bilinear form on $M(k)^{*}\times M$ as follows
\begin{eqnarray}
\langle u',v\rangle=u'(p(u))\;\;\;\mbox{for any }u'\in M(k)^{*},v\in M.
\end{eqnarray}
Then
\begin{eqnarray}
J(M)=\{v\in M|\langle u',yv\rangle=0\;\mbox{for any }u'\in
M(0)^{*},y\in U({\bf g})\}.
\end{eqnarray}
By definition, $S(M)$ is a lower-truncated ${\bf Z}$-graded
$g(V)$-module with  radical being zero.

{\bf Evaluation $g(V)$-modules $M_{z}$ 2.2.10.}
Let $M$ be a $V$-module. Then it is clear that $M$ is a $g(V)$-module
of level one where $a(n)$ acts on $M$ as $a_{n}$ for any $a\in V,n\in
{\bf Z}$. For any nonzero complex number $z$, let ${\bf C}_{z}$ be the
evaluation module for the associative algebra ${\bf C}[t,t^{-1}]$ with
$t$ acting as a scalar $z$.
Then from Examples 2.1.5 ${\bf C}_{z}$ is a module for vertex algebra
${\bf C}[t,t^{-1}]$,
so that ${\bf C}_{z}\otimes M$ is a $\hat{V}$-module (where
$\hat{V}$ is the affinization of $V$). Therefore ${\bf C}_{z}\otimes M$
is a $g(V)=g_{0}(\hat{V})$-module (by Lemma 2.2.1). From (2.2.1) we have:
\begin{eqnarray}
a(m)\cdot (1\otimes u)=\sum_{i=0}^{\infty}\left(\begin{array}{c}m\\i
\end{array}\right)z^{m-i}(1\otimes a_{i}u)\;\;\;\mbox{for }a\in V,u\in M.
\end{eqnarray}
Therefore ${\bf C}_{z}\otimes M$ becomes a $g(V)$-module of level
zero. Denote this
$g(V)$-module by $M_{z}$ (for generic $z$). If $z=\alpha$ is a
specific complex number, we use the notation $M_{z=\alpha}$.

Notice that
$\displaystyle{\sum_{i=0}^{\infty}\left(\begin{array}{c}m\\i\end{array}
\right)z^{m-i}a(i)}$
is an infinite sum in $g(V)$ although it is a finite sum in $M$ after
applied to each vector $u$ of $M$. Due to this reason, we may consider
a certain  completion of $g(V)$. By considering the tensor product
vertex algebra ${\bf C}((t))\otimes V$ we obtain a Lie algebra
$g_{0}({\bf C}((t))\otimes V)$ (from Lemma 2.2.1). It is clear that this
Lie algebra is the completion of $g(V)$ with respect to a certain
topology for $g(V)$. We denote this Lie algebra by $\bar{g}(V)$.

For any
$f(t)=\sum_{m\ge K}c_{m}t^{m}\in {\bf C}((t))$, since the
following sum:
\begin{eqnarray}
\sum_{m\ge K}c_{m}\left(\sum_{i=0}^{\infty}
\left(\begin{array}{c}m\\i\end{array}\right)z^{m-i}t^{i}\right)
=\sum_{i=0}^{\infty}\left(\sum_{m\ge K}\left(\begin{array}{c}m\\i\end{array}
\right)c_{m}z^{m-i}\right)t^{i}
\end{eqnarray}
may not be a well-defined element of ${\bf C}((t))$, we cannot extend
an evaluation $g(V)$-module $M_{z}$ to be a $\bar{g}(V)$-module.

Next we define a linear map $\Delta_{z}$ as follows:
\begin{eqnarray}
\Delta_{z}:& &{\bf C}[t,t^{-1}]\otimes V\rightarrow ({\bf C}((t))\otimes
V)\otimes ({\bf C}((t))\otimes V);\nonumber\\
& &f(t)\otimes a\mapsto 1\otimes (f(t)\otimes a)+
(f(z+t)\otimes a)\otimes 1.
\end{eqnarray}

{\bf Proposition 2.2.11.} {\it $\Delta_{z}$ induces
an associative algebra
homomorphism from $U({\bf g}(V))$ to $U(\bar{{\bf g}}(V))\otimes
U(\bar{{\bf g}}(V))$.}

{\bf Proof.} Define the linear map $\Delta_{z}^{1}$ from
${\bf C}[t,t^{-1}]\otimes V$ to ${\bf C}((t))\otimes V$ as follows:
\begin{eqnarray}
\Delta_{z}^{1}(f(t)\otimes a)=f(z+t)\otimes a\;\;\;\mbox{for }f(t)\in
{\bf C}[t,t^{-1}],a\in V.
\end{eqnarray}
Then $\Delta_{z}=\Delta_{z}^{1}\otimes 1+1\otimes {\rm id}$. Therefore
it suffices to prove that $\Delta_{z}^{1}$ induces a Lie algebra
homomorphism from $g(V)$ to $\bar{g}(V)$. Let $\psi_{z}$ be the
algebra homomorphism from ${\bf C}[t,t^{-1}]$ to ${\bf C}((t))$
defined by: $\psi_{z}(f(t))=f(z+t)$ for $f(t)\in {\bf C}((t))$. From
Examples 2.1.5 $\psi_{z}$ is a vertex algebra homomorphism from ${\bf
C}[t,t^{-1}]$ to ${\bf C}((t))$, so that $\psi_{z}\otimes {\rm id}$ is
a vertex algebra homomorphism from ${\bf C}[t,t^{-1}]$ to ${\bf
C}((t))\otimes V$. By definition $\Delta_{z}^{1}=\psi_{z}\otimes {\rm
id}$. Therefore $\Delta_{z}^{1}$ induces a Lie algebra homomorphism from
$g(V)$ to $\bar{g}(V)$. $\;\;\;\;\Box$

{\bf Remark 2.2.12.} The Hopf-like algebra $(U(g(V)),U(\bar{{\bf g}}(V)),
\Delta_{z})$ is implicitly used in the construction of tensor product
for the representations of the vertex operator algebra $V$ (see [HL], [KL]
or [MS] for example).

{\bf Remark 2.2.13.} For any Lie algebra ${\bf g}$, $-{\rm id}$ is an
anti-automorphism. Let $\sigma$ be the corresponding
anti-automorphism of $U({\bf g}(V))$ for any vertex algebra. Let $V$
be a vertex operator algebra. Then it follows from FHL's
contragredient module theory [FHL] that the linear map
\begin{eqnarray}
 \theta :& &{\bf C}[t,t^{-1}]\otimes V\rightarrow {\bf C}[t,t^{-1}]\otimes V;
\nonumber\\
& &Y_{t}(a,x)\mapsto
Y_{t}(e^{xL(1)}(-x^{-2})^{L(0)}a,x^{-1})\;\;\;\mbox{for }a\in V
\end{eqnarray}
induces an anti-automorphism of ${\bf g}(V)$. It follows from [FHL]
that $\theta ^{2}={\rm id}$.
We will use the same
letter for the anti-automorphism for the Lie algebra ${\bf g}(V)$ and
for the universal enveloping algebra $U({\bf g}(V))$.


\newpage

\chapter{Zhu's $A(V)$-theory}
In [Zhu], Zhu has
constructed an associative
algebra $A(V)$ for each vertex operator algebra $V$ and established a
one-to-one correspondence between the set of equivalence classes of
lower-truncated ${\bf Z}$-graded irreducible weak $V$-modules and the
set of equivalence classes of irreducible $A(V)$-modules. The introduction of
$A(V)$-algebra is both conceptually and practically important in
vertex operator algebra theory.

Roughly speaking, Zhu's algebra $A(V)$ comes from all
weight-zero components of vertex operators.
Philosophically, Zhu's $A(V)$-theory is analogous to the well-known
lowest-weight (or degree) module theory for graded Lie algebras and graded
associative algebras in classical algebra theory. More specifically,
let $g=\oplus_{n\in {\bf Z}}g_{n}$ be a ${\bf Z}$-graded Lie algebra.
For any $g_{0}$-module $U$, using the triangular decomposition
$g=g_{+}\oplus g_{0}\oplus
g_{-}$, where ${\bf g}_{\pm}=\oplus_{n=1}^{\infty}{\bf g}_{\pm n}$,
we have the induced module: $M(U)=U({\bf
g})\otimes_{U(g_{0}+g_{-})}U$, which is namely called the generalized
Verma module (a term introduced by Lepowsky [Lep]).
Then $M(U)$ is a lower-truncated ${\bf Z}$-graded ${\bf g}$-module with
lowest-degree subspace $U$ and $M(U)$ satisfies a certain universal
property. Let ${\bf J}$ be
the maximal graded submodule satisfying the condition: $J\cap U=0$ and let
$L(U)$ be the corresponding quotient module. If $U$ is an irreducible
$g_{0}$-module, $L(U)$ is a graded irreducible module (i.e., there is
no nontrivial graded submodule). Then  there is a 1-1
correspondence between the set of equivalence classes of lower-truncated
${\bf Z}$-graded irreducible $g$-modules and the set of equivalence classes
of irreducible $g_{0}$-modules.

In a parallel way, let $A=\oplus_{n\in {\bf Z}}A_{n}$ be a ${\bf Z}$-graded
associative algebra with identity. Set
$A_{\pm}=\oplus_{n=1}^{\infty}A_{\pm n}$ and
$I_{0}=\sum_{n=1}^{\infty}A_{n}A_{-n}$.
Let  $A^{o}$ the quotient algebra of $A_{0}$ modulo the two-sided ideal
$I_{0}$. Let
$M=\oplus_{n=0}^{\infty}M(n+h)$ be any lower-truncated ${\bf Z}$-graded
$A$-module. Then
$I_{0}M(h)=0$. Therefore, $M(h)$ is an $A^{o}$-module in the natural
way. Conversely, let $U$ be any $A^{o}$-module. Since $A_{-}$ is
an ideal of the subalgebra $A_{0}+A_{-}$, we may naturally consider $U$ as a
module for $A_{0}+A_{-}$.
Form the induced module:
$M(U)=A\otimes_{(A_{0}+A_{-})}U$.
Then one can prove that $M(U)$ is a graded $A$-module with
lowest-degree subspace $U$. Therefore there is a 1-1
correspondence between the set of equivalence classes of lower-truncated
${\bf Z}$-graded irreducible $A$-modules and the set of equivalence classes
of $A^{0}$-modules.

For an $A(V)$-module $U$, to construct a module for the vertex
operator algebra $V$, Zhu first constructs a ``free'' space $F(U)$ generated
by all symbols $a(m)$ (linear in $a\in V$) from $U$. Then he
inductively defines a linear function on
 $U^{*}\otimes_{{\bf C}}F(U)$ and he proves  the
rationality, the commutativity and the associativity in terms of
``matrix-coefficients.'' Then he proves that
a certain quotient space of this free space is a (weak) $V$-module.

{}From Chapter 2 (see also [FFR]) we have a Lie algebra $g(V)$
associated to each vertex
algebra $V$. We also have an associative algebra $U(V)$ constructed by
Frenkel and
Zhu [FZ]. The structure of Lie algebra $g(V)$ is pretty clear, but
the lower-truncated ${\bf Z}$-graded $g(V)$-module theory is not
equivalent to the $V$-module theory because
the commutator relations (2.1.3) (the defining relations of $g(V)$) is
weaker than the Jacobi
identity. On the other hand, Frenkel and Zhu's algebra $U(V)$ [FZ] reflects
the whole Jacobi relations so that the theory of lower-truncated ${\bf
Z}$-graded $U(V)$-modules is equivalent the theory of lower-truncated
${\bf Z}$-graded weak $V$-modules, but $U(V)$ is too abstract to use.

In this chapter by making use of the Lie algebra $g(V)$ we  combine
the formal
variable technique with the  notion of generalized Verma module
to reformulate Zhu's $A(V)$-theory [Zhu]. More specifically, We first
formulate a functor
$\Omega$ from the category of (weak) $V$-modules to the category of
$A(V)$-modules. Then we construct a functor $L$ from the category of
$A(V)$-modules to the category of lower-truncated ${\bf Z}$-graded
(weak) $V$-modules. We also define the notion of ``generalized Verma (weak)
module'' for a vertex operator algebra.

\section{Zhu's $A(V)$-algebra and the functor $\Omega$}
In this section, we shall review Zhu's construction of $A(V)$-algebra
for a vertex operator algebra $V$ and formulate a functor $\Omega$
from the category of weak $V$-modules to the category of $A(V)$-modules.

{\bf Definition 3.1.1}. A {\it vertex operator algebra} is a ${\bf
Z}$-graded vertex algebra $V=\oplus_{n\in {\bf Z}}V_{(n)}$ together with
a distinguished vector $\omega$ (the {\it Virasoro element}) satisfying the
following conditions:

(V6)\hspace{0.25cm} $\dim V_{(n)}<\infty$ for all $n\in {\bf Z}$ and
$V_{(n)}=0$ for $n$ sufficiently small;

(V7)\hspace{0.25cm} The {\it Virasoro algebra relations} hold:
\begin{eqnarray}
[L(m),L(n)]=(m-n)L(m+n)={(m^{3}-m)\over 12}\delta_{m+n,0}({\rm rank}V),
\end{eqnarray}
where $Y(\omega,x)=\sum_{n\in {\bf Z}}L(n)x^{-n-2}$ and ${\rm rank}V$
is a constant called the {\it rank} of $V$;

(V8)\hspace{0.25cm} $L(0)a=na=({\rm wt}a)a$ for $a\in V_{(n)}$;

(V9)\hspace{0.25cm} $Y(L(-1)a,z)=\displaystyle{{d\over dx}}Y(a,x)$ for
any $a\in V$.

This completes the definition of vertex operator algebra.

{\bf Definition 3.1.2.} Let $V$ be a vertex operator algebra. A module
for $V$ viewed as a vertex algebra is called a {\it weak} module for $V$
as a vertex operator algebra. A lower truncated ${\bf Z}$-{\it
graded} weak $V$-module is a weak
$V$-module $M$ such there there exists a ${\bf Z}$-graded decomposition
$M=\oplus_{n\in {\bf Z}}M(n)$ satisfying the following conditions:
\begin{eqnarray}
& &M(n)=0\;\;\;\mbox{for }n \mbox{ suffiently small};\\
& &a_{m}M(n)\subseteq M(n+r-m-1)\;\;\;\mbox{for }a
\in V_{(r)},m,n,r\in {\bf Z}.
\end{eqnarray}
(This notion is a variant of Frenkel and Zhu's notion of ${\bf
N}$-graded module [FZ], where ${\bf N}$ is the set of all nonnegative
integers. Note that our definition allows grading-shifts.)

A {\it module} for a vertex operator
algebra $V$ is a weak $V$-module $M$ together with a ${\bf C}$-graded
decomposition  $M=\oplus_{\alpha\in {\bf
C}}M_{(\alpha)}$ satisfying the following conditions:

(M5)\hspace{0.25cm} $L(0)u=hu$ for any $u\in M_{(h)}, h\in {\bf C}$.

(M6)\hspace{0.25cm} $\dim M_{(\alpha)}<\infty$ for any $\alpha\in {\bf C}$ and
$M_{(\alpha+n)}=0$ for $n\in {\bf Z}$ sufficiently small.

If $M$ satisfies all
the conditions except (M6), we call $M$ a {\it generalized} $V$-module
(a notion introduced by Huang and Lepowsky [HL0]).

{}From Chapter 2 we have a Lie algebra $g(V)_{0}=V/(L(-1)+L(0))V$
from all weight-zero components of vertex operators, which is exactly
the Lie algebra with generating space $V$ and with certain
defining relations (Remark 2.2.8).

To get a product for an associative algebra, we naturally come
to the associativity for vertex operator algebras. Let $a$ and $b$ be two
homogeneous elements of $V$ and let $u$ be a vector of the lowest
weight subspace of a $V$-module $M$. Then we have
\begin{eqnarray}
& &x_{0}^{-1}\delta\left(\frac{x_{1}-x_{2}}{x_{0}}\right)Y(a,x_{1})Y(b,x_{2})u
-x_{0}^{-1}\delta\left(\frac{-x_{2}+x_{1}}{x_{0}}\right)Y(b,x_{2})Y(a,x_{1})u
\nonumber\\
&=&x_{2}^{-1}\delta\left(\frac{x_{1}-x_{0}}{x_{2}}\right)Y(Y(a,x_{0})b,x_{2})u.
\end{eqnarray}
In order to make the second term vanish to get an associative formula,
we take ${\rm Res}_{x_{1}}x_{1}^{{\rm wt }a}$ of the Jacobi
identity (3.1.4). Then we obtain
\begin{eqnarray}
& &{\rm Res}_{x_{1}}x_{0}^{-1}\delta\left(\frac{x_{1}-x_{2}}{x_{0}}\right)
x_{1}^{{\rm wt }a}Y(a,x_{1})Y(b,x_{2})u\nonumber\\
&=&(x_{2}+x_{0})^{{\rm wt }a}Y(Y(a,x_{0})b,x_{2})u,
\end{eqnarray}
or, equivalently
\begin{eqnarray}
(x_{0}+x_{2})^{{\rm wt }a}Y(a,x_{0}+x_{2})Y(b,x_{2})u
=(x_{2}+x_{0})^{{\rm wt }a}Y(Y(a,x_{0})b,x_{2})u.
\end{eqnarray}
In order to make the left side term be $a_{{\rm wt }a-1}b_{{\rm wt
}b-1}u$, we take ${\rm Res}_{x_{0}}{\rm
Res}_{x_{2}}x_{0}^{-1}x_{2}^{{\rm wt }b-1}$ of the formula (3.1.5). Then
we obtain
\begin{eqnarray}
& &a_{{\rm wt }a-1}b_{{\rm wt}b-1}u\nonumber\\
&=&{\rm Res}_{x_{0}}{\rm Res}_{x_{2}}x_{0}^{-1}x_{2}^{{\rm
wt}b-1}Y(Y(a,x_{0})b,x_{2})u\nonumber\\
&=&{\rm Res}_{x_{2}}\sum_{i=0}^{\infty}\left(\begin{array}{c}{\rm
wt}a\\i\end{array}\right)x_{2}^{{\rm wt}b-1}Y(a_{i-1}b,x_{2})u\nonumber\\
&=&{\rm Res}_{x}\frac{(1+x)^{{\rm wt}a}}{x}o(Y(a,x)b)u,
\end{eqnarray}
where $o(a)=a_{{\rm wt}a-1}$. Then
\begin{eqnarray}
o(a)o(b)={\rm Res}_{x}\frac{(1+x)^{{\rm wt}a}}{x}o(Y(a,x)b).
\end{eqnarray}
This gives the product formula for Zhu's algebra $A(V)$. Furthermore,
we can easily get
\begin{eqnarray}
o(L(-1)a+L(0)a)o(b)={\rm Res}_{x}\frac{(1+x)^{{\rm wt}a}}{x^{2}}o(Y(a,x)b).
\end{eqnarray}
Since
\begin{eqnarray}
o(L(-1)a+L(0)a)=0\;\;\;\mbox{ for any }a\in V,
\end{eqnarray}
this gives a necessary relation for $A(V)$.

Now let us start Zhu's construction of $A(V)$. Let $V$ be a vertex operator
algebra. For any homogeneous
element $a\in V$ and for any $b\in V$, following [Zhu] we define
\begin{eqnarray}
a* b={\rm Res}_{x}\frac{(1+x)^{{\rm wt a}}}{x}Y(a,x)b.
\end{eqnarray}
Then extend this product bilinearly to the whole space $V$. Let $O(V)$
be the subspace of $V$ linearly spanned by the elements of type
\begin{eqnarray}
{\rm Res}_{x}\frac{(1+x)^{{\rm wt a}}}{x^{2}}Y(a,x)b\;\;\;\mbox{for
homogeneous elements }a,b\in V.
\end{eqnarray}
Set $A(V)=V/O(V)$. For any weak $V$-module $M$ we define
\begin{eqnarray}
\Omega (M)=\{u\in M|g(V)_{-}u=0\}.
\end{eqnarray}

{\bf Theorem 3.1.3 [Zhu].}  a) {\it The defined bilinear operation $*$ on $V$
 induces
a bilinear operation on $A(V)$ such that $A(V)$ is an associative
algebra with identity ${\bf 1}$ and with $\omega$ as a central element.}

b) {\it There is an anti-automorphism $\phi$ of $A(V)$ such that $\phi
(a)=e^{L(1)}(-1)^{L(0)}a$.}

c) {\it For any weak $V$-module $M$, $\Omega (M)$ is an $A(V)$-module.}

Then we have a functor $\Omega$ from the category of weak $V$-modules to the
category of $A(V)$-modules.
Since $(L(-1)+L(0))V\subseteq O(V)$, $A(V)$ is a quotient vector space
of $g(V)_{0}$. Then $\Omega (M)$ is an $A(V)$-module and also a
$g(V)_{0}$-module.

As a direct consequence of Zhu's theorem we have:

{\bf Corollary 3.1.4.} {\it Zhu's algebra $A(V)$ is the associative
algebra with underlying
space $V$ and with (3.1.10) and (3.1.11) providing its defining relations .}

\section{The functor $L$, the generalized Verma modules and Zhu's 1-1
correspondence}
In this section,
we shall combine the formal variable technique with the notion
of generalized Verma module for Lie algebras to give
a slightly different construction of a lower-truncated (weak)
$V$-module from an $A(V)$-module. This construction gives rise to the
functor $L$. We also define a notion of
``generalized Verma (weak) $V$-module.''

For an $A(V)$-module $U$, we first view $U$ as a ${\bf
g}(V)_{0}$-module through the homomorphism
from ${\bf g}(V)_{0}$ to $A(V)_{Lie}$. Then we
construct a generalized Verma $g(V)$-module $M(U)$, which is a
lower-truncated ${\bf Z}$-graded module
with $U$ as its lowest-degree homogeneous subspace. Then the
commutativity ``without involving matrix-coefficients''  and
the $L(-1)$-derivative property automatically hold. Next we consider
the quotient $g(V)$-module $L(U)$ of $M(U)$ by modulo the maximal graded
submodule $J$ such that $J\cap U=0$. From our
analysis in Section 3.1 for the product formula of $A(V)$, $A(V)$-module
structure on $U$
is an associative relation. Using this initial data and an induction we
prove the associativity for vertex operators acting on the whole space
$L(U)$. Combining the commutativity with the associativity we obtain
the Jacobi identity, so that $L(U)$ is a (weak) $V$-module.

 Throughout this section, $V$ will be a fixed vertex operator algebra
and $g=g_{+}\oplus g_{0}\oplus g_{-}$ will be the corresponding Lie algebra.

{\bf Lemma 3.2.1 } {\it Let $A(V)_{Lie}$ be the Lie algebra of $A(V)$.
Then the identity endomorphism of
$V$ induces a Lie algebra homomorphism from $g(V)_{0}$ onto
$A(V)_{Lie}$. }

{\bf Proof.} From Lemma 2.1.3 [Zhu], for any $a,b\in V$ we have:
\begin{eqnarray}
a*b -b*a={\rm Res}_{x}(1+x)^{{\rm wt}a-1}Y(a,x)b \;\;\;mod\; O(V).
\end{eqnarray}
Then $a*b -b*a=[a,b]\;\;\;mod\; O(V)$.$\;\;\;\;\Box$

By Lemma 3.2.1, for any $A(V)$-module $U$, we may naturally
consider $U$ as a Lie algebra $g(V)_{0}$-module. Furthermore, we may
naturally consider $U$ as a module for $P=(g(V)_{0}\oplus g(V)_{-})$.
let $M(U)=U(g(V))\otimes_{U(P)} U$ be the generalized Verma module.
Then $M(U)=U(g_{+})\otimes U=\oplus_{m=0}^{\infty}U(g_{+})_{m}U$, so
that we may consider $M(U)$ as a ${\bf Z}$-graded $g$-module with $U$
as its lowest-degree subspace.
Let $J$ be the maximal
graded submodule such that $J\cap U=0$ and let $L(U)$
be the corresponding quotient module of $M(U)$.

{\bf Proposition 3.2.2.} {\it For any homogeneous elements $a,b\in V$ and for
any $u'\in U^{*}, u\in U,i\in {\bf Z}_{+}$, we have:}
\begin{eqnarray}
\langle u',(x_{0}+x_{2})^{{\rm wt }a+i}Y(a,x_{0}+x_{2})Y(b,x_{2})u\rangle
=\langle u',(x_{2}+x_{0})^{{\rm wt }a+i}Y(Y(a,x_{0})b,x_{2})u\rangle.
\nonumber\\
& &\mbox{}
\end{eqnarray}

{\bf Proof.} By the analysis in Section 3.1, we see that
the $A(V)$-module structure on $U$ is equivalent to
\begin{eqnarray}
& &{\rm Res}_{x_{0}}{\rm
Res}_{x_{2}}x_{0}^{-1}(x_{0}+x_{2})^{{\rm wt }a}x_{2}^{{\rm wt }b-1}
Y(a,x_{0}+x_{2})Y(b,x_{2})u\nonumber\\
&=&{\rm Res}_{x_{0}}{\rm
Res}_{x_{2}}x_{0}^{-1}(x_{0}+x_{2})^{{\rm wt }a}x_{2}^{{\rm wt }b-1}
Y(Y(a,x_{0})b,x_{2})u.
\end{eqnarray}
Since $\langle u',L(U)(n)\rangle=0$ for $n\ne 0$, we have:
\begin{eqnarray}
& &{\rm Res}_{x_{0}}x_{0}^{-1}(x_{0}+x_{2})^{{\rm wt }a}
\langle u',Y(a,x_{0}+x_{2})Y(b,x_{2})u\rangle \nonumber\\
&=&{\rm Res}_{x_{0}}
x_{0}^{-1}(x_{0}+x_{2})^{{\rm wt }a}
\langle u',Y(Y(a,x_{0})b,x_{2})u\rangle.
\end{eqnarray}
Since $L(U)$ is a $g$-module, for any $i,j\in {\bf Z}_{+}$,
by the commutator formula we have:
\begin{eqnarray}
& &{\rm Res}_{x_{0}}x_{0}^{i}(x_{0}+x_{2})^{{\rm wt} a+j}
Y(a,x_{0}+x_{2})Y(b,x_{2})u\nonumber\\
&=&{\rm Res}_{x_{0}}{\rm Res}_{x_{1}}x_{0}^{i}x_{1}^{{\rm wt} a+j}
x_{0}^{-1}\delta\left(\frac{x_{1}-x_{2}}{x_{0}}\right)
Y(a,x_{1})Y(b,x_{2})u\nonumber\\
&=&{\rm Res}_{x_{0}}{\rm Res}_{x_{1}}x_{0}^{i}x_{1}^{{\rm wt} a+j}
x_{0}^{-1}\delta\left(\frac{x_{1}-x_{2}}{x_{0}}\right)
Y(a,x_{1})Y(b,x_{2})u\nonumber\\
& &-{\rm Res}_{x_{0}}{\rm Res}_{x_{1}}x_{0}^{i}x_{1}^{{\rm wt} a+j}
x_{0}^{-1}\delta\left(\frac{-x_{2}+x_{1}}{x_{0}}\right)
Y(b,x_{2})Y(a,x_{1})u\nonumber\\
&=&{\rm Res}_{x_{1}}(x_{1}-x_{2})^{i}x_{1}^{{\rm wt} a+j}
[Y(a,x_{1}),Y(b,x_{2})]u\nonumber\\
&=&{\rm Res}_{x_{0}}{\rm Res}_{x_{1}}x_{0}^{i}x_{1}^{{\rm wt} a+j}
x_{0}^{-1}\delta\left(\frac{x_{1}-x_{0}}{x_{2}}\right)Y(Y(a,x_{0})b,x_{2})u
\nonumber\\
&=&{\rm Res}_{x_{0}}{\rm Res}_{x_{1}}x_{0}^{i}x_{1}^{{\rm wt} a+j}
x_{1}^{-1}\delta\left(\frac{x_{2}+x_{0}}{x_{1}}\right)Y(Y(a,x_{0})b,x_{2})u
\nonumber\\
&=&{\rm Res}_{x_{0}}x_{0}^{i}(x_{2}+x_{0})^{{\rm wt} a+i}
Y(Y(a,x_{0})b,x_{2})u.
\end{eqnarray}
Since
\begin{eqnarray}
& &{\rm Res}_{x_{0}}x_{0}^{-1}(x_{0}+x_{2})^{{\rm wt} a+j}
\langle u',Y(a,x_{0}+x_{2})Y(b,x_{2})u\rangle\nonumber\\
&=&\sum_{r=0}^{\infty}\left(\begin{array}{c}j\\r\end{array}\right)
x_{0}^{r-1}x_{2}^{j-r}(x_{0}+x_{2})^{{\rm wt} a}
\langle u',Y(a,x_{0}+x_{2})Y(b,x_{2})u\rangle\nonumber\\
&=&\sum_{r=0}^{\infty}\left(\begin{array}{c}j\\r\end{array}\right)
x_{0}^{r-1}x_{2}^{j-r}(x_{2}+x_{0})^{{\rm wt} a}
\langle u',Y(Y(a,x_{0})b,x_{2})u\rangle\nonumber\\
&=&{\rm Res}_{x_{0}}x_{0}^{-1}(x_{2}+x_{0})^{{\rm wt} a+j}
\langle u',Y(Y(a,x_{0})b,x_{2})u\rangle,
\end{eqnarray}
then for any $i,j\in {\bf Z}_{+}$ we have:
\begin{eqnarray}
& &{\rm Res}_{x_{0}}x_{0}^{-1+i}(x_{0}+x_{2})^{{\rm wt} a+j}
\langle u',Y(a,x_{0}+x_{2})Y(b,x_{2})u\rangle\nonumber\\
&=&{\rm Res}_{x_{0}}x_{0}^{-1+i}(x_{2}+x_{0})^{{\rm wt} a+j}
\langle u',Y(Y(a,x_{0})b,x_{2})u\rangle.
\end{eqnarray}
Suppose that $k$ is a positive integer such that
\begin{eqnarray}
& &{\rm Res}_{x_{0}}x_{0}^{n}(x_{0}+x_{2})^{{\rm wt} a+j}
\langle u',Y(a,x_{0}+x_{2})Y(b,x_{2})u\rangle\nonumber\\
&=&{\rm Res}_{x_{0}}x_{0}^{n}(x_{2}+x_{0})^{{\rm wt} a+j}
\langle u',Y(Y(a,x_{0})b,x_{2})u\rangle
\end{eqnarray}
for any $a,b\in V, u'\in U^{*},u\in U,j\in {\bf Z}_{+}$ and for any $n\ge -k$.
Then
\begin{eqnarray}
& &{\rm Res}_{x_{0}}x_{0}^{-k}(x_{0}+x_{2})^{{\rm wt} a+j+1}
\langle u',Y(L(-1)a,x_{0}+x_{2})Y(b,x_{2})u\rangle\nonumber\\
&=&{\rm Res}_{x_{0}}x_{0}^{-k}(x_{2}+x_{0})^{{\rm wt} a+j+1}
\langle u',Y(Y(L(-1)a,x_{0})b,x_{2})u\rangle.
\end{eqnarray}
Since
\begin{eqnarray}
& &{\rm Res}_{x_{0}}x_{0}^{-k}(x_{0}+x_{2})^{{\rm wt} a+j+1}
\langle u',Y(L(-1)a,x_{0}+x_{2})Y(b,x_{2})u\rangle\nonumber\\
&=&-{\rm Res}_{x_{0}}\left({\partial\over\partial x_{0}}
x_{0}^{-k}(x_{0}+x_{2})^{{\rm wt} a+j+1}\right)
\langle u',Y(a,x_{0}+x_{2})Y(b,x_{2})u\rangle\nonumber\\
&=&{\rm Res}_{x_{0}}k
x_{0}^{-k-1}(x_{0}+x_{2})^{{\rm wt}a+j+1}
\langle u',Y(a,x_{0}+x_{2})Y(b,x_{2})u\rangle\nonumber\\
& &-{\rm Res}_{x_{0}}({\rm wt} a+j+1)x_{0}^{-k}(x_{0}+x_{2})^{{\rm wt} a+j}
\langle u',Y(a,x_{0}+x_{2})Y(b,x_{2})u\rangle\nonumber\\
&=&{\rm Res}_{x_{0}}k
x_{0}^{-k-1}x_{2}(x_{0}+x_{2})^{{\rm wt} a+j}
\langle u',Y(a,x_{0}+x_{2})Y(b,x_{2})u\rangle\nonumber\\
& &+{\rm Res}_{x_{0}}k
x_{0}^{-k}x_{2}(x_{0}+x_{2})^{{\rm wt} a+j}
\langle u',Y(a,x_{0}+x_{2})Y(b,x_{2})u\rangle\nonumber\\
& &-{\rm Res}_{x_{0}}({\rm wt} a+j+1)x_{0}^{-k}(x_{2}+x_{0})^{{\rm wt} a+j}
\langle u',Y(Y(a,x_{0})b,x_{2})u\rangle\nonumber\\
&=&{\rm Res}_{x_{0}}k
x_{0}^{-k-1}x_{2}(x_{0}+x_{2})^{{\rm wt} a+j}
\langle u',Y(a,x_{0}+x_{2})Y(b,x_{2})u\rangle\nonumber\\
& &+{\rm Res}_{x_{0}}k
x_{0}^{-k}x_{2}(x_{2}+x_{0})^{{\rm wt} a+j}
\langle u',Y(Y(a,x_{0})b,x_{2})u\rangle\nonumber\\
& &-{\rm Res}_{x_{0}}({\rm wt} a+j+1)x_{0}^{-k}(x_{2}+x_{0})^{{\rm wt} a+j}
\langle u',Y(Y(a,x_{0})b,x_{2})u\rangle,
\end{eqnarray}
and
\begin{eqnarray}
& &{\rm Res}_{x_{0}}x_{0}^{-k}(x_{2}+x_{0})^{{\rm wt} a+j+1}
\langle u',Y(Y(L(-1)a,x_{0})b,x_{2})u\rangle\nonumber\\
&=&-{\rm Res}_{x_{0}}\left({\partial\over\partial x_{0}}x_{0}^{-k}
(x_{2}+x_{0})^{{\rm wt} a+j+1}\right)\langle u',Y(Y(a,x_{0})b,x_{2})u\rangle
\nonumber\\
&=&{\rm Res}_{x_{0}}kx_{0}^{-k-1}(x_{2}+x_{0})^{{\rm wt} a+j+1}
\langle u',Y(Y(a,x_{0})b,x_{2})u\rangle\nonumber\\
& &-{\rm Res}_{x_{0}}({\rm wt} a+j+1)x_{0}^{-k}(x_{2}+x_{0})^{{\rm wt} a+j}
\langle u',Y(Y(a,x_{0})b,x_{2})u\rangle\nonumber\\
&=&{\rm Res}_{x_{0}}kx_{2}x_{0}^{-k-1}(x_{2}+x_{0})^{{\rm wt} a+j}
\langle u',Y(Y(a,x_{0})b,x_{2})u\rangle\nonumber\\
& &+{\rm Res}_{x_{0}}kx_{0}^{-k}(x_{2}+x_{0})^{{\rm wt} a+j}
\langle u',Y(Y(a,x_{0})b,x_{2})u\rangle\nonumber\\
& &-{\rm Res}_{x_{0}}({\rm wt} a+j+1)x_{0}^{-k}(x_{2}+x_{0})^{{\rm wt} a+j}
\langle u',Y(Y(a,x_{0})b,x_{2})u\rangle,
\end{eqnarray}
then (3.2.8) holds for $-k-1$. Thus by mathematical induction (3.2.8)
holds for any integer $k$. Therefore (3.2.2) holds.$\;\;\;\;\Box$

{\bf Proposition 3.2.3.} {\it Let $M$ be a restricted ${\bf
g}(V)$-module of level one and let $U$ be a subspace of $M$ satisfying the
following conditions:}

$(a)\;\;\;\; M=U({\bf g}(V))U$;

$(b)\;\;\;\;${\it For any }$a\in V, u\in U$, {\it the following
associativity holds for the pair } $(a,u)$, {\it i.e., there is a positive
integer }$k$ {\it such that }
\begin{eqnarray}
(x_{0}+x_{2})^{k}Y(a,x_{0}+x_{2})Y(b,x_{2})u=(x_{0}+x_{2})^{k}
Y(Y(a,x_{0})b,x_{2})u
\end{eqnarray}
{\it for any $b\in V$. Then $M$ is a weak $V$-module.}

{\bf Proof.} Since the vacuum property and the $L(-1)$-derivative
property have already been built to the $g(V)$-module structure on
$M$, we only need to prove the
Jacobi identity. Since the Jacobi identity is equivalent to the
commutativity and the associativity ``without involving
matrix-coefficients'' (Proposition 2.1.3), and the
commutativity has been built
to the ${\bf g}(V)$-module structure, thus we only need to prove the
associativity.

Let $W$ be the subspace of $M$ consisting of vectors
$u$ satisfying the associativity condition (b) for any $a\in V$.
Let $u\in W, c\in V$ and let $n$ be any integer. For any
$a\in V$, let $N$ be a positive integer such that $c_{i}a=0$ for $i\ge
N$. Since $u\in W$, there is a positive integer $k$ such that
\begin{eqnarray}
(x_{0}+x_{2})^{k}Y(c_{i}a,x_{0}+x_{2})Y(b,x_{2})c_{n}u
&=&(x_{0}+x_{2})^{k}Y(Y(c_{i}a,x_{0})b,x_{2})u,\\
(x_{0}+x_{2})^{k}Y(a,x_{0}+x_{2})Y(c_{i}b,x_{2})c_{n}u
&=&(x_{0}+x_{2})^{k}Y(Y(a,x_{0})c_{i}b,x_{2})u
\end{eqnarray}
for any $b\in V$ and for any nonnegative integer $i$. Choose $K$ such
that $K>k, k+N-n$. Then
\begin{eqnarray}
& &(x_{0}+x_{2})^{K}Y(a,x_{0}+x_{2})Y(b,x_{2})c_{n}u\nonumber\\
&=&(x_{0}+x_{2})^{K}c_{n}Y(Y(a,x_{0})b,x_{2})u\nonumber\\
& &-\sum_{i=0}^{\infty}\left(\begin{array}{c}n\\i\end{array}\right)
(x_{0}+x_{2})^{n-i}(x_{0}+x_{2})^{K}Y(c_{i}a,x_{0}+x_{2})Y(b,x_{2})u\nonumber\\
& &-\sum_{i=0}^{\infty}\left(\begin{array}{c}n\\i\end{array}\right)
x_{2}^{n-i}(x_{0}+x_{2})^{K}Y(a,x_{0}+x_{2})Y(c_{i}b,x_{2})u\nonumber\\
&=&(x_{0}+x_{2})^{K}c_{n}Y(Y(a,x_{0})b,x_{2})u\nonumber\\
& &-\sum_{i=0}^{\infty}\left(\begin{array}{c}n\\i\end{array}\right)
x_{2}^{n-i}(x_{0}+x_{2})^{K}Y(Y(a,x_{0})c_{i}b,x_{2})u\nonumber\\
& &-\sum_{i=0}^{\infty}\left(\begin{array}{c}n\\i\end{array}\right)
(x_{0}+x_{2})^{n-i}(x_{0}+x_{2})^{K}Y(Y(c_{i}a,x_{0})b,x_{2})u\nonumber\\
&=&(x_{0}+x_{2})^{K}Y(Y(a,x_{0})b,x_{2})c_{n}u\nonumber\\
& &+\sum_{i=0}^{\infty}\left(\begin{array}{c}n\\i\end{array}\right)
x_{2}^{n-i}(x_{0}+x_{2})^{K}Y(c_{i}Y(a,x_{0})b,x_{2})u\nonumber\\
& &-\sum_{i=0}^{\infty}\left(\begin{array}{c}n\\i\end{array}\right)
x_{2}^{n-i}(x_{0}+x_{2})^{K}Y(Y(a,x_{0})c_{i}b,x_{2})u\nonumber\\
& &-\sum_{i=0}^{\infty}\left(\begin{array}{c}n\\i\end{array}\right)
(x_{0}+x_{2})^{n-i}(x_{0}+x_{2})^{K}Y(Y(c_{i}a,x_{0})b,x_{2})u\nonumber\\
&=&(x_{0}+x_{2})^{K}Y(Y(a,x_{0})b,x_{2})c_{n}u\nonumber\\
& &+\sum_{j=0}^{\infty}\sum_{i=0}^{j}\left(\begin{array}{c}n\\j\end{array}
\right)\left(\begin{array}{c}j\\i\end{array}\right)x_{0}^{j-i}x_{2}^{n-j}
(x_{0}+x_{2})^{K}Y(Y(c_{i}a,x_{0})b,x_{2})u\nonumber\\
& &-\sum_{i=0}^{\infty}\left(\begin{array}{c}n\\i\end{array}\right)
(x_{0}+x_{2})^{n-i}(x_{0}+x_{2})^{K}Y(Y(c_{i}a,x_{0})b,x_{2})u\nonumber\\
&=&(x_{0}+x_{2})^{K}Y(Y(a,x_{0})b,x_{2})c_{n}u\nonumber\\
& &+\sum_{j=0}^{\infty}\sum_{i=0}^{j}\left(\begin{array}{c}n\\i\end{array}
\right)\left(\begin{array}{c}n-i\\j-i\end{array}\right)x_{0}^{j-i}x_{2}^{n-j}
(x_{0}+x_{2})^{K}Y(Y(c_{i}a,x_{0})b,x_{2})u\nonumber\\
& &-\sum_{i=0}^{\infty}\left(\begin{array}{c}n\\i\end{array}\right)
(x_{0}+x_{2})^{n-i}(x_{0}+x_{2})^{K}Y(Y(c_{i}a,x_{0})b,x_{2})u\nonumber\\
&=&(x_{0}+x_{2})^{K}Y(Y(a,x_{0})b,x_{2})c_{n}u.
\end{eqnarray}
Then the associativity for $(a,c_{n}u)$ holds.
It follows from (a) and
(b) that $M=W$. Therefore, $M$ is a weak $V$-module.$\;\;\;\;\Box$

The following Proposition is similar to Proposition 3.2.3.

{\bf Proposition 3.2.4.} {\it Let $M=\sum_{n\in {\bf N}}M(h+n)$
be a lower-truncated ${\bf Z}$-graded ${\bf g}(V)$-module of level
one. Let $p$ be the projection map from $M$ onto $M(h)$ and define}
\begin{eqnarray}
\langle u',u\rangle =u'(p(u))\;\;\;\mbox{{\it for }}u'\in
M(h)^{*},u\in M.
\end{eqnarray}
{\it Suppose that the following conditions hold:}

$(a)\;\;\;\; M=U({\bf g}(V))M(h);$

$(b)\;\;\;\; \mbox{{\it For any }}a\in V, u\in M(h), \mbox{ {\it
there is a positive integer }}k\mbox{ {\it such that }}$
\begin{eqnarray}
\langle u',(x_{0}+x_{2})^{k}Y(a,x_{0}+x_{2})Y(b,x_{2})u\rangle
=\langle u',(x_{0}+x_{2})^{k}Y(Y(a,x_{0})b,x_{2})u\rangle
\end{eqnarray}
{\it for any $b\in V, u'\in M(h)^{*}$. Then for any $a\in V, u\in
M,u'\in M(h)^{*}$, there is a positive integer $k$ such that (3.2.17) holds.}

{\bf Proof} Let $W$ consist of each $u\in M$ such that for any $a\in
V, u'\in M(h)^{*}$, there is a positive integer $k$ satisfying (3.2.17)
for any $b\in V$. Then it is equivalent to prove that $W=M$. First $W$
contains $M(h)$ by (b). It follows from PBW theorem and (a) that
$M=U(g(V)_{+})M(h)$. Then it is sufficient to prove that
$g(V)_{+}W\subseteq W$. Let $u\in W$ and  $c\in V_{(m)}, n\in {\bf Z}$
such that $\deg c_{n}=m-n-1>0$. Let $a,b\in V, u'\in M(h)^{*}$. Then
we have:
\begin{eqnarray}
& &\langle u', (x_{0}+x_{2})^{k}Y(a,x_{0}+x_{2})Y(b,x_{2})c_{n}u\rangle
\nonumber\\
&=&\langle u', (x_{0}+x_{2})^{k}c_{n}Y(a,x_{0}+x_{2})Y(b,x_{2})u\rangle
\nonumber\\
& &-\sum_{i=0}^{\infty}\left(\begin{array}{c}n\\i\end{array}\right)
(x_{0}+x_{2})^{k+n-i}\langle u',Y(c_{i}a,x_{0}+x_{2})Y(b,x_{2})u
\rangle\nonumber\\
& &-\sum_{i=0}^{\infty}\left(\begin{array}{c}n\\i\end{array}\right)
x_{2}^{n-i}(x_{0}+x_{2})^{k}\langle u',Y(a,x_{0}+x_{2})Y(c_{i}b,x_{2})u\rangle
\nonumber\\
&=&-\sum_{i=0}^{\infty}\left(\begin{array}{c}n\\i\end{array}\right)
(x_{0}+x_{2})^{k+n-i}\langle u',Y(Y(c_{i}a,x_{0})b,x_{2})u
\rangle\nonumber\\
& &-\sum_{i=0}^{\infty}\left(\begin{array}{c}n\\i\end{array}\right)
x_{2}^{n-i}(x_{0}+x_{2})^{k}\langle u',Y(Y(a,x_{0})c_{i}b,x_{2})u\rangle
\nonumber\\
&=&-\sum_{i=0}^{\infty}\left(\begin{array}{c}n\\i\end{array}\right)
(x_{0}+x_{2})^{k+n-i}\langle u',Y(Y(c_{i}a,x_{0})b,x_{2})u
\rangle\nonumber\\
& &-\sum_{i=0}^{\infty}\left(\begin{array}{c}n\\i\end{array}\right)
x_{2}^{n-i}(x_{0}+x_{2})^{k}\langle u',Y(c_{i}Y(a,x_{0})b,x_{2})u\rangle
\nonumber\\
& &+\sum_{i=0}^{\infty}\sum_{j=0}^{\infty}
\left(\begin{array}{c}n\\i\end{array}\right)
\left(\begin{array}{c}i\\j\end{array}\right)
x_{2}^{n-i}(x_{0}+x_{2})^{k}x_{0}^{i-j}\langle
u',Y(Y(c_{j}a,x_{0})b,x_{2})u\rangle \nonumber\\
&=&(x_{0}+x_{2})^{k}\langle u',Y(Y(a,x_{0})b,x_{2})c_{n}u\rangle
-(x_{0}+x_{2})^{k}\langle u',c_{n}Y(Y(a,x_{0})b,x_{2})u\rangle
\nonumber\\
&=&(x_{0}+x_{2})^{k}\langle u',Y(Y(a,x_{0})b,x_{2})c_{n}u\rangle.
\end{eqnarray}
Thus $g(V)_{+}W\subseteq W$. Therefore, the proof is complete.
$\;\;\;\;\Box$

{\bf Proposition 3.2.5.} {\it Let $M$ be a ${\bf g}(V)$-module
satisfying the assumption of Proposition 3.2.3. Then for any $y\in
U({\bf g}(V)), a\in V, u\in M$, there is a positive integer $k$ such that}
\begin{eqnarray}
\langle u',(x_{0}+x_{2})^{k}y\cdot Y(a,x_{0}+x_{2})Y(b,x_{2})u\rangle
=\langle u',(x_{0}+x_{2})^{k}y\cdot Y(Y(a,x_{0})b,x_{2})u\rangle
\end{eqnarray}
{\it for any $b\in V, u'\in M(h)^{*}$.}

{\bf Proof.} Let $L$ be the subspace of $U({\bf g}(V))$ consisting of
each $y$ satisfying (3.2.19). Let $y\in L$, let $c$ be any homogeneous
element of $V$ and let $n$ be any integer. Then
\begin{eqnarray}
& &\langle u',yc_{n}Y(a,x_{0}+x_{2})Y(b,x_{2})u\rangle (x_{0}+x_{2})^{k}
\nonumber\\
&=&\sum_{i=0}^{\infty}\left(\begin{array}{c}n\\i\end{array}\right)
(x_{0}+x_{2})^{k+n-i}\langle u',yY(c_{i}a,x_{0}+x_{2})Y(b,x_{2})u\rangle
\nonumber\\
& &+\sum_{i=0}^{\infty}\left(\begin{array}{c}n\\i\end{array}\right)
x_{2}^{n-i}(x_{0}+x_{2})^{k}\langle u',yY(a,x_{0}+x_{2})Y(c_{i}b,x_{2})u
\rangle\nonumber\\
& &+(x_{0}+x_{2})^{k}\langle u',yY(a,x_{0}+x_{2})Y(b,x_{2})c_{n}u\rangle.
\end{eqnarray}
Using the same method we used in the proof of Proposition 3.2.3, we find
that $yc_{n}\in L$. Since $U({\bf g}(V))$ is generated by all such
$c_{n}$'s, $L=U({\bf g}(V))$. Then the proof is complete.$\;\;\;\;\Box$

The following theorem is a different formulation for Zhu's construction of a
${\bf Z}$-graded $V$-module with the lowest-degree subspace $U$ from
an $A(V)$-module $U$.

{\bf Theorem 3.2.6.} {\it Let $U$ be any $A(V)$-module. Then
$L(U)$ is a lower-truncated ${\bf Z}$-graded weak $V$-module.}

{\bf Proof.} It follows from Proposition 3.2.4 and [L1] or the
Proposition 3.1.1 [FHL]  that
\begin{eqnarray}
& &x_{0}^{-1}\delta\left(\frac{x_{1}-x_{2}}{x_{0}}\right)
\langle u',y\cdot Y(a,x_{1})Y(b,x_{2})v\rangle\nonumber\\
& &-x_{0}^{-1}\delta\left(\frac{x_{2}-x_{1}}{-x_{0}}\right)
\langle u',y\cdot Y(b,x_{2})Y(a,x_{1})v\rangle\nonumber\\
&=&x_{2}^{-1}\delta\left(\frac{x_{1}-x_{0}}{x_{2}}\right)
\langle u',y\cdot Y(Y(a,x_{0})b,x_{2})v\rangle
\end{eqnarray}
for any $u'\in U, y\in U({\bf g}(V)),v\in M(U)$ and for any elements
$a,b\in V$. By the definition of $L(U)$, we obtain the Jacobi identity
for a $V$-module. Therefore, $L(U)$ is a weak $V$-module.$\;\;\;\;\Box$

Theorem 3.2.6 gives a functor $L$ from the category of $A(V)$-modules to
the category of lower-truncated ${\bf Z}$-graded weak $V$-modules.

{\bf Remark 3.2.7.} Let $U$ be an $A(V)$-module and let $W$ be the
subspace of $M(U)$ linearly spanned by the coefficients in the following
expressions:
\begin{eqnarray}
(x_{0}+x_{2})^{{\rm
wt}a-1}Y(a,x_{0}+x_{2})Y(b,x_{2})u-(x_{2}+x_{0})^{{\rm wt a}-1}
Y(Y(a,x_{0})b,x_{2})u
\end{eqnarray}
for any homogeneous elements $a,b\in V$ and for any $u\in U$. Let
$\bar{M}(U)$ be the quotient $g(V)$-module of $M(U)$ modulo $U({\bf g}(V))W$.
It follows from Theorem 3.2.6 that $U({\bf g}(V))W$ is a graded
${\bf g}(V)$-submodule of $J$. Then by Proposition 2.1.3
$\bar{M}(U)$ is a lower-truncated ${\bf Z}$-graded weak $V$-module. It
is clear that
$\bar{M}(U)$ satisfies the following universal property:

{\bf Corollary 3.2.8.} {\it Let $M$ be
any weak $V$-module and let $\phi$ be an $A(V)$-module homomorphism
from $U$ to $\Omega (M)$. Then there exists a unique $V$-module homomorphism
$\bar{\phi}$ from $\bar{M}(U)$ to $M$ extending $\phi$.}

{}From Corollary 3.2.8, $\bar{M}(U)$ is a universal object. From this we
call $\bar{M}(U)$ the {\it generalized Verma (weak) $V$-module} with
lowest-degree subspace $U$.

{\bf Proposition 3.2.9.} {\it Let $U$ be an $A(V)$-module on which
$L(0)$ acts semisimply. Then $\Omega (L(U))=U$.}

{\bf Proof.} Suppose $U$ is a direct sum of $A(V)$-modules
$U^{\alpha}$ $(\alpha \in \Phi)$.
Then from the construction of $L(U)$ we have:
$L(U)=\oplus _{\alpha\in \Phi}L(U^{\alpha})$. Thus it suffices to prove this
proposition for $U$ on which $L(0)$ acts as a scalar $h$. If $L(0)$
acts on $U$ as a scalar $h$, $L(0)$ acts on
$U(g_{+})_{n}U$ as a scalar $n+h$ for any $n\in {\bf N}$ so that any
submodule of $L(U)$ is a graded submodule.
Since $L(0)$ acts semisimply
on $\Omega(L(U))$, we have:$\Omega(L(U))=\oplus_{n=0}^{\infty}\Omega
(L(U))_{n+h}$, where
\begin{eqnarray}
\Omega (L(U))_{n+h}=\{u\in \Omega(L(U))|L(0)u=(n+h)u\}.
\end{eqnarray}
By the definition of $L(U)$ we have $\Omega (L(U))_{0}=U$. If
$\Omega(L(U))_{n+h}\ne 0$ for some positive integer $n$, then
$U(g)\Omega(L(U))_{n+h}=U(g_{+})\Omega(L(U))_{n+h}$ is a $V$-submodule
of $L(U)$ with lowest weight $n+h$. This is against the definition of $L(U)$.
Thus $\Omega (L(U))_{n+h}=0$ for $n>0$, so that $\Omega
(L(U))=U$.$\;\;\;\;\Box$

{\bf Lemma 3.2.10.} {\it Let $U$ be an irreducible $A(V)$-module on which
$L(0)$ acts as a scalar $h$. Then $L(U)$ is an irreducible generalized
$V$-module with lowest weight $h$.}

{\bf Proof.} Since $L(0)$ acts on $U$ as a scalar $h$, $L(0)$ acts on
$U(g_{+})_{n}U$ as a scalar $n+h$ for any $n\in {\bf N}$ so that any
submodule of $L(U)$ is a graded submodule. Then the irreducibility of
$L(U)$ follows from the irreducibility of $U$ and from the definition
of $L(U)$.$\;\;\;\;\Box$

{\bf Corollary 3.2.11.} {\it Let $U$ be a completely reducible
$A(V)$-module on which $L(0)$ acts semisimply. Then $L(U)$ is a
completely reducible generalized $V$-module.}

{\bf Lemma 3.2.12.} {\it Let $M$ be an irreducible generalized
$V$-modules with lowest weight $h$. Then
$\Omega(M)=M_{(h)}$ is an irreducible $A(V)$-module.}

{\bf Proof.} Suppose that $M$ is an irreducible $V$-module with lowest weight
$h$. It follows from the proof of Proposition 3.2.9 that $\Omega
(M)=M_{(h)}$. Let $U$ be any nonzero $A(V)$-submodule of $M_{(h)}$.
Then $U(g)U=U(g_{+})U$ is a nonzero $V$-submodule of $M$ with
lowest-weight subspace $U$. Then $M=U(g)U$, so that $U=M$. Thus $\Omega (M)$
is an irreducible $A(V)$-module.$\;\;\;\;\Box$

{\bf Corollary 3.2.13.} {\it Let $V$ be any vertex operator algebra and let
$M$ is a completely reducible $V$-module. Then $L\Omega (M)=M$.}

{\bf Proof.} Suppose $M=\oplus _{\alpha \in \Phi}M^{\alpha}$,
where $M^{\alpha}$ is irreducible.
Since $\Omega M= \oplus _{\alpha \in \Phi}\Omega(M^{\alpha})$,
$L\Omega M=\oplus _{\alpha \in \Phi}L\Omega(M^{\alpha})$. Thus it
suffices to prove that $L\Omega (M)=M$ for any irreducible $V$-module $M$.
This follows from Lemmas 3.2.10 and 3.2.12 immediately.
$\;\;\;\;\Box$

{\bf Definition 3.2.14 [Zhu].} A vertex operator algebra $V$ is said to
be {\it rational} if any lower-truncated ${\bf Z}$-graded weak
$V$-module $M$ is completely reducible.

{\bf Theorem 3.2.15 [Zhu].} {\it Let $V$ be a rational vertex operator
algebra. Then $A(V)$ is semisimple.}

If $V$ is
rational, then $\bar{M}(U)$ is completely reducible. In particular, if
$U$ is an irreducible $A(V)$-module, $\bar{M}(U)$ is an irreducible
$V$-module so that $\bar{M}(U)=L(U)$. (To be safe, we should say that
$\bar{M}(U)$ is a
generalized $V$-module because we don't know if its homogeneous
subspaces are finite-dimensional).

{\bf Remark 3.2.16.} As has been noticed by [Lian], the
construction of $A(V)$-algebra works for general graded vertex
algebras. Let $(V,Y,1,d)$ be a graded vertex algebra constructed from
a ${\bf Z}$-graded commutative associative algebra $A$ (with identity)
with a derivation $d$. Since $a_{n}=0$ for any $a\in V, n\in {\bf Z}_{+}$,
for any homogeneous elements $a,b\in A$, we have:
\begin{eqnarray}
{\rm Res}_{x}\frac{(1+x)^{\deg a}}{x}Y(a,x)b&=&a_{-1}b=ab;\\
{\rm Res}_{x}\frac{(1+x)^{\deg a}}{x^{2}}Y(a,x)b
&=&a_{-2}b+(\deg a)a_{-1}b\nonumber\\
&=&(da)_{-1}b+(\deg a)a_{-1}b\nonumber\\
&=&(da+d(0)a)b.
\end{eqnarray}
Then $O(V)=A(d+d(0))A$, so that $A(V)=A/A(d+d(0))A$ (the quotient
algebra of $A$ modulo the ideal $A(d+d(0))A$). For an $A(V)$-module
$U$, since $g(V)$ is an
abelian Lie algebra, it follows from the construction of $L(U)$ that
$g_{+}U(g_{+})U=0$ so that $L(U)=U$. Therefore $Y(a,x)=ax^{-\deg a
}$ acting on $L(U)$ for any $a\in V$. For the $A(V)$-module $U$, we
may consider $U$ as an $A$-module. Then from Examples 2.1.5 we have a
vertex algebra module structure on $U$. But this module structure is
different from the last one because $Y(a,x)u=\left(e^{xd}a\right)u$
may have infinitely many terms.

\newpage

\chapter{Frenkel and Zhu's fusion rule formula}

The notion of intertwining operator and the notion of fusion rule
have been defined by Frenkel, Huang and Lepowsky [FHL]. The notion of
intertwining operator is a generalization of the notion of $V$-module
so that for a
$V$-module $(M,Y_{M})$, the linear map $Y_{M}(\cdot,x)$ is a special
intertwining operator of type
$\left(\begin{array}{c}M\\V,M\end{array}\right)$.
In [FZ], Zhu's $A(V)$ theory has been generalized to a $A(M)$ theory
to determine intertwining operators among three arbitrary irreducible
$V$-modules.

In Chapter 3 we have seen that Zhu's $A(V)$ algebras comes from the weight-zero
components of vertex operators.  Let $V$ be a vertex operator
algebra, $M^{i}$ $(i=1,2,3)$ be three $V$-modules and let $I(\cdot,x)$
be an intertwining operator of type
$\left(\begin{array}{c}M^{3}\\M^{1},M^{2}\end{array}\right)$. In this
general situation, we cannot talk about weight-zero homomorphisms from
$M^{2}$ to $M^{3}$ because the weights of $M^{2}$ may not be congruent
to the weights of $M^{3}$ modulo ${\bf Z}$. But if each $M^{i}$ is
irreducible, we may talk about degree-zero homomorphisms from $M^{2}$
to $M^{3}$. Similar to $A(V)$, for an irreducible $V$-module $M$,
$A(M)$ is resulted from the consideration of degree-zero components of
intertwining operators. This time, $A(M)$ is a $A(V)$-bimodule for an
$V$-module $M$.

Roughly speaking, in Zhu's $A(V)$-theory, for
a $V$-module $M$, the $V$-module structure on
$M$ is uniquely determined by a linear map from $A(V)\otimes M_{(h)}$ to
$M_{(h)}$, which is characterized by an $A(V)$-module structure on
$M_{(h)}$. Conversely, given an $A(V)$-module $U$ or a linear map from
$A(V)\otimes U$ to $U$ satisfying certain conditions, by Zhu's
construction theorem [Zhu] or Theorem 3.2.6 we can obtain a $V$-module
$M$ with lowest-weight subspace $U$.
In Frenkel and Zhu's $A(M)$-theory, an intertwining
operator of type
$\left(\begin{array}{c}M^{3}\\M^{1},M^{2}\end{array}\right)$ is
uniquely determined by a linear map from $A(M^{1})\otimes
M^{2}_{(h_{2})}$ to $M^{3}_{(h_{3})}$. This linear map can be
characterized as an $A(V)$-module homomorphism from $A(M^{1})\otimes
_{A(V)}M^{2}_{(h_{2})}M^{3}_{(h_{3})}$. Conversely, for an given
$A(V)$-module homomorphism from $A(M^{1})\otimes
_{A(V)}M^{2}_{(h_{2})}$ to $M^{3}_{(h_{3})}$ or
a linear map
from $A(M^{1})\otimes M^{2}_{(h_{2})}$ to $M^{3}_{(h_{3})}$ satisfying
certain conditions, Theorem 1.5.3 [FZ] says that we can have an intertwining
operator of the corresponding type. It was believed that Theorem 1.5.3
[FZ] could be proved by using the
similar method used in Zhu's $A(V)$-theory [Zhu].
As a matter of fact, the situation for
intertwining operators of arbitrary type is quite different from
(and much more complicated than)
the situation for modules and Theorem 1.5.3 [FZ]
is not true in general without assuming certain conditions. In
appendix A we will give a counterexample to show this.

In this chapter we first review Frenkel and Zhu's construction of the
$A(V)$-bimodule $A(M)$ for any $V$-module $M$. Then we prove
that Theorem 1.5.3 [FZ] is true if $M^{2}$ and $(M^{3})'$ are
generalized Verma
modules for $V$. Especially, if $V$ is rational, Theorem 1.5.3 [FZ] is true.
Our proof
is a sort of generalization of Tsuchiya and Kanie's approach for WZW
model [TK]. It is not
unexpected that our tensor product construction in Chapter 5 enters
the proof.

Throughout this chapter, $V$ will be a fixed vertex operator algebra
and $g(V)$ or briefly $g$ will be the corresponding Lie algebra with
the triangular decomposition: $g=g_{+}\oplus g_{0}\oplus g_{-}$.

\section{Intertwining operators and fusion rules}
In [FHL], they introduced the notions of intertwining operators and
fusion rules. They also introduced the
notions of adjoint and transpose intertwining
operators under the assumption that the three modules
have integral weights and proved certain symmetric properties for
fusion rules under the same condition. These notions and results are
fundamental in vertex operator algebra theory.

In this section we shall first recall the definition of an
intertwining operator of a certain type for three $V$-modules [FHL].
Then we define the transpose and the adjoint (later renamed as
contragredient [HL2]) intertwining
operators for three arbitrary irreducible $V$-modules without assuming
the integrability condition and we prove that they are intertwining
operators. Then we prove the symmetric
property for fusion rules. These slightly
generalize Frenkel, Huang and Lepowsky's results in [FHL]. The same
results have also been obtained in [HL2] by using a different approach.

{\bf Definition 4.1.1.} Let $M^{1}$, $M^{2}$ and $M^{3}$ be three
$V$-modules. An {\it intertwining operator} of type
$\left(\begin{array}{c}M^{3}\\ M^{1}, M^{2}\end{array}\right)$
is a linear map
\begin{eqnarray}
I(\cdot,x):& & M^{1}\rightarrow (\mbox{Hom}(M^{2},M^{3}))\{x\},\nonumber\\
& &u\mapsto I(u,x)=\sum_{\alpha \in {\bf C}}u_{\alpha}x^{-\alpha-1}
\end{eqnarray}
satisfying the following conditions:

(I1)\hspace{0.25cm} For any fixed $u \in M^{1},v \in M^{2},\alpha\in
{\bf C}$, $u_{\alpha+n}v= 0$ for $m\in {\bf Z}$ sufficiently large;

(I2)\hspace{0.25cm} $I(L(-1)u,x)v=\displaystyle{{d\over dx}}I(u,x)v
\mbox{  for }u \in M^{1},v \in M^{2};$

(I3)\hspace{0.25cm} For $a \in V,u \in M^{1},v \in M^{2}$, the following
Jacobi identity holds:
\begin{eqnarray}
& &x^{-1}_{0}\delta
\left(\frac{x_{1}-x_{2}}{x_{0}}\right)Y(a,x_{1})I(u,x_{2})v-x^{-1}_{0}\delta
 \left(\frac{x_{2}-x_{1}}{-x_{0}}\right)I(u,x_{2})Y(a,x_{1})v\nonumber\\
&=&x^{-1}_{2}\delta\left(\frac{x_{1}-x_{0}}{x_{2}}\right)
I(Y(a,x_{0})u,x_{2})v.
\end{eqnarray}

Denote by
$I\left(\begin{array}{c}M^{3}\\M^{1},M^{2}\end{array}\right)$
the vector space of all
intertwining operators of the indicated type and we call the dimension of
this vector space the {\it fusion rule} of the corresponding type.

Let $M=\oplus_{\alpha\in {\bf C}}M_{(\alpha)}$ be any $V$-module. For
any complex number $h$ set $M^{\alpha}=\oplus _{n\in {\bf Z}}M_{(\alpha+n)}$.
Then $M^{\alpha}$ is a submodule of $M$ for any $\alpha\in {\bf C}$
and $M^{\alpha}=M^{\beta}$ if and only if $\alpha-\beta\in {\bf Z}$.
Then we have a canonical decomposition for $M$:
\begin{eqnarray}
M=\oplus_{\bar{h}\in {\bf C}/{\bf Z}}M^{\bar{h}}.
\end{eqnarray}
Let $S$ be the set of equivalence classes of $V$-modules such that all
weights are congruent modulo ${\bf Z}$. Then any $V$-module is a
direct sum of $V$-modules from $S$. Let $M$ be a $V$-module in $S$ with
 lowest weight $h$. Then $M=\oplus_{n\in {\bf
N}}M_{(h+n)}$ so that $M$ is a lower-truncated ${\bf Z}$-graded
module. Set $M(n)=M_{(n+h)}$ for
any $n\in {\bf N}$. For convenience, following [FZ], we call $M(n)$
the {\it degree} $n$ subspace of $M$ and we define
\begin{eqnarray}
{\rm deg}u=n\;\;\;\;\mbox{for any }u\in M(n)=M_{(n+h)}, n\in {\bf N}.
\end{eqnarray}
The following proposition is quoted from [FZ].

{\bf Proposition 4.1.2}. {\it Let
$M^{i}=\bigoplus_{n=0}^{\infty}M^{i}(n)$ $(i=1,2,3)$ be three
$V$-modules in $S$ such that $L(0)|_{M^{i}(n)}=(h_{i}+n){\rm id}$
$(i=1,2,3)$ and let
$I(\cdot,x)$ be an intertwining operator of type
$\left(\begin{array}{c}M^{3}\\ M^{1},M^{2}\end{array}\right)$.
Then }
\begin{eqnarray}
I^{o}(u,x):=x^{h_{1}+h_{2}-h_{3}}I(u,x)\in ({\rm
Hom}(M^{2},M^{3}))[[x,x^{-1}]]. \end{eqnarray}
Set $I^{o}(u,x)=\sum_{n\in {\bf Z}}u(n)x^{-n-1}$. {\it Then
for every homogeneous $u\in M^{1},m,n \in {\bf N}$},
\begin{eqnarray}
u(n)M^{2}(m)\subseteq M^{3}(m+{\rm deg}\;u-n-1). \end{eqnarray}
{\it In particular},
\begin{eqnarray}
u({\rm deg}\;u+m+i)M^{2}(m)=0\;\;\;\mbox{{\it for all }} i \ge 0.
\end{eqnarray}

The following lemma is a simple corollary of Schur's lemma.

{\bf Lemma 4.1.3.} {\it If $M$ is an irreducible $V$-module, then}
$I\left(\begin{array}{c}M\\V,M\end{array}\right)={\bf
C}Y_{M}(\cdot,x)$. {\it In
particular, if $V$ is simple, then}
$I\left(\begin{array}{c}V\\V,V\end{array}\right)={\bf C}Y(\cdot,x)$.

{\bf Proof}. Let $I(\cdot,x)$ be any intertwining operator of type
$\left(\begin{array}{c}M\\V,M\end{array}\right)$. Since
$\displaystyle{{d\over dx}}I({\bf 1},x)=I(L(-1){\bf 1},x)=0$, $I({\bf
1},x)$ is a constant. Since $a_{i}{\bf 1}=0$ for $a\in V, i\in {\bf
Z}_{+}$, by taking
${\rm Res}_{x_{0}}$ of the Jacobi identity (4.1.2), we get $[I({\bf
1},x_{1}),Y(a,x_{2})]=0$ for all $a\in V$. Then $I({\bf 1},x)$ is a
$V$-module endomorphism of $M$.
By Schur's lemma, $I({\bf 1},x)$ is a scalar $\alpha$id. Set
$I^{1}(\cdot,x)=\alpha Y_{M}(\cdot,x)-I(\cdot,x)$. Then
$I^{1}(\cdot,x)$ is an intertwining operator of type
$\left(\begin{array}{c}M\\V,M\end{array}\right)$ and
$I^{1}({\bf 1},x)M=0$. Since ${\rm Ann}_{V}M=\{ a\in V|
I^{1}(a,x)M=0\}$ is an ideal of $V$ containing ${\bf 1}$,
$I^{1}(\cdot,x)=0$. Thus $I(\cdot,x)\in {\bf C}Y_{M}(\cdot,x)$.$
\;\;\;\;\Box$.

Next we shall define the
notion of adjoint and transpose intertwining operators without
assuming the integrability condition assumed in [FHL].

Let $M^{i}$ $(i=1,2,3)$ be $V$-modules from the set $S$ with lowest
weights $h_{i}$ $(i=1,2,3)$, respectively. Let $I(\cdot,x)$ be an
intertwining operator of type
$\left(\begin{array}{c}M^{3}\\M^{1},M^{2}\end{array}\right)$. Then by
[FHL] or [FZ], we have
\begin{eqnarray}
I(u^{1},x)=\sum_{n\in h+{\bf Z}}u^{1}_{n}x^{-n-1}=x^{-h}\sum_{n\in {\bf
Z}}u^{1}(n)x^{-n-1}
\end{eqnarray}
for any $u^{1}\in M^{1}$, where $h=h_{1}+h_{2}-h_{3}$. Therefore
$u^{1}(m)$ is a homomorphism from $M^{2}$ to $M^{3}$ of degree $({\rm
deg}u^{1}-m-1)$ for any homogeneous element $u^{1}$ of $M^{1}$.
Then $I^{o}(\cdot,x)=x^{h}I(\cdot,x)$ (defined in Proposition 4.1.2)
involves only integral
powers of $x$ and still satisfies the Jacobi identity and the
truncation condition.
The {\it transpose operator} $I^{t}(\cdot,x)$ is defined by:
\begin{eqnarray}
I^{t}(\cdot,x):& &M^{2}\otimes M^{1}\rightarrow M^{3}\{x\}\nonumber\\
& &I^{t}(u^{2},x)u^{1}=x^{-h}e^{xL(-1)}I^{o}(u^{1},-x)u^{2}
\end{eqnarray}
for $u^{1}\in M^{1},u^{2}\in M^{2}$,
where $h=h_{1}+h_{2}-h_{3}$ is defined as before. The {\it adjoint operator}
$I'(\cdot,x)$ is defined by:
\begin{eqnarray}
I'(\cdot,x)&:&M^{1}\otimes (M^{3})' \rightarrow (M^{2})'\{x\}\nonumber\\
& &\langle I'(u^{1},x)(u^{3})',u^{2}\rangle =x^{-2h_{1}}\langle
(u^{3})',I^{o}(e^{xL(1)}(-x^{-2})
^{L(0)-h_{1}}u^{1},x^{-1})u^{2}\rangle \end{eqnarray}
for $u^{1}\in M^{1},u^{2}\in M^{2},(u^{3})'\in (M^{3})'$.

{\bf Proposition 4.1.4.} {\it The transpose operator} $I^{t}(\cdot,x)$
{\it and the adjoint operator} $I'(\cdot,x)$ {\it are intertwining
operators of corresponding types.}

{\bf Proof.} The proof of Theorem 5.5.1 [FHL] works if we make a slight
change. First it is
easy to see that $I^{o}(\cdot,x)$ satisfies all the axioms of an
intertwining operator except the $L(-1)$-derivative property. It
follows from the ${\it S}_{3}$-symmetry that $x^{-h}I^{t}(\cdot,x)$
satisfies the Jacobi identity. Then $I^{t}(\cdot,x)$ satisfies the
Jacobi identity. Since
\begin{eqnarray}
I(L(-1)u^{1},x)u^{2}&=&{d\over dx}I(u^{1},x)u^{2}\nonumber\\
&=&-h x^{-h-1}I^{o}(u^{1},x)u^{2}+x^{-h}{d\over
dx}I^{o}(u^{1},x)u^{2},\end{eqnarray}
we obtain
\begin{eqnarray}
I^{o}(L(-1)u^{1},x)u^{2}= -hx^{-1}I^{o}(u^{1},x)u^{2}+{d\over
dx}I^{o}(u^{1},x)u^{2}.\end{eqnarray}
Or equivalently
\begin{eqnarray}
I^{o}(L(-1)u^{1},-x)u^{2}=h x^{-1}I^{o}(u^{1},-x)u^{2}-{d\over
dx}I^{o}(u^{1},-x)u^{2}.\end{eqnarray}
Then
\begin{eqnarray}
& &{d\over dx}I^{t}(u^{2},x)u^{1}\nonumber\\&=&h x^{-h
-1}e^{xL(-1)}I^{o}(u^{1},-x)u^{2}\nonumber\\
& &+x^{-h}e^{xL(-1)}L(-1)I^{o}(u^{1},-x)u^{2}+x^{-h}e^{xL(-1)}{d\over
dx}I^{o}(u^{1},-x)u^{2}\nonumber\\
&=&x^{-h}e^{xL(-1)}(L(-1)I^{o}(u^{1},-)u^{2}-I^{o}(L(-1)u^{1},-x)u^{2})\\
&=&x^{-h}e^{xL(-1)}I^{o}(u^{1},-x)L(-1)u^{2}\nonumber\\
&=&I^{t}(L(-1)u^{2},x)u^{1}.\end{eqnarray}
Therefore $I^{t}(\cdot,x)$ is an intertwining operator of type
$\left(\begin{array}{c}M^{3}\\M^{2},M^{1}\end{array}\right)$.

Similarly, one can check the $L(-1)$-derivative property for
the adjoint operator $I'(\cdot,x)$. For the Jacobi identity we only
need to follow FHL's original proof using the $S_{3}$-symmetry
for $I(\cdot,x)$ and the conjugation formula for $V$. $\;\;\;\;\Box $

{\bf Corollary 4.1.5.} {\it Let} $M^{i}$ (i=1,2,3) {\it be
$V$-modules from $S$. Then}
\begin{eqnarray}
& &\dim I\left(\begin{array}{c}M^{3}\\M^{1},M^{2}\end{array}\right)=
\dim I\left(\begin{array}{c}M^{3}\\M^{2},M^{1}\end{array}\right),\\
& &\dim I\left(\begin{array}{c}M^{3}\\M^{1},M^{2}\end{array}\right)=
\dim I\left(\begin{array}{c}(M^{2})'\\M^{1},(M^{3})'\end{array}\right).
\end{eqnarray}

{\bf Proof.} It directly follows from
$(I^{t})^{t}(\cdot,x)=I(\cdot,x)$ and
$(I')'(\cdot,x)=I(\cdot,x)$.$\;\;\;\;\Box$

{\bf Remark 4.1.6.} Let $M^{i}$ $(i=1,2,3)$ be any
$V$-modules. Then $M^{i}=\oplus_{\bar{h}\in {\bf C}/{\bf Z}}
(M^{i})^{\bar{h}}$. Let $I(\cdot,x)$ be an intertwining
operator of type $\left(\begin{array}{c}M^{3}\\M^{1},M^{2}\end{array}\right)$.
Then
\begin{eqnarray}
I(\cdot,x)=\oplus_{\bar{\alpha},\bar{\beta},\bar{\gamma}\in {\bf
C}/{\bf Z}}I_{\bar{\alpha},\bar{\beta}}^{\bar{\gamma}}(\cdot,x)
\end{eqnarray}
where
$I_{\bar{\alpha},\bar{\beta}}^{\bar{\gamma}}(\cdot,x)$ is an
intertwining operator of type
$\left(\begin{array}{c}(M^{3})^{\bar{\gamma}}\\(M^{1})^{\bar{\alpha}},
(M^{2})^{\bar{\beta}}\end{array}\right)$.
Define
\begin{eqnarray}
I^{t}(\cdot,x)=\oplus_{\bar{\alpha},\bar{\beta},\bar{\gamma}\in {\bf
C}/{\bf
Z}}\left(I_{\bar{\alpha},\bar{\beta}}^{\bar{\gamma}}\right)^{t}(\cdot,x).
\end{eqnarray}
Then
$I^{t}(\cdot,x)$ is an intertwining operator of type
$\left(\begin{array}{c}M^{3}\\M^{2},M^{1}\end{array}\right)$. (Since
the decompositions for $M^{i}$ are canonical, $I^{t}(\cdot,x)$ is
well-defined.) Therefore, Corollary 4.1.5 is true for any modules.

\section{Frenkel and Zhu's formula for fusion rule}
In this section we shall first review Frenkel and Zhu's construction of the
$A(V)$-bimodule $A(M)$ for any $V$-module $M$. Then we prove that Frenkel
and Zhu's theorem is true under a certain condition.

Let $M$ be a $V$-module. Recall from [FZ] that $O(M)$ is the subspace
of $M$ linearly spanned by all elements:
\begin{eqnarray}
{\rm Res}_{x}{\frac{(1+x)^{{\rm wt}a}}{x^{2}}}Y(a,x)u\;\;\;\;\mbox{for
any homogeneous }a\in V,u\in M.
\end{eqnarray}
The left and right actions $''\cdot''$ of $V$ on $M$ are defined
as follows:
\begin{eqnarray}
a\cdot u&=&{\rm Res}_{x}{\frac{(1+x)^{{\rm wt}a}}{x}}Y(a,x)u,\\
u\cdot a&=&{\rm Res}_{x}{\frac{(1+x)^{{\rm wt}a-1}}{x}}Y(a,x)u
\end{eqnarray}
for any homogeneous element $a\in V$ and for any $u\in M$.
Set $A(M)=M/O(M)$.

{\bf Proposition 4.2.1 [FZ].} {\it For any
$V$-module $M$, $A(M)$ is an $A(V)$-bimodule under the defined left
and right action.}

{\bf Remark 4.2.2.} For any homogeneous element $a\in V$ and for $u\in
M$, we have:
\begin{eqnarray}
a\cdot u-u\cdot a&=&{\rm Res}_{x}(1+x)^{{\rm wt}a-1}Y(a,x)u\nonumber\\
&=&\sum_{i=0}^{\infty}\left(\begin{array}{c}{\rm wt }a-1\\i\end{array}
\right)a_{i}u\nonumber\\
&=&a\circ u
\end{eqnarray}
where $a\circ u$ denotes the action of $a\in g_{0}$ on $u\in M_{z=1}$
(the evaluation module).

Let $M^{i}$ $(i=1,2,3)$ be three irreducible $V$-modules and let
$I(\cdot,x)$ be an intertwining operator of type $\left(\begin{array}{c}
M^{3}\\M^{1},M^{2}\end{array}\right)$. Then we define the following
bilinear map
\begin{eqnarray}
\pi (I): & &A(M^{1})\otimes_{A(V)}M^{2}(0)\rightarrow M^{3}(0);\nonumber\\
& &u^{1}\otimes u^{2}\mapsto o(u^{1})u^{2}\;\;\;\;\mbox{for }u^{1}\in
A(M^{1}), u^{2}\in M^{2}(0).
\end{eqnarray}
Then it has been proved [FZ] that $\pi (I)$ is an $A(V)$-module homomorphism.

{\bf Lemma 4.2.3.} {\it The defined linear map is an injective map from the
intertwining operator space $I\left(\begin{array}{c}
M^{3}\\M^{1},M^{2}\end{array}\right)$ to
${\rm Hom}_{A(V)}\left(A(M^{1})\otimes_{A(V)}M^{2}(0),M^{3}(0)\right)$.}

{\bf Proof.} It is equivalent to prove that $\pi(I)=0$ implies $I=0$.
If $\pi (I)=0$, then
\begin{eqnarray}
\langle u', I(u^{1},x)u^{2}\rangle=0\;\;\;\mbox{for any }u'\in
(M^{3})',u^{1}\in M^{1},u^{2}\in M^{2}(0).
\end{eqnarray}
Let $W\subseteq M^{2}$ consist of $u$ such that
\begin{eqnarray}
\langle u', I(u^{1},x)u\rangle=0\;\;\;\mbox{for any }u'\in
(M^{3})',u^{1}\in M^{1}.
\end{eqnarray}
Then $M^{2}(0)\subseteq W$. Let $u\in W, a\in V, n\in {\bf Z}$ such that ${\rm
wt }a_{n}>0$. Then
\begin{eqnarray}
& &\langle u', I(u^{1},x)a_{n}u\rangle\nonumber\\
&=&\langle u',a_{n}\left(I(u^{1},x)u\right)\rangle
+\sum_{i=0}^{\infty}\left(\begin{array}{c}n\\i\end{array}\right)
x^{n-i}\langle u',I(a_{i}u^{1},x)u\rangle\nonumber\\
&=&0.
\end{eqnarray}
Then $a_{n}u\in W$.
Since $W$ contains $M^{2}(0)$ and $M^{2}(0)$ generates $M^{2}$ by
$U({\bf g}(V)_{+})$, thus $W=M^{2}$, so that (4.2.6) holds for any
$u'\in (M^{3}(0))^{*}, u^{1}\in M^{1}, u^{2}\in M^{2}$.
Similarly, since $(M^{3}(0))^{*}$ generates $(M^{3})'$ by
$U({\bf g}(V))$, (4.2.6) holds for any $u'\in (M^{3})', u^{1}\in
M^{1}, u^{2}\in M^{2}$. Therefore $I=0.\;\;\;\;\Box$

The following is Theorem 1.5.3 [FZ] with a minor correction.

{\bf Theorem 4.2.4.} {\it Let $M^{i}$ $(i=1,2,3)$ be three irreducible
$V$-modules such that
$M^{2}$ and $(M^{3})'$ are universal or generalized Verma modules
for $V$. Then $\pi$ is a linear isomorphism from
$I\left(\begin{array}{c}M^{3}\\M^{1},M^{2}\end{array}\right)$ onto
${\rm Hom}_{A(V)}(A(M^{1})\otimes _{A(V)}M^{2}(0),M^{3}(0))$.}

The proof of Theorem 4.2.4 will take the rest of this section. Our proof
is a generalization of Tsuchiya and Kanie's proof for WZW model [TK].
But we do not have singular vectors to use instead we have to use the
associativity (which is equivalent to the existence of singular vectors
for WZW model).

Let $\psi_{0}$ be an $A(V)$-module homomorphism from
$A(M^{1})\otimes_{A(V)}M^{2}(0)$ to $M^{3}(0)$. Lifting $\psi_{0}$
through the natural linear projection map from $M^{1}\otimes M^{2}(0)$ to
$A(M^{1})\otimes_{A(V)}M^{2}(0)$, we obtain a linear map $\psi_{1}$ from
$M^{1}\otimes M^{2}(0)$ to $M^{3}(0)$.

{\bf Lemma 4.2.5.} {\it The linear map $\psi_{1}$ satisfies the following
conditions:}
\begin{eqnarray}
(a)& &\psi_{1} \;\mbox{{\it is a }}{\bf g}_{0}\mbox{{\it -module
homomorphism from }}M^{1}_{z=1}\otimes M^{2}(0)\;\mbox{{\it to }}M^{3}(0);\\
(b)& &\psi_{1} (O(M^{1})\otimes M^{2}(0))=0;\\
(c)& &o(a)\psi_{1}(u^{1}\otimes u^{2})={\rm Res}_{x}
\frac{(1+x)^{{\rm wt }a}}{x}\psi_{1} (Y(a,x)u^{1}\otimes u^{2})
\end{eqnarray}
{\it for any homogeneous element $a$ of $V$ and for any $u^{1}\in
M^{1},u^{2}\in M^{2}(0)$.}

{\bf Proof.} (b) and (c) simply mean that $\psi_{0}$ is an
$A(V)$-module homomorphism from $A(M^{1})\otimes_{{\bf C}}M^{2}(0)$ to
$M^{3}(0)$, where $A(M^{1})\otimes _{{\bf C}}M^{2}(0)$ is viewed as an
$A(V)$-module with the left action on the first factor. For any $a\in
V$, by (c) and Remark 4.2.2 we have:
\begin{eqnarray}
a_{{\rm wt }a-1}\psi_{1}(u^{1}\otimes u^{2})&=&o(a)\psi_{1}(u^{1}\otimes u^{2})
\nonumber\\
&=&\psi_{1}(a\cdot u^{1}\otimes u^{2})\nonumber\\
&=&\psi_{1}(a\circ u^{1}\otimes u^{2})+\psi_{1}(u^{1}\cdot a\otimes u^{2})
\nonumber\\
&=&\psi_{1}(a\circ u^{1}\otimes u^{2})+\psi_{1}(u^{1}\otimes o(a)u^{2})
\nonumber\\
&=&\psi_{1}(a_{{\rm wt }a-1}(u^{1}\otimes u^{2})).
\end{eqnarray}
This proves that $\psi_{1}$ is a $g_{0}$-module homomorphism from
$M^{1}_{z=1}\otimes M^{2}(0)$ to $M^{3}(0)$. $\;\;\;\;\Box$

It follows from
Lemma 2.2.6 that ${\bf C}[t,t^{-1}]\otimes M^{1}$ is a $g(V)$-module
of level zero. For any homogeneous $u^{1}\in M^{1}$ and for any
integer $n$ we define
\begin{eqnarray}
\deg (t^{n}\otimes u^{1})=\deg u^{1}-n-1 \;\;
\left({\rm wt} u^{1}-h_{1}-n-1\right).
\end{eqnarray}
Then ${\bf C}[t,t^{-1}]\otimes M^{1}$ becomes a
${\bf Z}$-graded $g$-module.  We consider
the generalized Verma $g$-module (of level one)
$M(M^{2}(0))$ as a ${\bf Z}$-graded $g$-module with the
decomposition $M(M^{2}(0))=\oplus_{n=0}^{\infty}U(g_{+})_{n}M^{2}(0)$. Then
the tensor product $g$-module (of level one) $F={\bf
C}[t,t^{-1}]\otimes M^{1}\otimes M(M^{2}(0))$ is a ${\bf Z}$-graded
$g$-module, where
\begin{eqnarray}
F(n)=\oplus_{k,r\in {\bf Z}_{+}}t^{k+r-n-1}\otimes M^{1}(k)\otimes
U(g_{+})_{r}M^{2}(0).
\end{eqnarray}

We consider $(M^{3}(0))^{*}$ as a $g_{0}$-module by defining:
\begin{eqnarray}
(af)(u)=f(\theta (a)u)\;\;\;\mbox{ for }a\in g_{0},f\in
(M^{3}(0))^{*}, u\in M^{3}(0),
\end{eqnarray}
where $\theta$ is the anti-automorphism of $g$ defined in Section 2.2.
 That is,
\begin{eqnarray}
\langle af, u\rangle =\langle f,\theta (a)u\rangle,
\end{eqnarray}
where $\langle \cdot,\cdot\rangle$ is the natural pair between
$(M^{3}(0))^{*}$ and $M^{3}(0)$.

Next we define a bilinear form on $(M^{3}(0))^{*}\times F$ as follows:
First for homogeneous $u^{1}\in M^{1}, u^{2}\in M^{2}(0), u'\in
(M^{3}(0))^{*}$ we define
\begin{eqnarray}
\langle u',t^{n}\otimes u^{1}\otimes u^{2}\rangle
=\delta_{n-\deg u^{1}+1,0}\langle u',\psi_{1}(u^{1}\otimes u^{2})\rangle.
\end{eqnarray}
Then for $y\in g_{+}U(g_{+})$ we inductively define
\begin{eqnarray}
\langle u',t^{n}\otimes u^{1}\otimes yu^{2}\rangle
=\langle u',\sigma (y)(t^{n}\otimes u^{1})\otimes u^{2}\rangle.
\end{eqnarray}

{\bf Proposition 4.2.6.} {\it The defined bilinear form on
$(M^{3}(0))^{*}\times F$ satisfies the following conditions:}
\begin{eqnarray}
& &\langle u',F(n)\rangle=0\;\;\mbox{{\it if }}n\ne 0;\\
& &\langle u',yu\rangle=0;\\
& &\langle u',y_{0}u\rangle=\langle \theta (y_{0})u',u\rangle
\end{eqnarray}
{\it for any $ u'\in (M^{3}(0))^{*}, y\in g_{+}U(g),y_{0}\in U(g_{0}),u\in F$.}

{\bf Proof.} Let $u^{1}\in M^{1}$ be any homogeneous element, let
$u'\in (M^{3}(0))^{*}, u\in M^{2}(0)$ and let $n$ be an integer such
that $\deg (t^{n}\otimes u^{1}\otimes u)\;(=\deg u^{1}-n-1)\ne 0$. Then
by definition  we have:
\begin{eqnarray}
\langle u', t^{n}\otimes u^{1}\otimes u\rangle=0.
\end{eqnarray}
Let $u^{1}\in M^{1}$ be any homogeneous element, let
$u'\in (M^{3}(0))^{*}, y\in U(g_{+})_{m}\;(m>0), u\in M^{2}(0)$ and
let $n$ be an integer such
that $\deg (t^{n}\otimes u^{1}\otimes yu)\;(=\deg u^{1}-n-1+m)\ne 0$.
Since $\deg \sigma(y)(t^{n}\otimes u^{1})\otimes u=\deg (t^{n}\otimes
u^{1}\otimes yu)\ne 0$, we have:
\begin{eqnarray}
\langle u', t^{n}\otimes u^{1}\otimes yu\rangle
=\langle u', \sigma(y)(t^{n}\otimes u^{1})\otimes u\rangle=0.
\end{eqnarray}
This proves (4.2.19).

Let $a\in V$ be any homogeneous element of $V$ and let
$n$ be any integer such that $\deg a_{n}={\rm wt }a-n-1>0$. Then
\begin{eqnarray}
& &\langle u',a_{n}(t^{m}\otimes u^{1}\otimes yu)\rangle\nonumber\\
&=&\langle u',a_{n}(t^{m}\otimes u^{1})\otimes yu\rangle
+\langle u',t^{m}\otimes u^{1}\otimes a_{n}yu\rangle\nonumber\\
&=&\langle u',\sigma (y)a_{n}(t^{m}\otimes u^{1})\otimes u\rangle
+\langle u',\sigma (a_{n}y)t^{m}\otimes u^{1}\otimes u\rangle\nonumber\\
&=&\langle u',\sigma (y)a_{n}(t^{m}\otimes u^{1})\otimes u\rangle
-\langle u',\sigma (y)a_{n}t^{m}\otimes u^{1}\otimes u\rangle\nonumber\\
&=&0
\end{eqnarray}
for $y\in U(g_{+}), u'\in (M^{3}(0))^{*}, u\in M^{2}(0),m\in {\bf Z}$.
Since $g_{+}$ is linearly spanned by such $a_{n}$'s, the second
property follows.

For the third property, let $a\in V, n\in {\bf Z},u^{1}\in M^{1}, u\in
M^{2}(0)$. Then
\begin{eqnarray}
& &\langle u',\bar{a}(t^{n}\otimes u^{1}\otimes u)\rangle\nonumber\\
&=&\sum_{i=0}^{\infty}\left(\begin{array}{c}{\rm wt }a-1\\i\end{array}\right)
\langle u',t^{n+{\rm wt}a-i-1}\otimes a_{i}u^{1}\otimes u\rangle
+\langle u',t^{n}\otimes u^{1}\otimes a_{{\rm wt}a-1}u\rangle\nonumber\\
&=&\sum_{i=0}^{\infty}\left(\begin{array}{c}{\rm wt }a-1\\i\end{array}\right)
\delta_{n,\deg u^{1}-1}\langle u',\psi_{0}(a_{i}u^{1}\otimes u)\rangle
\nonumber\\
& &+\delta_{n,\deg u^{1}-1}\langle u',\psi_{0}(u^{1}\otimes a_{{\rm wt}a-1}u)
\rangle\nonumber\\
&=&\delta_{n,\deg u^{1}-1}\langle u',\psi_{0}(\bar{a}(u^{1}\otimes u))\rangle
\nonumber\\
&=&\delta_{n,\deg u^{1}-1}\langle u',\bar{a}\psi_{0}(u^{1}\otimes u)\rangle
\nonumber\\
&=&\delta_{n,\deg u^{1}-1}\langle \theta (\bar{a})u',\psi_{0}(u^{1}\otimes u)
\rangle\nonumber\\
&=&\langle \theta (\bar{a})u',t^{n}\otimes u^{1}\otimes u\rangle.
\end{eqnarray}
Furthermore, for any $y\in g_{+}U(g_{+})$, we have:
\begin{eqnarray}
& &\langle u',\bar{a}(t^{n}\otimes u^{1}\otimes yu)\rangle\nonumber\\
&=&\langle u',\bar{a}(t^{n}\otimes u^{1})\otimes yu+t^{n}\otimes u^{1}
\otimes \bar{a}yu\rangle\nonumber\\
&=&\langle u',\sigma (y)\bar{a}(t^{n}\otimes u^{1})\otimes u\rangle
+\langle u',t^{n}\otimes u^{1}
\otimes (y\bar{a}u+[\bar{a},y]u)\rangle\nonumber\\
&=&\langle u',\sigma (y)\bar{a}(t^{n}\otimes u^{1})\otimes u\rangle\nonumber\\
& &+\langle u',\sigma(y)(t^{n}\otimes u^{1})\otimes \bar{a}u\rangle
+\langle u',\sigma([\bar{a},y])(t^{n}\otimes u^{1})\otimes u\rangle
\nonumber\\
&=&\langle u',\sigma(y)(t^{n}\otimes u^{1})\otimes \bar{a}u\rangle
+\langle u',\bar{a}\sigma(y)(t^{n}\otimes u^{1})\otimes u\rangle\nonumber\\
&=&\langle u',\bar{a}(\sigma(y)(t^{n}\otimes u^{1})\otimes u)\rangle
\nonumber\\
&=&\langle \theta (\bar{a})u',\sigma(y)(t^{n}\otimes u^{1})\otimes u\rangle
\nonumber\\
&=&\langle \theta (\bar{a})u',t^{n}\otimes u^{1})\otimes yu\rangle.
\;\;\;\;\Box
\end{eqnarray}

Next we define
\begin{eqnarray}
I=\{u\in F|\langle u', yu\rangle =0\;\;\mbox{for any }u'\in
(M^{3}(0))^{*}, y\in U(g)\}.
\end{eqnarray}
In other words, $I$ is
the maximal ${\bf g}(V)$-submodule contained in the right kernel of
the bilinear form on $(M^{3}(0))^{*}\times F$.

{\bf Lemma 4.2.7.} {\it If $n<0$, then $F(n)\subseteq I$.}

{\bf Proof.} By definition we need to prove
\begin{eqnarray}
\langle u',y\cdot u\rangle =0\;\;\;\mbox{for any }u'\in
(M^{3}(0))^{*},y\in U(g),u\in F(n).
\end{eqnarray}
By PBW theorem we have: $U(g)=U(g_{+})U(g_{0})U(g_{-})$. From
Proposition 4.2.6 it is enough to prove (4.2.28) for $y\in
U(g_{0})U(g_{-})$. Because $yu$ has a negative degree, by Proposition
4.2.6, (4.2.28) is true. Therefore $F(n)\subseteq I$ for any
$n<0$.$\;\;\;\;\Box$

For $u^{1}\in M^{1}$, set
\begin{eqnarray}
I_{t}(u^{1},x)=\sum_{n\in {\bf Z}}(t^{n}\otimes u^{1})x^{-n-1-h},
\end{eqnarray}
where $h=h_{1}+h_{2}-h_{3}$ is defined as before. Then we have:
\begin{eqnarray}
\langle u', I_{t}(u^{1},x)\otimes u^{2}\rangle
=x^{-\deg u^{1}-\deg u^{2}-h}
\langle u', t^{\deg u^{1}+\deg u^{2}-1}\otimes u^{1}\otimes u^{2}\rangle
\end{eqnarray}
for $u'\in (M^{3}(0))^{*}$ and for homogeneous $u^{1}\in
M^{1},u^{2}\in M(M^{2}(0))$.

{\bf Lemma 4.2.8.} {\it For $u'\in (M^{3}(0))^{*},u^{1}\in M^{1}, u^{2}\in
M(M^{2}(0))$, we have}
\begin{eqnarray}
\langle u',I_{t}(L(-1)u^{1},x)\otimes u^{2}\rangle
={d\over dx}\langle u',I_{t}(u^{1},x)\otimes u^{2}\rangle.
\end{eqnarray}

{\bf Proof.} Without losing generality we may assume that both $u^{1}$
and $u^{2}$ are homogeneous. Noticing that $L(0)=\omega_{1}$, we get
\begin{eqnarray}
& &L(0)(t^{n}\otimes u^{1}\otimes u^{2})\nonumber\\
&=&L(0)(t^{n}\otimes u^{1})\otimes u^{2}+t^{n}\otimes u^{1}\otimes L(0)u^{2}
\nonumber\\
&=&t^{n+1}\otimes L(-1)u^{1}\otimes u^{2}+t^{n}\otimes L(0)u^{1}\otimes u^{2}
+t^{n}\otimes u^{1}\otimes L(0)u^{2}\nonumber\\
&=&t^{n+1}\otimes L(-1)u^{1}\otimes u^{2}+({\rm wt }u^{1}+{\rm wt
}u^{2})(t^{n}\otimes u^{1}\otimes u^{2}).
\end{eqnarray}
Then
\begin{eqnarray}
& &t^{n}\otimes L(-1)u^{1}\otimes u^{2}\nonumber\\
&=&L(0)(t^{n-1}\otimes u^{1}\otimes u^{2})
-({\rm wt }u^{1}+{\rm wt }u^{2})(t^{n-1}\otimes u^{1}\otimes u^{2}).
\end{eqnarray}
Therefore
\begin{eqnarray}
& &\langle u',I_{t}(L(-1)u^{1},x)\otimes u^{2}\rangle\nonumber\\
&=&\langle u',t^{\deg u^{1}+\deg u^{2}}\otimes L(-1)u^{1}\otimes u^{2}
\rangle x^{-\deg u^{1}-\deg u^{2}-h-1}\nonumber\\
&=&\langle u',L(0)(t^{\deg u^{1}+\deg u^{2}-1}\otimes u^{1}\otimes u^{2})
\rangle x^{-\deg u^{1}-\deg u^{2}-h-1}\nonumber\\
& &-({\rm wt }u^{1}+{\rm wt }u^{2})
\langle u',t^{\deg u^{1}+\deg u^{2}-1}\otimes u^{1}\otimes u^{2}\rangle
x^{-\deg u^{1}-\deg u^{2}-h-1} \nonumber\\
&=&\langle L(0)u',t^{\deg u^{1}+\deg u^{2}-1}\otimes u^{1}\otimes u^{2}\rangle
x^{-\deg u^{1}-\deg u^{2}-1-h}\nonumber\\
& &-({\rm wt }u^{1}+{\rm wt }u^{2})\langle u',t^{\deg u^{1}+\deg u^{2}-1}
\otimes u^{1}\otimes u^{2}\rangle x^{-\deg u^{1}-\deg u^{2}-h-1}\nonumber\\
&=&h_{3}\langle u',t^{\deg u^{1}+\deg u^{2}-1}\otimes u^{1}\otimes u^{2}
\rangle x^{-\deg u^{1}-\deg u^{2}-h-1}\nonumber\\
& &-({\rm wt }u^{1}+{\rm wt }u^{2})\langle u',t^{\deg u^{1}+\deg u^{2}-1}
\otimes u^{1}\otimes u^{2}\rangle x^{-\deg u^{1}-\deg u^{2}-h-1}\nonumber\\
&=&(-\deg u^{1}-\deg u^{2}-h)\langle u',t^{\deg u^{1}+\deg u^{2}-1}
\otimes u^{1}\otimes u^{2}\rangle x^{-\deg u^{1}-\deg u^{2}-h-1}\nonumber\\
&=&{d\over dx}\langle u',t^{\deg u^{1}+\deg u^{2}-1}
\otimes u^{1}\otimes u^{2}\rangle x^{-\deg u^{1}-\deg u^{2}-h}\nonumber\\
&=&{d\over dx}\langle u',I_{t}(u^{1},x)\otimes u^{2}\rangle. \;\;\;\;\Box
\end{eqnarray}

{\bf Lemma 4.2.9.} {\it For $y\in U(g), u'\in (M^{3}(0))^{*}, u^{1}\in M^{1},
u\in M(M^{2}(0))$, we have}
\begin{eqnarray}
\langle u',yI_{t}(L(-1)u^{1},x)\otimes u\rangle
={d\over dx}\langle u',yI_{t}(u^{1},x)\otimes u\rangle.
\end{eqnarray}

{\bf Proof.} Let $L$ be the subspace of $U(g)$ consisting of each $x$ such that
(4.2.35) holds
for any $u'\in (M^{3}(0))^{*},u^{1}\in M^{1},u\in M(M^{2}(0))$. Then
$1\in L$ (by Lemma 4.2.8).
Let $y\in L$, let $a$ be any homogeneous element of $V$ and let $n$ be any
integer. Then
\begin{eqnarray}
& &\langle u',ya_{n}I_{t}(L(-1)u^{1},x)\otimes u\rangle\nonumber\\
&=&\sum_{i=0}^{\infty}\left(\begin{array}{c}n\\i\end{array}\right)
x^{n-i}\langle u', yI_{t}(a_{i}L(-1)u^{1},x)\otimes u\rangle\nonumber\\
&=&\sum_{i=0}^{\infty}\left(\begin{array}{c}n\\i\end{array}\right)
x^{n-i}\langle u', yI_{t}(L(-1)a_{i}u^{1},x)\otimes u\rangle\nonumber\\
& &+\sum_{i=0}^{\infty}\left(\begin{array}{c}n\\i\end{array}\right)
x^{n-i}\langle u', yI_{t}(ia_{i-1}u^{1},x)\otimes u\rangle\nonumber\\
&=&\sum_{i=0}^{\infty}\left(\begin{array}{c}n\\i\end{array}\right)
x^{n-i}{d\over dx}\langle u', yI_{t}(a_{i}u^{1},x)\otimes u\rangle\nonumber\\
& &+\sum_{i=0}^{\infty}\left(\begin{array}{c}n\\i\end{array}\right)
(n-i)x^{n-i-1}\langle u', yI_{t}(a_{i}u^{1},x)\otimes u\rangle\nonumber\\
&=&\sum_{i=0}^{\infty}\left(\begin{array}{c}n\\i\end{array}\right)
x^{n-i}{d\over dx}\langle u', yI_{t}(a_{i}u^{1},x)\otimes u\rangle\nonumber\\
& &+\sum_{i=0}^{\infty}\left(\begin{array}{c}n\\i\end{array}\right)
{dx^{n-i}\over dx}\langle u', yI_{t}(a_{i-1}u^{1},x)\otimes u\rangle\nonumber\\
&=&{d\over dx}\langle u',ya_{n}I_{t}(u^{1},x)\otimes u\rangle.
\end{eqnarray}
Then $Lg\subseteq L$. Thus $L=U(g)$. $\;\;\;\;\Box$.

{\bf Proposition 4.2.10.} {\it For $y\in U(g), u'\in (M^{3}(0))^{*},
u^{1}\in M^{1}, u\in M(M^{2}(0))$, we have}
\begin{eqnarray}
\langle u',y\left(I_{t}(L(-1)u^{1},x)\otimes u\right)\rangle
={d\over dx}\langle u',y\left(I_{t}(u^{1},x)\otimes u\right)\rangle,
\end{eqnarray}
{\it or, equivalently}
\begin{eqnarray}
I_{t}(L(-1)u^{1},x)\otimes u ={d\over dx}I_{t}(u^{1},x)\otimes u
\;\;\;\mbox{mod }I[[x,x^{-1}]].
\end{eqnarray}

{\bf Proof.} Since
\begin{eqnarray}
y\left(I_{t}(L(-1)u^{1},x)\otimes u\right)=yI_{t}(u^{1},x)\otimes u
+I_{t}(u^{1},x)\otimes yu,
\end{eqnarray}
the proposition follows from Lemmas 4.2.8 and 4.2.9 immediately.$\;\;\;\;\Box$

{\bf Proposition 4.2.11.} {\it For any $a\in V,u'\in
(M^{3}(0))^{*},u^{1}\in M^{1},u^{2}\in M(M^{2}(0))$, we have}
\begin{eqnarray}
& &x_{0}^{-1}\delta\left(\frac{x_{1}-x_{2}}{x_{0}}\right)
\langle u',Y(a,x_{1})\left(I_{t}(u^{1},x_{2})\otimes u^{2}\right)
\rangle\nonumber\\
& &-x_{0}^{-1}\delta\left(\frac{-x_{2}+x_{1}}{x_{0}}\right)
\langle u',I_{t}(u^{1},x_{2})\otimes Y(a,x_{1})u^{2}\rangle\nonumber\\
&=&x_{2}^{-1}\delta\left(\frac{x_{1}-x_{0}}{x_{2}}\right)
\langle u',I_{t}(Y(a,x_{0})u^{1},x_{2})\otimes u^{2}\rangle.
\end{eqnarray}

{\bf Proof.} Since the Jacobi identity is equivalent to the
commutativity and the associativity (without involving
``matrix-coefficients'') and the commutativity has already been built
to the $g(V)$-module structure, it is sufficient for us to prove the
following associativity:
\begin{eqnarray}
& &\langle u',(x_{0}+x_{2})^{{\rm wt }a}Y(a,x_{0}+x_{2})
\left(I_{t}(u^{1},x_{2})\otimes u^{2}\right)\rangle\nonumber\\
&=&\langle u',(x_{2}+x_{0})^{{\rm wt }a}I_{t}(Y(a,x_{0})u^{1},x_{2})
\otimes u^{2}\rangle.
\end{eqnarray}
If $u^{2}\in M^{2}(0)$, the proof is exactly the same as the proof of
Proposition 3.2.3. For a general vector $u^{2}\in M(M^{2}(0))$ it
follows from the proof of Proposition 3.2.5.$\;\;\;\;\Box$

{\bf Proposition 4.2.12.} {\it For any $u'\in (M^{3}(0))^{*},a\in V,
u^{1}\in M^{1},u^{2}\in M(M^{2}(0)),y\in U(g)$, we have}
\begin{eqnarray}
& &x_{0}^{-1}\delta\left(\frac{x_{1}-x_{2}}{x_{0}}\right)
\langle u',y\cdot Y(a,x_{1})\left(I_{t}(u^{1},x_{2})\otimes u^{2}\right)\rangle
\nonumber\\
& &-x_{0}^{-1}\delta\left(\frac{-x_{2}+x_{1}}{x_{0}}\right)
\langle u',y\cdot (I_{t}(u^{1},x_{2})\otimes Y(a,x_{1})u^{2})\rangle\nonumber\\
&=&x_{2}^{-1}\delta\left(\frac{x_{1}-x_{0}}{x_{2}}\right)
\langle u',y\cdot (I_{t}(Y(a,x_{0})u^{1},x_{2})\otimes u^{2})\rangle,
\end{eqnarray}
{\it or, equivalently}
\begin{eqnarray}
& &x_{0}^{-1}\delta\left(\frac{x_{1}-x_{2}}{x_{0}}\right)
Y(a,x_{1})\left(I_{t}(u^{1},x_{2})\otimes u^{2}\right)
-x_{0}^{-1}\delta\left(\frac{x_{2}-x_{1}}{-x_{0}}\right)
I_{t}(u^{1},x_{2})\otimes Y(a,x_{1})u^{2}\nonumber\\
&=&x_{2}^{-1}\delta\left(\frac{x_{1}-x_{0}}{x_{2}}\right)
I_{t}(Y(a,x_{0})u^{1},x_{2})\otimes u^{2}\;\;\;\mbox{mod }
I[[x_{0},x_{1},x_{2},x_{0}^{-1},x_{1}^{-1},x_{2}^{-1}]].
\end{eqnarray}

{\bf Proof.} The proof is exactly the same as the proof of Proposition
3.2.6, so that we will just sketch the proof.
Let $L$ be the subspace of $U(g)$ consisting of each y such that
(4.2.42) holds.
By Lemma 4.2.8 and Proposition 4.2.10, $U(g_{+}), U(g_{0})\subseteq L$. The
by PBW theorem we only need to prove that $Lg_{-}\subseteq L$.
$\;\;\;\;\Box$

{\bf Proposition 4.2.13.} {\it For any $u'\in (M^{3}(0))^{*}, a,b\in V,
u^{1}\in M^{1},u^{2}\in M^{2}(0)$, we have}
\begin{eqnarray}
& &x_{0}^{-1}\delta\left(\frac{x_{1}-x_{2}}{x_{0}}\right)
\langle u',I_{t}(u^{1},x)\otimes Y(a,x_{1})Y(b,x_{2})u^{2}\rangle\nonumber\\
& &-x_{0}^{-1}\delta\left(\frac{-x_{2}+x_{1}}{x_{0}}\right)
\langle u',I_{t}(u^{1},x)\otimes Y(b,x_{2})Y(a,x_{1})u^{2}\rangle\nonumber\\
&=&x_{2}^{-1}\delta\left(\frac{x_{1}-x_{0}}{x_{2}}\right)
\langle u',I_{t}(u^{1},x)\otimes Y(Y(a,x_{0})b,x_{2})u^{2}\rangle.
\end{eqnarray}

{\bf Proof.} Similar to Proposition 3.2.3, it is sufficient to prove
the following associativity:
\begin{eqnarray}
& &(x_{0}+x_{2})^{{\rm wt}a}\langle u',
I_{t}(u^{1},x)\otimes Y(a,x_{0}+x_{2})Y(b,x_{2})u^{2}\rangle\nonumber\\
&=&(x_{2}+x_{0})^{{\rm wt}a}\langle u',
I_{t}(u^{1},x)\otimes Y(Y(a,x_{0})b,x_{2})u^{2}\rangle,
\end{eqnarray}
or, equivalently
\begin{eqnarray}
& &{\rm Res}_{x_{0}}x_{0}^{n}(x_{0}+x_{2})^{{\rm wt}a}\langle u',
I_{t}(u^{1},x)\otimes Y(a,x_{0}+x_{2})Y(b,x_{2})u^{2}\rangle\nonumber\\
&=&{\rm Res}_{x_{0}}x_{0}^{n}(x_{2}+x_{0})^{{\rm wt}a}\langle u',
I_{t}(u^{1},x)\otimes Y(Y(a,x_{0})b,x_{2})u^{2}\rangle
\end{eqnarray}
for any integer $n$. Using the same technique as we used in the proof
Proposition 3.2.3 we see that it suffices to prove the special case:
\begin{eqnarray}
& &{\rm Res}_{x_{0}}x_{0}^{-1}(x_{0}+x_{2})^{{\rm wt}a}\langle u',
I_{t}(u^{1},x)\otimes Y(a,x_{0}+x_{2})Y(b,x_{2})u^{2}\rangle\nonumber\\
&=&{\rm Res}_{x_{0}}x_{0}^{-1}(x_{2}+x_{0})^{{\rm wt}a}\langle u',
I_{t}(u^{1},x)\otimes Y(Y(a,x_{0})b,x_{2})u^{2}\rangle.
\end{eqnarray}
By definition we have:
\begin{eqnarray}
& &{\rm Res}_{x_{0}}x_{0}^{-1}(x_{0}+x_{2})^{{\rm wt}a}\langle u',
I_{t}(u^{1},x)\otimes Y(a,x_{0}+x_{2})Y(b,x_{2})u^{2}\rangle\nonumber\\
&=&\sum_{m,n\in {\bf X}}{\rm Res}_{x_{0}}x_{0}^{-1}(x_{0}+x_{2})^{{\rm
wt}a-m-1}x_{2}^{-n-1}\langle u', I_{t}(u^{1},x)\otimes a_{m}b_{n}u^{2}
\rangle\nonumber\\
&=&\sum_{m\le {\rm wt}a-1}\sum_{n\le {\rm wt}b-1}{\rm Res}_{x_{0}}
x_{0}^{-1}(x_{0}+x_{2})^{{\rm wt}a-m-1}x_{2}^{-n-1}\langle u',
I_{t}(u^{1},x)\otimes a_{m}b_{n}u^{2}\rangle\nonumber\\
& &\left(\mbox{since}\; {\rm Res}_{x_{0}}x_{0}^{-1}
(x_{0}+x_{2})^{{\rm wt}a-m-1}=0\;{\mbox{ if }}m>{\rm wt}a-1,
a_{m}b_{n}=0\;{\mbox{ if }}n>{\rm wt}b-1\right)\nonumber\\
&=&\sum_{m\le {\rm wt}a-1}\sum_{n\le {\rm wt}b-1}
x_{2}^{{\rm wt}a-m-n-2}\langle u',
I_{t}(u^{1},x)\otimes a_{m}b_{n}u^{2}\rangle\nonumber\\
&=&\langle u', I_{t}(u^{1},x)\otimes a_{{\rm wt}a-1}b_{{\rm
wt}b-1}u^{2}\rangle x_{2}^{-{\rm wt}b}\nonumber\\
& &+\sum_{m< {\rm wt}a-1}\langle u', I_{t}(u^{1},x)\otimes a_{m}b_{{\rm
wt}b-1}u^{2}\rangle x_{2}^{{\rm wt}a-{\rm wt}b-m-1}\nonumber\\
& &+\sum_{n< {\rm wt}b-1}\langle u', I_{t}(u^{1},x)\otimes a_{{\rm
wt}a-1}b_{n}u^{2}\rangle x_{2}^{-n-1}\nonumber\\
& &+\sum_{m< {\rm wt}a-1}\sum_{n< {\rm wt}b-1}
\langle u', I_{t}(u^{1},x)\otimes a_{m}b_{n}u^{2}\rangle
x_{2}^{{\rm wt}a-m-n-2}\nonumber\\
&=&\langle u', I_{t}(u^{1},x)\otimes a_{{\rm wt}a-1}b_{{\rm
wt}b-1}u^{2}\rangle x_{2}^{-{\rm wt}b}\nonumber\\
& &-\sum_{m< {\rm wt}a-1}\sum_{i=0}^{\infty}
\left(\begin{array}{c}m\\i\end{array}\right)
\langle u', I_{t}(a_{i}u^{1},x)\otimes b_{{\rm wt}b-1}u^{2}\rangle
x_{2}^{{\rm wt}a-{\rm wt}b-m-1}x^{m-i}\nonumber\\
& &+\sum_{n< {\rm wt}b-1}\langle u', I_{t}(u^{1},x)\otimes b_{n}a_{{\rm
wt}a-1}u^{2}\rangle x_{2}^{-n-1}\nonumber\\
& &+\sum_{n< {\rm wt}b-1}\sum_{i=0}^{\infty}
\left(\begin{array}{c}{\rm wt}a-1\\i\end{array}\right)
\langle u', I_{t}(u^{1},x)\otimes (a_{i}b)_{{\rm wt}a+n-i-1}u^{2}\rangle
x_{2}^{-n-1}\nonumber\\
& &-\sum_{m< {\rm wt}a-1}\sum_{n< {\rm wt}b-1}\sum_{i=0}^{\infty}
\left(\begin{array}{c}m\\i\end{array}\right)
\langle u', I_{t}(a_{i}u^{1},x)\otimes b_{n}u^{2}\rangle
x_{2}^{{\rm wt}a-m-n-2}x^{m-i}\nonumber\\
&=&\langle u', I_{t}(u^{1},x)\otimes a_{{\rm wt}a-1}b_{{\rm
wt}b-1}u^{2}\rangle x_{2}^{-{\rm wt}b}\nonumber\\
& &-\sum_{m< {\rm wt}a-1}\sum_{i=0}^{\infty}
\left(\begin{array}{c}m\\i\end{array}\right)
x_{2}^{{\rm wt}a-{\rm wt}b-m-1}x^{m-{\rm wt}a-\deg u^{1}-h+1}
 \langle u', \psi_{0}(a_{i}u^{1}\otimes b_{{\rm wt}b-1}u^{2}\rangle
\nonumber\\
& &-\sum_{n< {\rm wt}b-1}\sum_{i=0}^{\infty}
\left(\begin{array}{c}n\\i\end{array}\right)x_{2}^{-n-1}x^{n-i}
\langle u', I_{t}(b_{i}u^{1},x)\otimes a_{{\rm wt}a-1}u^{2}\rangle
\nonumber\\
& &-\sum_{n< {\rm wt}b-1}\sum_{i=0}^{\infty}\sum_{j=0}^{\infty}
\left(\begin{array}{c}{\rm wt}a-1\\i\end{array}\right)
\left(\begin{array}{c}{\rm wt}a+n-i-1\\j\end{array}\right)
x_{2}^{-n-1}x^{{\rm wt}a+n-i-j-1}\cdot\nonumber\\
& &\cdot \langle u', I_{t}((a_{i}b)_{j}u^{1},x)\otimes u^{2}\rangle
\nonumber\\
& &+\sum_{m< {\rm wt}a-1}\sum_{n< {\rm wt}b-1}\sum_{i=0}^{\infty}
\sum_{j=0}^{\infty}\left(\begin{array}{c}m\\i\end{array}\right)
\left(\begin{array}{c}n\\j\end{array}\right)
x_{2}^{{\rm wt}a-m-n-2}x^{m+n-i-j}\cdot\nonumber\\
& &\cdot \langle u', I_{t}(b_{j}a_{i}u^{1},x)\otimes u^{2}\rangle
\nonumber\\
&=&\langle u', I_{t}(u^{1},x)\otimes a_{{\rm wt}a-1}b_{{\rm
wt}b-1}u^{2}\rangle x_{2}^{-{\rm wt}b}\nonumber\\
& &-\sum_{m< {\rm wt}a-1}{\rm Res}_{x_{1}}(1+x_{1})^{m}
x_{2}^{{\rm wt}a-{\rm wt}b-m-1}x^{m-{\rm wt}a-\deg u^{1}-h+1}\cdot\nonumber\\
& &\cdot \langle u',\psi_{0}(Y(a,x_{1})u^{1}\otimes b_{{\rm wt}b-1}u^{2})
\rangle \nonumber\\
& &-\sum_{n< {\rm wt}b-1}\sum_{i=0}^{\infty}
\left(\begin{array}{c}n\\i\end{array}\right)
x_{2}^{-n-1}x^{n-{\rm wt}b-\deg u^{1}-h+1}
\langle u', \psi_{0}(b_{i}u^{1}\otimes a_{{\rm wt}a-1}u^{2})\rangle
\nonumber\\
& &-\sum_{n< {\rm wt}b-1}\sum_{i=0}^{\infty}\sum_{j=0}^{\infty}
\left(\begin{array}{c}{\rm wt}a-1\\i\end{array}\right)
\left(\begin{array}{c}{\rm wt}a+n-i-1\\j\end{array}\right)
x_{2}^{-n-1}x^{n-{\rm wt}b-\deg u^{1}-h+1}\cdot\nonumber\\
& &\cdot \langle u', \psi_{0}((a_{i}b)_{j}u^{1}\otimes u^{2})\rangle
\nonumber\\
& &+\sum_{m< {\rm wt}a-1}\sum_{n< {\rm wt}b-1}\sum_{i=0}^{\infty}
\sum_{j=0}^{\infty}\left(\begin{array}{c}m\\i\end{array}\right)
\left(\begin{array}{c}n\\j\end{array}\right)
x_{2}^{{\rm wt}a-m-n-2}\cdot\nonumber\\
& &\cdot x^{m+n+2-{\rm wt}a-{\rm wt}b-\deg u^{1}-h}
\langle u', \psi_{0}(b_{j}a_{i}u^{1}\otimes u^{2})\rangle \nonumber\\
&=&\langle u', I_{t}(u^{1},x)\otimes a_{{\rm wt}a-1}b_{{\rm
wt}b-1}u^{2}\rangle x_{2}^{-{\rm wt}b}\nonumber\\
& &-{\rm Res}_{x_{1}}\sum_{m< {\rm wt}a-1}
(1+x_{1})^{m}
x_{2}^{{\rm wt}a-{\rm wt}b-m-1}x^{m-{\rm wt}a-\deg u^{1}-h+1}\cdot\nonumber\\
& &\cdot \langle u',\psi_{0}(Y(a,x_{1})u^{1}\otimes b_{{\rm wt}b-1}u^{2})
\rangle \nonumber\\
& &-{\rm Res}_{x_{3}}\sum_{n< {\rm wt}b-1}
(1+x_{3})^{n}
x_{2}^{-n-1}x^{n-{\rm wt}b-\deg u^{1}-h+1}\cdot\nonumber\\
& &\cdot \langle u',\psi_{0}(Y(b,x_{3})u^{1}\otimes a_{{\rm wt}a-1}u^{2})
\rangle \nonumber\\
& &-\sum_{n< {\rm wt}b-1}\sum_{i=0}^{\infty}{\rm Res}_{x_{3}}
\left(\begin{array}{c}{\rm wt}a-1\\i\end{array}\right)x_{2}^{-n-1}
x^{n-{\rm wt}b-\deg u^{1}-h+1}\cdot\nonumber\\
& &\cdot (1+x_{3})^{{\rm wt}a+n-i-1}
\langle u', \psi_{0}(Y(a_{i}b,x_{3})u^{1}\otimes u^{2})\rangle
\nonumber\\
& &+\sum_{m< {\rm wt}a-1}\sum_{n< {\rm wt}b-1}{\rm Res}_{x_{1}}
{\rm Res}_{x_{3}}x_{2}^{{\rm wt}a-m-n-2}x^{m+n+2-{\rm wt}a-{\rm wt}b-
\deg u^{1}-h}\nonumber\\
& &\cdot (1+x_{1})^{m}(1+x_{3})^{n}\langle u', \psi_{0}(Y(b,x_{3})Y(a,x_{1})
u^{1}\otimes u^{2})\rangle \nonumber\\
&=&\langle u', I_{t}(u^{1},x)\otimes a_{{\rm wt}a-1}b_{{\rm
wt}b-1}u^{2}\rangle x_{2}^{-{\rm wt}b}\nonumber\\
& &-{\rm Res}_{x_{1}}\frac{(1+x_{1})^{{\rm wt}a-1}x_{2}^{1-{\rm
wt}b}x^{-\deg u^{1}-h}}{x(1+x_{1})-x_{2}}
\langle u',\psi_{0}(Y(a,x_{1})u^{1}\otimes b_{{\rm wt}b-1}u^{2})
\rangle \nonumber\\
& &-{\rm Res}_{x_{3}}\frac{(1+x_{3})^{{\rm wt}b-1}x_{2}^{1-{\rm
wt}b}x^{-\deg u^{1}-h}}{x(1+x_{3})-x_{2}}
\langle u',\psi_{0}(Y(b,x_{3})u^{1}\otimes a_{{\rm wt}a-1}u^{2})
\rangle \nonumber\\
& &-\sum_{n< {\rm wt}b-1}{\rm Res}_{x_{0}}{\rm Res}_{x_{3}}x_{2}^{-n-1}
x^{n-{\rm wt}b-\deg u^{1}-h+1}(1+x_{3})^{{\rm wt}a+n-1}\cdot\nonumber\\
& &\cdot \left(1+\frac{x_{0}}{1+x_{3}}\right)^{{\rm wt}a-1}
\langle u', \psi_{0}(Y(Y(a,x_{0})b,x_{3})u^{1}\otimes u^{2})\rangle
\nonumber\\
& &+{\rm Res}_{x_{1}}{\rm Res}_{x_{3}}
\frac{x_{2}^{2-{\rm wt}b}x^{-\deg u^{1}-h}(1+x_{1})^{{\rm wt}a-1}
(1+x_{3})^{{\rm wt}b-1}}{(x(1+x_{1})-x_{2})(x(1+x_{3})-x_{2})}\cdot\nonumber\\
& &\cdot \langle u', \psi_{0}(Y(b,x_{3})Y(a,x_{1})
u^{1}\otimes u^{2})\rangle \nonumber\\
&=&\langle u', I_{t}(u^{1},x)\otimes a_{{\rm wt}a-1}b_{{\rm
wt}b-1}u^{2}\rangle x_{2}^{-{\rm wt}b}\nonumber\\
& &-{\rm Res}_{x_{1}}{\rm Res}_{x_{3}}\frac{(1+x_{1})^{{\rm wt}a-1}
x_{2}^{1-{\rm
wt}b}x^{-\deg u^{1}-h}}{x(1+x_{1})-x_{2}}\frac{(1+x_{3})^{{\rm wt}b-1}}{x_{3}}
\cdot\nonumber\\
& &\cdot\langle u',\psi_{0}(Y(b,x_{3})Y(a,x_{1})u^{1}\otimes u^{2})
\rangle \nonumber\\
& &-{\rm Res}_{x_{1}}{\rm Res}_{x_{3}}\frac{(1+x_{3})^{{\rm wt}b-1}
x_{2}^{1-{\rm
wt}b}x^{-\deg u^{1}-h}}{x(1+x_{3})-x_{2}}\frac{(1+x_{1})^{{\rm wt}a-1}}{x_{1}}
\cdot\nonumber\\
& &\langle u',\psi_{0}(Y(a,x_{1})Y(b,x_{3})u^{1}\otimes u^{2})
\rangle\nonumber\\
& &-{\rm Res}_{x_{0}}{\rm Res}_{x_{3}}\frac{x_{2}^{1-{\rm wt}b}
x^{-\deg u^{1}-h}(1+x_{3})^{{\rm wt}b-2}(1+x_{3}+x_{0})^{{\rm wt}a-1}}
{1-\frac{x_{2}}{x(1+x_{3})}}\cdot\nonumber\\
& &\cdot\langle u', \psi_{0}(Y(Y(a,x_{0})b,x_{3})u^{1}\otimes u^{2})\rangle
\nonumber\\
& &+{\rm Res}_{x_{1}}{\rm Res}_{x_{3}}
\frac{x_{2}^{2-{\rm wt}b}x^{-\deg u^{1}-h}(1+x_{1})^{{\rm wt}a-1}
(1+x_{3})^{{\rm wt}b-1}}{(x(1+x_{1})-x_{2})(x(1+x_{3})-x_{2})}\cdot\nonumber\\
& &\cdot \langle u', \psi_{0}(Y(b,x_{3})Y(a,x_{1})
u^{1}\otimes u^{2})\rangle \nonumber\\
&=&\langle u', I_{t}(u^{1},x)\otimes a_{{\rm wt}a-1}b_{{\rm
wt}b-1}u^{2}\rangle x_{2}^{-{\rm wt}b}\nonumber\\
& &-{\rm Res}_{x_{1}}{\rm Res}_{x_{3}}\frac{(1+x_{3})^{{\rm wt}b-1}
x_{2}^{1-{\rm
wt}b}x^{-\deg u^{1}-h}}{x(1+x_{3})-x_{2}}\frac{(1+x_{1})^{{\rm wt}a-1}}{x_{1}}
\cdot\nonumber\\
& &\langle u',\psi_{0}(Y(a,x_{1})Y(b,x_{3})u^{1}\otimes u^{2})
\rangle\nonumber\\
& &+{\rm Res}_{x_{1}}{\rm Res}_{x_{3}}
\frac{x_{2}^{1-{\rm wt}b}x^{-\deg u^{1}-h}(1+x_{1})^{{\rm wt}a-1}
(1+x_{3})^{{\rm wt}b}(x_{2}-x)}{x_{3}(x(1+x_{1})-x_{2})(x(1+x_{3})-x_{2})}
\cdot\nonumber\\
& &\cdot \langle u', \psi_{0}(Y(b,x_{3})Y(a,x_{1})
u^{1}\otimes u^{2})\rangle \nonumber\\
& &-{\rm Res}_{x_{0}}{\rm Res}_{x_{3}}\frac{x_{2}^{1-{\rm wt}b}
x^{-\deg u^{1}-h}(1+x_{3})^{{\rm wt}b-1}(1+x_{3}+x_{0})^{{\rm wt}a-1}}
{x(1+x_{3})-x_{2}}\cdot\nonumber\\
& &\cdot\langle u', \psi_{0}(Y(Y(a,x_{0})b,x_{3})u^{1}\otimes u^{2})\rangle.
\end{eqnarray}

On the other hand, we have:
\begin{eqnarray}
& &{\rm Res}_{x_{0}}x_{0}^{-1}(x_{2}+x_{0})^{{\rm wt}a}
\langle u', I_{t}(u^{1},x)\otimes Y(Y(a,x_{0})b,x_{2})u^{2}\rangle\nonumber\\
&=&\sum_{i=0}^{\infty}\left(\begin{array}{c}{\rm wt}a\\i\end{array}\right)
\langle u', I_{t}(u^{1},x)\otimes Y(a_{i-1}b,x_{2})u^{2}\rangle
x_{2}^{{\rm wt}a-i}\nonumber\\
&=&\sum_{i=0}^{\infty}\sum_{n\le {\rm wt}a_{i-1}b-1}
\left(\begin{array}{c}{\rm wt}a\\i\end{array}\right)
\langle u', I_{t}(u^{1},x)\otimes (a_{i-1}b)_{n}u^{2}\rangle
x_{2}^{{\rm wt}a-n-i-1}\nonumber\\
&=&\sum_{i=0}^{\infty}\left(\begin{array}{c}{\rm wt}a\\i\end{array}\right)
\langle u', I_{t}(u^{1},x)\otimes o(a_{i-1}b)u^{2}\rangle
x_{2}^{-{\rm wt}b}\nonumber\\
& &+\sum_{i=0}^{\infty}\sum_{n< {\rm wt}a_{i-1}b-1}
\left(\begin{array}{c}{\rm wt}a\\i\end{array}\right)
\langle u', I_{t}(u^{1},x)\otimes (a_{i-1}b)_{n}u^{2}\rangle
x_{2}^{{\rm wt}a-n-i-1}\nonumber\\
&=&\langle u', I_{t}(u^{1},x)\otimes o(a*b)u^{2}\rangle
x_{2}^{-{\rm wt}b}\nonumber\\
& &-\sum_{i=0}^{\infty}\sum_{n< {\rm wt}a_{i-1}b-1}\sum_{j=0}^{\infty}
\left(\begin{array}{c}{\rm wt}a\\i\end{array}\right)
\left(\begin{array}{c}n\\j\end{array}\right)x_{2}^{{\rm wt}a-n-i-1}x^{n-j}
\cdot\nonumber\\
& &\cdot \langle u', I_{t}((a_{i-1}b)_{j}u^{1},x)\otimes u^{2}\rangle
\nonumber\\
&=&\langle u', I_{t}(u^{1},x)\otimes o(a*b)u^{2}\rangle
x_{2}^{-{\rm wt}b}\nonumber\\
& &-\sum_{i=0}^{\infty}\sum_{n< {\rm wt}a_{i-1}b-1}\sum_{j=0}^{\infty}
\left(\begin{array}{c}{\rm wt}a\\i\end{array}\right)
\left(\begin{array}{c}n\\j\end{array}\right)x_{2}^{{\rm
wt}a-n-i-1}x^{n-{\rm wt}a-{\rm wt}b+i+1-\deg u^{1}-h}
\cdot\nonumber\\
& &\cdot \langle u', \psi_{0}((a_{i-1}b)_{j}u^{1}\otimes u^{2})\rangle
\nonumber\\
&=&\langle u', I_{t}(u^{1},x)\otimes o(a*b)u^{2}\rangle
x_{2}^{-{\rm wt}b}\nonumber\\
& &-{\rm Res}_{x_{3}}\sum_{i=0}^{\infty}\sum_{n< {\rm wt}a_{i-1}b-1}
\left(\begin{array}{c}{\rm wt}a\\i\end{array}\right)x_{2}^{{\rm
wt}a-n-i-1}x^{n-{\rm wt}a-{\rm wt}b+i+1-\deg u^{1}-h}
\cdot\nonumber\\
& &\cdot (1+x_{3})^{n}\langle u', \psi_{0}(Y(a_{i-1}b,x_{3})u^{1}\otimes u^{2})
\rangle\nonumber\\
&=&\langle u', I_{t}(u^{1},x)\otimes o(a*b)u^{2}\rangle
x_{2}^{-{\rm wt}b}\nonumber\\
& &-{\rm Res}_{x_{3}}\sum_{i=0}^{\infty}\sum_{j=0}^{\infty}
\left(\begin{array}{c}{\rm wt}a\\i\end{array}\right)x_{2}^{1-{\rm wt}b+j}
x^{-1-j-\deg u^{1}-h}\cdot\nonumber\\
& &\cdot (1+x_{3})^{{\rm wt}a+{\rm wt}b-i-j-2}
\langle u', \psi_{0}(Y(a_{i-1}b,x_{3})u^{1}\otimes u^{2})
\rangle\nonumber\\
&=&\langle u', I_{t}(u^{1},x)\otimes o(a*b)u^{2}\rangle
x_{2}^{-{\rm wt}b}\nonumber\\
& &-{\rm Res}_{x_{3}}\sum_{i=0}^{\infty}
\left(\begin{array}{c}{\rm wt}a\\i\end{array}\right)
\frac{x_{2}^{1-{\rm wt}b}x^{-\deg u^{1}-h}(1+x_{3})^{{\rm wt}a+{\rm wt }b-i-1}}
{x(1+x_{3})-x_{2}}\cdot\nonumber\\
& &\cdot\langle u', \psi_{0}(Y(a_{i-1}b,x_{3})u^{1}\otimes u^{2})
\rangle\nonumber\\
&=&\langle u', I_{t}(u^{1},x)\otimes o(a*b)u^{2}\rangle
x_{2}^{-{\rm wt}b}\nonumber\\
& &-{\rm Res}_{x_{0}}{\rm Res}_{x_{3}}\sum_{i=0}^{\infty}
\left(\begin{array}{c}{\rm wt}a\\i\end{array}\right)
\frac{x_{2}^{1-{\rm wt}b}x^{-\deg u^{1}-h}(1+x_{3})^{{\rm wt}a+{\rm wt }b-i-1}}
{x(1+x_{3})-x_{2}}x_{0}^{i-1}\cdot\nonumber\\
& &\cdot\langle u', \psi_{0}(Y(Y(a,x_{0})b,x_{3})u^{1}\otimes u^{2})
\rangle\nonumber\\
&=&\langle u', I_{t}(u^{1},x)\otimes o(a*b)u^{2}\rangle
x_{2}^{-{\rm wt}b}\nonumber\\
& &-{\rm Res}_{x_{0}}{\rm Res}_{x_{3}}
\frac{x_{2}^{1-{\rm wt}b}x^{-\deg u^{1}-h}(1+x_{3})^{{\rm wt }b-1}}
{x(1+x_{3})-x_{2}}\frac{(1+x_{3}+x_{0})^{{\rm wt}a}}{x_{0}}\cdot\nonumber\\
& &\cdot \langle u', \psi_{0}(Y(Y(a,x_{0})b,x_{3})u^{1}\otimes u^{2})
\rangle\nonumber\\
&=&\langle u', I_{t}(u^{1},x)\otimes o(a*b)u^{2}\rangle
x_{2}^{-{\rm wt}b}\nonumber\\
& &-{\rm Res}_{x_{0}}{\rm Res}_{x_{3}}
\frac{x_{2}^{1-{\rm wt}b}x^{-\deg u^{1}-h}(1+x_{3})^{{\rm wt }b-1}
(1+x_{3}+x_{0})^{{\rm wt}a-1}}{x(1+x_{3})-x_{2}}\cdot\nonumber\\
& &\cdot \langle u', \psi_{0}(Y(Y(a,x_{0})b,x_{3})u^{1}\otimes u^{2})
\rangle\nonumber\\
& &-{\rm Res}_{x_{0}}{\rm Res}_{x_{3}}
\frac{x_{2}^{1-{\rm wt}b}x^{-\deg u^{1}-h}(1+x_{3})^{{\rm wt }b}
(1+x_{3}+x_{0})^{{\rm wt}a-1}}
{(x(1+x_{3})-x_{2})x_{0}}\cdot\nonumber\\
& &\cdot \langle u', \psi_{0}(Y(Y(a,x_{0})b,x_{3})u^{1}\otimes u^{2})
\rangle\nonumber\\
&=&\langle u', I_{t}(u^{1},x)\otimes o(a*b)u^{2}\rangle
x_{2}^{-{\rm wt}b}\nonumber\\
& &-{\rm Res}_{x_{0}}{\rm Res}_{x_{3}}
\frac{x_{2}^{1-{\rm wt}b}x^{-\deg u^{1}-h}(1+x_{3})^{{\rm wt }b-1}
(1+x_{3}+x_{0})^{{\rm wt}a-1}}{x(1+x_{3})-x_{2}}\cdot\nonumber\\
& &\cdot \langle u', \psi_{0}(Y(Y(a,x_{0})b,x_{3})u^{1}\otimes u^{2})
\rangle\nonumber\\
& &-{\rm Res}_{x_{0}}{\rm Res}_{x_{1}}{\rm Res}_{x_{3}}
\frac{x_{2}^{1-{\rm wt}b}x^{-\deg u^{1}-h}(1+x_{3})^{{\rm wt }b}
(1+x_{3}+x_{0})^{{\rm wt}a-1}}
{(x(1+x_{3})-x_{2})x_{0}}\cdot\nonumber\\
& &\cdot x_{0}^{-1}\delta\left(\frac{x_{1}-x_{3}}{x_{0}}\right)
\langle u', \psi_{0}(Y(a,x_{1})Y(b,x_{3})u^{1}\otimes u^{2})
\rangle\nonumber\\
& &+{\rm Res}_{x_{0}}{\rm Res}_{x_{1}}{\rm Res}_{x_{3}}
\frac{x_{2}^{1-{\rm wt}b}x^{-\deg u^{1}-h}(1+x_{3})^{{\rm wt }b}
(1+x_{3}+x_{0})^{{\rm wt}a-1}}
{(x(1+x_{3})-x_{2})x_{0}}\cdot\nonumber\\
& &\cdot x_{0}^{-1}\delta\left(\frac{-x_{3}+x_{1}}{x_{0}}\right)
\langle u', \psi_{0}(Y(b,x_{3})Y(a,x_{1})u^{1}\otimes u^{2})
\rangle\nonumber\\
&=&\langle u', I_{t}(u^{1},x)\otimes o(a*b)u^{2}\rangle
x_{2}^{-{\rm wt}b}\nonumber\\
& &-{\rm Res}_{x_{0}}{\rm Res}_{x_{3}}
\frac{x_{2}^{1-{\rm wt}b}x^{-\deg u^{1}-h}(1+x_{3})^{{\rm wt }b-1}
(1+x_{3}+x_{0})^{{\rm wt}a-1}}{x(1+x_{3})-x_{2}}\cdot\nonumber\\
& &\cdot \langle u', \psi_{0}(Y(Y(a,x_{0})b,x_{3})u^{1}\otimes u^{2})
\rangle\nonumber\\
& &-{\rm Res}_{x_{1}}{\rm Res}_{x_{3}}
\frac{x_{2}^{1-{\rm wt}b}x^{-\deg u^{1}-h}(1+x_{3})^{{\rm wt }b}
(1+x_{1})^{{\rm wt}a-1}}
{(x(1+x_{3})-x_{2})(x_{1}-x_{3})}\cdot\nonumber\\
& &\cdot \langle u', \psi_{0}(Y(a,x_{1})Y(b,x_{3})u^{1}\otimes u^{2})
\rangle\nonumber\\
& &-{\rm Res}_{x_{1}}{\rm Res}_{x_{3}}
\frac{x_{2}^{1-{\rm wt}b}x^{-\deg u^{1}-h}(1+x_{3})^{{\rm wt }b}
(1+x_{1})^{{\rm wt}a-1}}
{(x(1+x_{3})-x_{2})(x_{3}-x_{1})}\cdot\nonumber\\
& &\cdot \langle u', \psi_{0}(Y(b,x_{3})Y(a,x_{1})u^{1}\otimes u^{2})
\rangle.
\end{eqnarray}

By comparing (4.2.48) with (4.2.49), we see that it is sufficient to prove
the following:
\begin{eqnarray}
& &{\rm Res}_{x_{1}}{\rm Res}_{x_{3}}\frac{(1+x_{3})^{{\rm
wt}b}(1+x_{1})^{{\rm wt}a-1}}{(x_{1}-x_{3})(x(1+x_{3})-x_{2})}
Y(a,x_{1})Y(b,x_{2})u^{1}\nonumber\\
&=&{\rm Res}_{x_{1}}{\rm Res}_{x_{3}}\frac{(1+x_{3})^{{\rm
wt}b-1}(1+x_{1})^{{\rm wt}a-1}}{x_{1}(x(1+x_{3})-x_{2})}
Y(a,x_{1})Y(b,x_{2})u^{1}\;\;\mbox{mod }O(M^{1});\\
& &{\rm Res}_{x_{1}}{\rm Res}_{x_{3}}\frac{(1+x_{3})^{{\rm
wt}b}(1+x_{1})^{{\rm wt}a-1}(x_{2}-x)}{(x_{3}-x_{1})(x(1+x_{3})-x_{2})}
Y(b,x_{3})Y(a,x_{1})u^{1}\nonumber\\
&=&-{\rm Res}_{x_{1}}{\rm Res}_{x_{3}}\frac{(1+x_{3})^{{\rm
wt}b}(1+x_{1})^{{\rm wt}a-1}}{x_{3}(x(1+x_{1})-x_{2})(x(1+x_{3})-x_{2})}
Y(b,x_{3})Y(a,x_{1})u^{1}\;\;\mbox{mod }O(M^{1}).\nonumber\\
& &\mbox{}
\end{eqnarray}
First we have:
\begin{eqnarray}
& &\frac{(1+x_{3})^{{\rm
wt}b}(1+x_{1})^{{\rm wt}a-1}}{(x_{1}-x_{3})(x(1+x_{3})-x_{2})}-
\frac{(1+x_{3})^{{\rm
wt}b-1}(1+x_{1})^{{\rm wt}a-1}}{x_{1}(x(1+x_{3})-x_{2})}\nonumber\\
&=&\frac{x_{3}(1+x_{3})^{{\rm
wt}b-1}(1+x_{1})^{{\rm wt}a}}{x_{1}(x_{1}-x_{3})(x(1+x_{3})-x_{2})}
\nonumber\\
&=&\sum_{i=0}^{\infty}\frac{(1+x_{1})^{{\rm wt}a}}{x_{1}^{2+i}}
\frac{x_{3}^{i+1}(1+x_{3})^{{\rm wt}b-1}}{(x(1+x_{3})-x_{2})},
\end{eqnarray}
and
\begin{eqnarray}
& &\frac{(1+x_{3})^{{\rm
wt}b}(1+x_{1})^{{\rm wt}a-1}(x_{2}-x)}{(x_{3}-x_{1})(x(1+x_{3})-x_{2})}+
\frac{(1+x_{3})^{{\rm
wt}b}(1+x_{1})^{{\rm wt}a-1}}{x_{3}(x(1+x_{1})-x_{2})(x(1+x_{3})-x_{2})}
\nonumber\\
&=&\frac{(1+x_{3})^{{\rm wt}b}(1+x_{1})^{{\rm wt}a-1}x_{1}}{x_{3}(x_{3}-x_{1})
(x(1+x_{1})-x_{2})}\nonumber\\
&=&\sum_{i=0}^{\infty}\frac{(1+x_{3})^{{\rm wt}b}}{x_{3}^{2+i}}
\frac{(1+x_{1})^{{\rm wt}a-1}x_{1}^{i+1}}{x(1+x_{1})-x_{2}}.
\end{eqnarray}
Since
\begin{eqnarray}
{\rm Res}_{x}\frac{(1+x)^{{\rm wt }c}}{x^{2+i}}Y(c,x)M^{1}\subseteq O(M^{1})
\end{eqnarray}
for any homogeneous element $c\in V$, (4.2.50) and (4.2.51)
follow immediately.$\;\;\;\;\;\Box$

{\bf Proposition 4.2.14.} {\it For any $u'\in (M^{3}(0))^{*}, a,b\in V,
u^{1}\in M^{1},u^{2}\in M(M^{2}(0))$, we have}
\begin{eqnarray}
& &x_{0}^{-1}\delta\left(\frac{x_{1}-x_{2}}{x_{0}}\right)
\langle u',I_{t}(u^{1},x)\otimes Y(a,x_{1})Y(b,x_{2})u^{2}\rangle\nonumber\\
& &-x_{0}^{-1}\delta\left(\frac{-x_{2}+x_{1}}{x_{0}}\right)
\langle u',I_{t}(u^{1},x)\otimes Y(b,x_{2})Y(a,x_{1})u^{2}\rangle\nonumber\\
&=&x_{2}^{-1}\delta\left(\frac{x_{1}-x_{0}}{x_{2}}\right)
\langle u',I_{t}(u^{1},x)\otimes Y(Y(a,x_{0})b,x_{2})u^{2}\rangle.
\end{eqnarray}

{\bf Proof.} It is similar to the proof of Proposition 3.2.5 and
Proposition 4.2.11.
Let $W$ be the subspace of $M(M^{2}(0))$ consisting of
each $u$ satisfying (4.2.55) for any $u'\in (M^{3}(0))^{*}, a,b\in V,
u^{1}\in M^{1}$. Then it is
equivalent to prove that $W=U(g_{+})M^{2}(0)$. By Proposition 4.2.12,
$M^{2}(0)\subseteq W$, so that it is enough to prove that
$g_{+}W\subseteq W$. Let $u\in W$, let $c\in V$ be any homogeneous
element and let $n$ be any integer less than ${\rm wt} c-1$ so that
$\deg c_{n}>0$. Then by choosing $k$ large enough we have:
\begin{eqnarray}
& &\langle u',I_{t}(u^{1},x)\otimes
Y(a,x_{0}+x_{2})Y(b,x_{2})c_{n}u\rangle (x_{0}+x_{2})^{k}\nonumber\\
&=&\sum_{i=0}^{\infty}\left(\begin{array}{c}n\\i\end{array}\right)
x_{2}^{n-i}\langle u',I_{t}(u^{1},x)\otimes
Y(a,x_{0}+x_{2})Y(c_{i}b,x_{2})u\rangle (x_{0}+x_{2})^{k}\nonumber\\
& &+\sum_{i=0}^{\infty}\left(\begin{array}{c}n\\i\end{array}\right)
(x_{0}+x_{2})^{k+n-i}\langle u',I_{t}(u^{1},x)\otimes
Y(c_{i}a,x_{0}+x_{2})Y(b,x_{2})u\rangle\nonumber\\
& &+\sum_{i=0}^{\infty}\left(\begin{array}{c}n\\i\end{array}\right)
x^{n-i}\langle u',I_{t}(c_{i}u^{1},x)\otimes
Y(a,x_{0}+x_{2})Y(b,x_{2})u\rangle (x_{0}+x_{2})^{k}\nonumber\\
& &+\langle u',c_{n}\left(I_{t}(u^{1},x)\otimes
Y(a,x_{0}+x_{2})Y(b,x_{2})u\right)\rangle (x_{0}+x_{2})^{k}\nonumber\\
&=&\sum_{i=0}^{\infty}\left(\begin{array}{c}n\\i\end{array}\right)
x_{2}^{n-i}\langle u',I_{t}(u^{1},x)\otimes
Y(Y(a,x_{0})c_{i}b,x_{2})u\rangle (x_{0}+x_{2})^{k}\nonumber\\
& &+\sum_{i=0}^{\infty}\left(\begin{array}{c}n\\i\end{array}\right)
(x_{0}+x_{2})^{k+n-i}\langle u',I_{t}(u^{1},x)\otimes
Y(Y(c_{i}a,x_{0}b,x_{2})u\rangle\nonumber\\
& &+\sum_{i=0}^{\infty}\left(\begin{array}{c}n\\i\end{array}\right)
x^{n-i}\langle u',I_{t}(c_{i}u^{1},x)\otimes
Y(Y(a,x_{0})b,x_{2})u\rangle (x_{0}+x_{2})^{k}\nonumber\\
&=&\langle u',I_{t}(u^{1},x)\otimes
Y(Y(a,x_{0})b,x_{2})c_{n}u\rangle (x_{0}+x_{2})^{k}.
\end{eqnarray}
Thus $c_{n}u\in W$. Then $g_{+}W\subseteq W$. Therefore
$W=U(g_{+})M^{2}(0)$. $\;\;\;\;\Box$

{\bf Proposition 4.2.15.} {\it For any $u'\in (M^{3}(0))^{*},y\in
U(g),a,b\in V, u^{1}\in M^{1},u^{2}\in M(M^{2}(0))$, we have}
\begin{eqnarray}
& &x_{0}^{-1}\delta\left(\frac{x_{1}-x_{2}}{x_{0}}\right)
\langle u',y\cdot (I_{t}(u^{1},x)\otimes Y(a,x_{1})Y(b,x_{2})u^{2})
\rangle\nonumber\\
& &-x_{0}^{-1}\delta\left(\frac{-x_{2}+x_{1}}{x_{0}}\right)
\langle u',y\cdot (I_{t}(u^{1},x)\otimes Y(b,x_{2})Y(a,x_{1})u^{2})
\rangle\nonumber\\
&=&x_{2}^{-1}\delta\left(\frac{x_{1}-x_{0}}{x_{2}}\right)
\langle u',y\cdot (I_{t}(u^{1},x)\otimes Y(Y(a,x_{0})b,x_{2})u^{2})\rangle.
\end{eqnarray}
{\it or, equivalently}
\begin{eqnarray}
& &x_{0}^{-1}\delta\left(\frac{x_{1}-x_{2}}{x_{0}}\right)
I_{t}(u^{1},x)\otimes Y(a,x_{1})Y(b,x_{2})u^{2}\nonumber\\
& &-x_{0}^{-1}\delta\left(\frac{-x_{2}+x_{1}}{x_{0}}\right)
I_{t}(u^{1},x)\otimes Y(b,x_{2})Y(a,x_{1})u^{2}\nonumber\\
&=&x_{2}^{-1}\delta\left(\frac{x_{1}-x_{0}}{x_{2}}\right)
I_{t}(u^{1},x)\otimes Y(Y(a,x_{0})b,x_{2})u^{2}\;\;\;\mbox{mod }
I[[x_{0},x_{1},x_{2},x_{0}^{-1},x_{1}^{-1},x_{2}^{-1}]].\nonumber\\
& &\mbox{}
\end{eqnarray}

{\bf Proof.} Let $L$ be the subspace of $U(g)$ consisting of each $y$
such that (4.2.57) holds. Then it follows from Lemma 4.2.7 that
$g_{+}U(g)\subseteq L$. By PBW theorem it is enough to prove that
$g_{0}\subseteq L$ and $Lg_{-}\subseteq L$. Let $y\in L$, let $c\in V$
be a homogeneous element and let $n$ be an integer such that $n\ge {\rm
wt }c-1$. Then
\begin{eqnarray}
& &\langle u',yc_{n}\left(I_{t}(u^{1},x)\otimes
Y(a,x_{0}+x_{2})Y(b,x_{2})u\right)\rangle (x_{0}+x_{2})^{k}\nonumber\\
&=&\sum_{i=0}^{\infty}\left(\begin{array}{c}n\\i\end{array}\right)
x^{n-i}\langle u',y\left(I_{t}(c_{i}u^{1},x)\otimes
Y(a,x_{0}+x_{2})Y(b,x_{2})u\right)\rangle (x_{0}+x_{2})^{k}\nonumber\\
& &+\sum_{i=0}^{\infty}\left(\begin{array}{c}n\\i\end{array}\right)
(x_{0}+x_{2})^{k+n-i}\langle u',y\left(I_{t}(u^{1},x)\otimes
Y(c_{i}a,x_{0}+x_{2})Y(b,x_{2})u\right)\rangle\nonumber\\
& &+\sum_{i=0}^{\infty}\left(\begin{array}{c}n\\i\end{array}\right)
x_{2}^{n-i}\langle u',y\left(I_{t}(u^{1},x)\otimes
Y(a,x_{0}+x_{2})Y(c_{i}b,x_{2})u\right)\rangle (x_{0}+x_{2})^{k}\nonumber\\
& &+\langle u',y\left(I_{t}(u^{1},x)\otimes
Y(a,x_{0}+x_{2})Y(b,x_{2})c_{n}u\right)\rangle (x_{0}+x_{2})^{k}\nonumber\\
&=&\sum_{i=0}^{\infty}\left(\begin{array}{c}n\\i\end{array}\right)
x^{n-i}\langle u',y\left(I_{t}(c_{i}u^{1},x)\otimes
Y(Y(a,x_{0})b,x_{2})u\right)\rangle (x_{0}+x_{2})^{k}\nonumber\\
& &+\sum_{i=0}^{\infty}\left(\begin{array}{c}n\\i\end{array}\right)
(x_{0}+x_{2})^{k+n-i}\langle u',y\left(I_{t}(u^{1},x)\otimes
Y(Y(c_{i}a,x_{0})b,x_{2})u\right)\rangle\nonumber\\
& &+\sum_{i=0}^{\infty}\left(\begin{array}{c}n\\i\end{array}\right)
x_{2}^{n-i}\langle u',y\left(I_{t}(u^{1},x)\otimes
Y(Y(a,x_{0})c_{i}b,x_{2})u\right)\rangle (x_{0}+x_{2})^{k}\nonumber\\
& &+\langle u',y\left(I_{t}(u^{1},x)\otimes
Y(Y(a,x_{0})b,x_{2})c_{n}u\right)\rangle (x_{0}+x_{2})^{k}\nonumber\\
&=&\langle u',yc_{n}\left(I_{t}(u^{1},x)\otimes
Y(Y(a,x_{0})b,x_{2})u\right)\rangle (x_{0}+x_{2})^{k}.
\end{eqnarray}
Thus $L(g_{-}+g_{0})\subseteq L$. Therefore $L=U(g)$.$\;\;\;\;\Box$

It follows from Proposition 4.2.15 that the bilinear form on
$(M^{3}(0))^{*}\times F$ induces a
bilinear form on $(M^{3}(0))^{*}\times {\bf C}[t,t^{-1}]\otimes
M^{1}\otimes \bar{M}(M^{2}(0))$. Now we extend this bilinear form to a
bilinear form on $M((M^{3}(0))^{*})\times {\bf C}[t,t^{-1}]\otimes
M^{1}\otimes \bar{M}(M^{2}(0))$ as follows:
\begin{eqnarray}
\langle yu', u\rangle=\langle u',\theta (y)u\rangle
\end{eqnarray}
for any $y\in U({\bf g}(V)), u'\in (M^{3}(0))^{*}$, where $\theta$
is the anti-automorphism of $U({\bf g}(V))$ or ${\bf g}(V)$ given in
Section 2.2.

{\bf Proposition 4.2.16.} {\it The ${\bf g}(V)$-module $T=\left({\bf
C}[t,t^{-1}]\otimes M^{1}\otimes \bar{M}(M^{2}(0))\right)/I$ is a weak
$V$-module.}

{\bf Proof.} This will follow from the proof of Theorem 5.2.9.$\;\;\;\;\Box$

It follows from Proposition 4.2.16 and FHL's contragredient module theory
that we have an induced bilinear form on $\bar{M}((M^{3}(0))^{*})\times
T$.

{\bf Proof of Theorem 4.2.4.} By the assumptions on $M^{1}$ and
$M^{3}$, we have a
bilinear form on $(M^{3})'\times {\bf C}[t,t^{-1}]\otimes M^{1}\otimes
M^{2}$. Then we have a bilinear map $\psi$ from ${\bf
C}[t,t^{-1}]\otimes M^{1}\otimes M^{2}$ to $M^{3}$. We define
\begin{eqnarray}
I(\cdot,x)\cdot :& &M^{1}\otimes M^{2}\rightarrow M^{3}\{x\};\nonumber\\
& &I(u^{1},x)u^{2}=\psi (I_{t}(u^{1},x)\otimes
u^{2})\;\;\;\mbox{ for }u^{1}\in M^{1},u^{2}\in M^{2}.
\end{eqnarray}
Then $I(\cdot,x)$ satisfies the $L(-1)$-derivative property
(Proposition 4.2.10) and the
Jacobi identity (Proposition 4.2.15).
Therefore, we have obtained an intertwining operator
$I(\cdot,x)$ corresponding to $\psi_{0}$.$\;\;\;\;\Box$

{\bf Remark 4.2.17.} A sufficient condition for Theorem 4.2.4 is the
quasi-rationality of the vertex operator algebra $V$ in the sense that
any module is completely reducible. In the Appendix A, we will give an
example to show that the conditions on $M^{2}$ and $M^{3}$ are necessary.

\newpage

\chapter{Tensor products and a construction}

In the classical Lie algebra level, the tensor product vector space of
two modules for a Lie algebra $g$ has a natural $g$-module structure
due to the Hopf algebra structure of the universal
enveloping algebra $U(g)$. An intertwining operator of type
$\left(\begin{array}{c}W_{3}\\W_{1},W_{2}\end{array}\right)$ can be
defined as a
$g$-module homomorphism from the tensor product module $W_{1}\otimes
W_{2}$ to $W_{3}$. It can also be defined (at least superficially)
without using the notion of tensor product as a linear map $\psi$ from $W_{1}$
to ${\rm Hom}_{{\bf C}}(W_{2},W_{3})$ such that
$$a\psi(w_{1})w_{2}=\psi(aw_{1})w_{2}+\psi(w_{1})aw_{2}\;\;\;\mbox{for
any }a\in g, w_{i}\in W_{i}.$$
In vertex operator algebra theory, the notion of intertwining operator
has been defined [FHL] as a certain generalization of the notion of module,
but there was initially no notion of
tensor product for modules for a vertex operator algebra.
This definition  of an
intertwining operator is analogous to the second definition of an
intertwining operator in the classical case.

To obtain an analogous construction of tensor product for modules for a
vertex operator algebra, we consider the classical case in the following way.
Suppose that we did not have the notion of tensor product but we
only had the notion of intertwining operator as a linear map
satisfying a certain condition (the second definition). Then the tensor product
module can be defined by using the notion of intertwining operator in
terms of a certain universal property and can be constructed by using
generating space and defining relations.
This consideration motivates our current approach to tensor product theory for
modules for a vertex operator algebra.
In this chapter we shall realize this idea for vertex operator
algebra theory to construct tensor product for modules for a vertex operator
algebra.

Throughout this chapter, $V$ will be a fixed vertex operator
algebra.

\section{ A definition for tensor product}
In this section we shall first formulate a definition of a tensor product in
terms of a certain universal property as an analogue of the notion of
the classical tensor product. Then we present some results
about abstract nonsense.

{\bf Definition 5.1.1.} Let $M^{1}$ and $M^{2}$ be two $V$-modules. A
 pair
$(M,F(\cdot,x))$, which consists of a $V$-module $M$ and an intertwining
operator $F(\cdot,x)$ of type
$\left(\begin{array}{c}M\\M^{1},M^{2}\end{array}\right)$, is called a
{\it tensor product} for the ordered pair $(M^{1},M^{2})$ if the
following universal
property holds: For any $V$-module $W$ and any intertwining
operator $I(\cdot,x)$ of type
$\left(\begin{array}{c}W\\M^{1},M^{2}\end{array}\right)$, there
exists a unique $V$-homomorphism $\psi$ from $M$ to $W$ such that
$I(\cdot,x)=\psi\circ F(\cdot,x)$. (Here $\psi$ extends canonically to a
linear map from $M\{x\}$ to $W\{x\}$.)

{\bf Remark 5.1.2.} Just as in the classical algebra theory, it follows
from the universal property that if there exists a tensor product
$(M,F(\cdot,x))$
for the ordered pair $(M^{1},M^{2})$, then it is unique up to
$V$-module isomorphism, i.e., if $(W,G(\cdot,x))$ is another tensor
product, then there is a $V$-module isomorphism $\psi$ from $M$ to $W$
such that $G=\psi \circ F$. Conversely, let $(M,F(\cdot,x))$ be a tensor
product for the ordered pair $(M^{1},M^{2})$ and let $\sigma$ be an
automorphism of the $V$-module $M$. Then $(M, \sigma\circ F(\cdot,x))$
is a tensor product for $(M^{1},M^{2})$.

{\bf Lemma 5.1.3.} {\it Let $(W,F(\cdot,x))$ is a tensor product for the
ordered pair $(M^{1},M^{2})$. Then $F(\cdot,x)$ is surjective in the
sense that all the coefficients of $F(u^{1},x)u^{2}$ for $u^{i}\in
M^{i}$ linearly span $W$.}

{\bf Proof.} Let $\overline{W}$ be the linear span of all the coefficients
of $F(u^{1},x)u^{2}$ for $u^{i}\in M^{i}$. Then $\overline{W}$ is a
submodule of $W$ and $F(\cdot,x)$ is an intertwining operator
of type
$\left(\begin{array}{c}\overline{W}\\M^{1},M^{2}\end{array}\right)$. Then
by Remark 5.1.2 it is equivalent to prove that $(\overline{W},F(\cdot,x))$
is also a tensor product for the ordered pair $(M^{1},M^{2})$.
It follows from the
universal property of $(W, F(\cdot,x))$ that there is a $V$-module
homomorphism $\psi$ from $W$ to $\overline{W}$ such that
\begin{eqnarray}
F(u^{1},x)u^{2}=\psi(F(u^{1},x)u^{2})\;\;\;\mbox{ for }u^{1}\in M^{1},
u^{2}\in M^{2}.
\end{eqnarray}
Then $\psi(w)=w$ for any $w\in \overline{W}$. Therefore we have:
$W=\overline{W}\oplus \ker \psi$. Let $M$ be any $V$-module and
$I(\cdot,x)$ be an intertwining operator of type
$\left(\begin{array}{c}M\\M^{1},M^{2}\end{array}\right)$. Then there
is a $V$-module homomorphism $\psi$ from $W$ to $M$ such that
$I(\cdot,x)=\psi\circ F(\cdot,x)$. Let $\bar{\psi}$  be the
restriction of $\psi$ on $\overline{W}$. Then  $I(\cdot,x)=\bar{\psi}\circ
F(\cdot,x)$. Suppose there is another $V$-module homomorphism $\phi$ from
$\overline{W}$ to $M$ such that  $I(\cdot,x)=\phi\circ F(\cdot,x)$. Then
 $I(\cdot,x)=(\phi\circ P)\circ F(\cdot,x)$, where $P$ is the projection map
from $W$ onto $\overline{W}$. It follows from the universal property of
$(W,F(\cdot,x))$ that $\psi =\phi\circ P$. Thus $\bar{\psi}=\phi$.
Therefore we have proved that $(\overline{W},F(\cdot,x))$ is a tensor
product for the ordered pair $(M^{1},M^{2})$. Then the proof is complete.
$\;\;\;\;\Box$

{\bf Remark 5.1.4.} Suppose that there exists a tensor product
$(W,F(\cdot,x))$ for the ordered pair $(M^{1},M^{2})$. Then it follows
from Lemma 5.1.3 that $W$ is a quotient space of the vector space
$S(\Phi)\otimes M^{1}\otimes M^{2}$, where $\Phi$
is the set of all powers of $x^{-1}$ with non identically zero coefficients
in the expression of $F(\cdot,x)$ and $S(\Phi)$ is the subspace
linearly spanned by $\Phi$ in the group algebra of ${\bf C}$ (as an
abelian group).

{\bf Proposition 5.1.5.} {\it If $(M,F(\cdot,x))$ is a tensor product for the
ordered pair $(M^{1},M^{2})$ of $V$-modules, then for any $V$-module
$M^{3}$, ${\rm Hom}_{V}(M,M^{3})$ is linearly isomorphic to the space
of intertwining operators of type
$\left(\begin{array}{c}M^{3}\\M^{1},M^{2}\end{array}\right)$.}

{\bf Proof.} Let $\phi$ be any $V$-homomorphism from $M$ to $M^{3}$.
Then $\phi F(\cdot,x)$ is an intertwining operator of type
$\left(\begin{array}{c}M^{3}\\M^{1},M^{2}\end{array}\right)$. Thus we
obtain a linear map $\pi$ from ${\rm Hom}_{V}(M,M^{3})$ to
$I\left(\begin{array}{c}M^{3}\\M^{1},M^{2}\end{array}\right)$ defined
by $\pi (\phi)=\phi F(\cdot,x)$. Since $F(\cdot,x)$ is surjective
(Lemma 5.1.3), $\pi$ is injective. On the other hand, the universal
property of $(W, F(\cdot,x))$ implies that $\pi$ is
surjective.$\;\;\;\;\Box$

{\bf Proposition 5.1.6.} {\it Let $M$ be any $V$-module. Then $(M,
Y_{M}(\cdot,x))$ is
a tensor product for $(V,M)$. Symmetrically, $(M,F(\cdot,x))$
is a tensor product for $(M,V)$ where $F(\cdot,x)$ is defined by
$F(u,x)a=e^{xL(-1)}Y_{M}(a,-x)u$ for $a\in V,u\in M$.}

{\bf Proof}. It follows from the $S_{3}$-symmetry [FHL] of the Jacobi
identity that $Y_{M}(\cdot,x)$ and $F(\cdot,x)$ are intertwining operators of
corresponding types.
Let $W$ be any $V$-module and let $I(\cdot,x)$ be any
intertwining operator of type
$\left(\begin{array}{c}W\\V,M\end{array}\right)$. Since $\displaystyle{{d\over
dx}}I({\bf 1},x)=I(L(-1){\bf 1},x)=0$, $I({\bf 1},x)$ is independent of
$x$. It follows from the
commutator formula that $I({\bf 1},x)$ commutes with any vertex
operator $Y(a,x)$ for $a\in V$. Then $I({\bf 1},x)$ is a
$V$-homomorphism from $M$ to $W$. Furthermore, we have
\begin{eqnarray}
& &I(a,x_{2})u\nonumber\\
&=&{\rm Res}_{x_{0}}x_{0}^{-1}I(Y(a,x_{0}){\bf 1},x_{2})u\nonumber\\
&=&{\rm Res}_{x_{1}}{\rm Res}_{x_{0}}x_{0}^{-1}x_{2}^{-1}\delta\left(
\frac{x_{1}-x_{0}}{x_{2}}\right)I({\bf 1},x_{2})Y_{M}(a,x_{1})u\nonumber\\
&=&{\rm Res}_{x_{0}}z^{-1}_{0}I({\bf 1},x_{2})Y_{M}(a,x_{2}+x_{0})u\nonumber\\
&=&I({\bf 1},x_{2})Y_{M}(a,x_{2})u\end{eqnarray}
for $a\in V,u\in M$. On the other hand, if $\psi$ is a
$V$-homomorphism from $M$ to $W$ such that $I(a,x)u=\psi (Y_{M}(a,x)u)$
for $a\in V,u\in M$, then $\psi (u)=I({\bf 1},x)u$. Thus the universal
property is satisfied. Therefore, $(M,Y_{M}(\cdot,x))$ is a tensor product for
$(V,M)$. Similarly, the universal property for
$(M,F(\cdot,x))$ can be proved.$\;\;\;\;\Box$

{\bf Proposition 5.1.7.} {\it If $(M,F(\cdot,x))$ is a tensor product for
the ordered pair $(M^{1},M^{2})$, then $(M,F^{t}(\cdot,x))$ is a
tensor product for the ordered pair $(M^{2},M^{1})$.}

{\bf Proof.} Let $M$ be any $V$-module and let $I(\cdot,x)$ be an
intertwining operator of type
$\left(\begin{array}{c}M\\M^{1},M^{2}\end{array}\right)$. It is easy
to see that there is a $V$-module homomorphism $\psi$ from $W$ to $M$
such that
$I^{t}(\cdot,x)=\psi \circ F(\cdot,x)$ if and only if $I(\cdot,x)=\psi \circ
F^{t}(\cdot,x)$. Then the proposition follows.$\;\;\;\;\Box$

{\bf Remark 5.1.8.} Let $M^{1i}$ $(i=1,\cdots, m)$ and $M^{2j}$
$(j=1,\cdots,n)$ be $V$-modules and let $(W^{ij},F^{ij}(\cdot,x))$ be
a tensor product for the ordered pair $(M^{1i},M^{2j})$. Let
$M^{1}=\oplus_{i=1}^{m}M^{1i}$, $M^{2}=\oplus_{j=1}^{n}M^{2j}$. Then
one can prove that $(\oplus W^{ij},\oplus F^{ij}(\cdot,x))$ is a
tensor product for the ordered pair $(M^{1},M^{2})$.

{\bf Remark 5.1.9.} Let $M^{1}$ and $M^{2}$ be two
$V$-modules and let $(W,F(\cdot,x))$ be a pair consisting of a $V$-module
$W$ and an intertwining operator $F(\cdot,x)$ of type
$\left(\begin{array}{c}W\\M^{1},M^{2}\end{array}\right)$. Let $M$ be a
direct sum of $V$-modules $M^{\alpha}$ $(\alpha \in \Phi)$. Suppose
that for any intertwining operator $I_{\alpha}(\cdot,x)$ of type
$\left(\begin{array}{c}M^{\alpha}\\M^{1},M^{2}\end{array}\right)$,
there exists a unique $V$-homomorphism $\psi_{\alpha}$ from $W$ to
$M^{\alpha}$ such that
$I_{\alpha}(\cdot,x)=\psi_{\alpha}\circ F(\cdot,x)$. Then one can
easily prove that for any intertwining
operator $I(\cdot,x)$ of type
$\left(\begin{array}{c}M\\M^{1},M^{2}\end{array}\right)$, there exists
a unique $V$-homomorphism $\psi$ from $W$ to $M$ such that
$I(\cdot,x)=\psi\circ F(\cdot,x)$.
Since any $V$-module $M$ is a direct sum of $V$-modules $M^{\alpha}$
$(\alpha\in \Phi)$ such that all weights of each $M_{\alpha}$ are
congruent modulo ${\bf Z}$, in Definition 5.1.1
it is sufficient to check the universal property for each $V$-module
$M$ whose weights are congruent modulo ${\bf Z}$. If $V$ is rational,
it is sufficient to check the universal property in Definition 5.1.1
for each irreducible $V$-module $M$.

\section{A construction of tensor product modules}
In this section we shall construct a tensor product
for two modules for a given rational vertex operator algebra $V$. We first
construct a tensor product for two $V$-modules from
$S$ (where $S$ is the set of equivalence classes of $V$-modules whose
weights are congruent modulo ${\bf Z}$), then extend the
definition to any $V$-modules.

Let $M^{1}$ and $M^{2}$ be two $V$-modules from $S$. Since the notion of vertex
operator algebra is a ``complex analogue'' of the notion of Lie
algebra, one naturally considers the tensor product vector space
${\bf C}[t,t^{-1}]\otimes M^{1}\otimes M^{2}$. Set
\begin{eqnarray}
F_{0}(M^{1},M^{2})={\bf C}[t,t^{-1}]\otimes M^{1}\otimes M^{2}.
\end{eqnarray}
For $u\in M^{1}$, we set
\begin{eqnarray}
Y_{t}(u,x)=\sum_{n\in {\bf Z}}(t^{n}\otimes u)x^{-n-1}.\end{eqnarray}
Then ${\bf C}[t,t^{-1}]\otimes M^{1}$ is linearly spanned by the
coefficients of all generating elements $Y_{t}(u,x)$.

Next, we define a ${\bf Z}$-grading for $F_{0}(M^{1},M^{2})$ as follows:
For $k \in {\bf
Z}; m, n \in {\bf N},u \in M^{1}(m),v \in M^{2}(n)$, define
\begin{eqnarray}
\deg\:(t^{k}\otimes u\otimes v)=m+n-k-1.\end{eqnarray}

Let $I(\cdot,x)$ be an intertwining operator of type
$\left(\begin{array}{c}M^{3}\\M^{1},M^{2}\end{array}\right)$, and let
$I^{o}(\cdot,x)$ be the normalization given in (4.1.5). Then for
any non-zero complex number $z$ we have
\begin{eqnarray}
& &x_{0}^{-1}\delta\left(\frac{x_{1}-zx_{2}}{x_{0}}\right)Y(a,x_{1})
I^{o}(u,zx_{2})
-x_{0}^{-1}\delta\left(\frac{-zx_{2}+x_{1}}{x_{0}}\right)I^{o}(u,zx_{2})
Y(a,x_{1})\nonumber\\
&=&(zx_{2})^{-1}\delta\left(\frac{x_{1}-x_{0}}{zx_{2}}\right)
I^{o}(Y(a,x_{0})u,zx_{2}).\end{eqnarray}
Taking ${\rm Res}_{x_{0}}$ of the Jacobi
identity above, we obtain the commutator formula:
\begin{eqnarray}
& &Y(a,x_{1})I^{o}(u,zx_{2})v\nonumber\\
&=&I^{o}(u,zx_{2})Y(a,x_{1})v+
{\rm
Res}_{x_{0}}(zx_{2})^{-1}\delta\left(\frac{x_{1}-x_{0}}{zx_{2}}\right)
I^{o}(Y(a,x_{0})u,zx_{2})v     \end{eqnarray}
for $a\in V, u\in M^{1}, v\in M^{2}$.

{}From now on, $z$ will be a fixed  non-zero complex number. Motivated
by the classical construction of tensor product,
we define an action of $\hat{V}={\bf C}[t,t^{-1}]\otimes V$ on
$F_{0}(M^{1},M^{2})$ as follows: for $a \in V,u \in M^{1},M^{2}$,
\begin{eqnarray}
& &Y_{t}(a,x_{1})(Y_{t}(u,x_{2})\otimes v)\nonumber\\
&=&Y_{t}(u,x_{2})\otimes
Y(a,x_{1})v+{\rm Res}_{x_{0}}(zx_{2})^{-1}\delta\left(\frac{x_{1}-x_{0}}
{zx_{2}}\right)Y_{t}(Y(a,x_{0})u,x_{2})\otimes v.\nonumber\\
& &\mbox{}
\end{eqnarray}

{\bf Proposition 5.2.1.} {\it Under the above defined action of $\hat{V}$,
$F_{0}(M^{1},M^{2})$ becomes a ${\bf Z}$-graded ${\bf g}(V)$-module of
level one, i.e.,}
\begin{eqnarray}
& &Y_{t}({\bf 1},z)={\rm id},\;\;Y_{t}(L(-1)a,x)={d \over
dx}Y_{t}(a,x)\;\;\;\mbox{{\it for} }a \in V; \\
& &\deg\:a(n)={\rm wt}\:a-n-1\;\;\;\mbox{{\it for each
homogeneous }}a \in V, n \in {\bf Z};\\
& &[Y_{t}(a,x_{1}),Y_{t}(b,x_{2})]={\rm Res}_{x_{0}}x_{2}^{-1}\delta\left(
\frac{x_{1}-x_{0}}{x_{2}}\right)Y_{t}(Y(a,x_{0})b,x_{2})
\end{eqnarray}
{\it for $a,b\in V$.}

{\bf Proof.} Writing (5.2.6) into components we have:
\begin{eqnarray}
& &(t^{m}\otimes a)(t^{n}\otimes u\otimes v)\nonumber\\
&=&t^{n}\otimes u\otimes a_{m}u\nonumber\\
& &+{\rm Res}_{x_{0}}{\rm Res}_{x_{1}}{\rm Res}_{x_{2}}
x_{1}^{m}x_{2}^{n}x_{1}^{-1}x_{1}^{-1}\delta\left(\frac{zx_{2}+x_{0}}{x_{1}}
\right)Y_{t}((Y(a,x_{0})u,x_{2})\otimes v  \nonumber\\
&=&t^{n}\otimes u\otimes a_{m}u\nonumber\\
& &+{\rm Res}_{x_{2}}\sum_{i=0}^{\infty}
\left(\begin{array}{c}m\\i\end{array}\right)
z^{m-i}x_{2}^{m+n-i}Y_{t}(a_{i}u,x_{2})\otimes v\nonumber\\
&=&t^{n}\otimes u\otimes a_{m}u
+\sum_{i=0}^{\infty}\left(\begin{array}{c}m\\i\end{array}\right)
z^{m-i}(t^{m+n-i}\otimes a_{i}u\otimes v).
\end{eqnarray}
It follows from Lemma 2.2.6 that (5.2.6) defines a $g(V)$-module
structure on ${\bf C}[t,t^{-1}]\otimes M^{1}\otimes M^{2}$, which is a
tensor product $g(V)$-module of level-zero $g(V)$-module ${\bf
C}[t,t^{-1}]\otimes M^{1}$ with the level-one $g(V)$-module $M^{2}$.
$\;\;\;\;\Box$

Let $\theta$ be the linear automorphism of $F_{0}(M^{1},M^{2})$ given
by the left multiplication by $t$, i.e.,
\begin{eqnarray}
\theta (t^{n} \otimes u \otimes v)=t^{n+1} \otimes u \otimes
v\;\;\;\mbox{for } n\in {\bf Z},u \in M^{1},v \in M^{2}.\end{eqnarray}
Or equivalently,
\begin{eqnarray}
\theta (Y_{t}(u,x) \otimes v)=zY_{t}(u,x) \otimes v.\end{eqnarray}
Then $\theta$ is homogeneous of degree $-1$.

{\bf Proposition 5.2.2.} {\it The linear map $\theta$ is a
$g(V)$-endomorphism of $F_{0}(M^{1},M^{2})$.}

{\bf Proof}. For $a \in V,n \in {\bf Z},u \in M^{1},v \in M^{2}$, by
(5.2.12) we have
\begin{eqnarray*}& &\theta (Y_{t}(a,x_{1})(Y_{t}(u,x_{2})\otimes v))\\
&=&\theta (Y_{t}(u,x_{2})\otimes
Y(a,x_{1})v+{\rm Res}_{x_{0}}(zx_{2})^{-1}\delta
\left(\frac{x_{1}-x_{0}}{zx_{2}}\right)Y_{t}(Y(a,x_{0})u,x_{2}) \otimes v)\\
&=&x_{2}Y_{t}(u,x_{2}) \otimes Y(a,x_{1})v+{\rm Res}_{x_{0}}(zx_{2})^{-1}\delta
\left(\frac {x_{1}-x_{0}}{zx_{2}}\right)x_{2}Y_{t}(Y(a,x_{0})u,x_{2})
\otimes v\\
&=&Y_{t}(a,x_{1})(x_{2}Y_{t}(u,x_{2}) \otimes v)\\
&=&Y_{t}(a,x_{1})\theta (Y_{t}(u,x_{2}) \otimes v).\;\;\;\;\Box\end{eqnarray*}

Define $J$ to be the $g(V)$-submodule of $F_{0}(M^{1},M^{2})$ generated by the
 following subspaces:
\begin{eqnarray}
t^{m+n+i} \otimes M^{1}(m) \otimes M^{2}(n)\;\;\;\mbox{for
}m,n,i \in {\bf N}.
\end{eqnarray}

{\bf Lemma 5.2.3}. {\it The submodule $J$ is graded and preserved by
the endomorphism $\theta$.}

{\bf Proof}. Because the generating subspaces (5.2.13) are homogeneous,
$J$ is graded. Similarly, because $\theta$ preserves the generating subspaces
(5.2.13), $\theta$ preserves $J$.$\;\;\;\;\Box$

Motivated by the $L(-1)$-derivative property relation (4.1.12), for any
complex number $\lambda$, we define $L_{\lambda}$ to be the subspace of
$F_{0}(M^{1},M^{2})$ linearly spanned by the following elements:
\begin{eqnarray}
t^{n}\otimes L(-1)u \otimes v+nz^{-1}t^{n-1}\otimes u \otimes v-\lambda
z^{-1}t^{n-1}\otimes u \otimes v \end{eqnarray}
or linearly spanned by the coefficients of $z^{n}$
in the following expressions:
\begin{eqnarray}
Y_{t}(L(-1)u,x)\otimes v-z^{-1}{d \over dx}Y_{t}(u,x)\otimes v-\lambda
(zx)^{-1}Y_{t}(u,x)\otimes v \end{eqnarray}
for $u \in M^{1}, v \in M^{2}, n \in {\bf Z}$.

{\bf Lemma 5.2.4}. {\it The subspace $L_{\lambda}$ is a graded
$g(V)$-submodule of $F_{0}(M^{1},M^{2})$ satisfying the condition
$\theta(L_{\lambda})=L_{\lambda +1}\;\;\;\mbox{for }\lambda
\in {\bf C}$. }

{\bf Proof}. $L_{\lambda}$ is graded because one can get a homogeneous
basis for $L_{\lambda}$ from (5.2.15).
For $n \in {\bf Z},u \in M^{1},M^{2}$ we have
\begin{eqnarray*}& &\theta (t^{n}\otimes L(-1)u \otimes
v+nz^{-1}t^{n-1}\otimes u \otimes
v-\lambda x^{-1}t^{n-1}\otimes u \otimes v)\\
&=&t^{n+1}\otimes L(-1)u\otimes v+nz^{-1}t^{n}\otimes u \otimes v-\lambda
z^{-1}t^{n}\otimes u\otimes v\\
&=&t^{n+1}\otimes L(-1)u\otimes v+(n+1)z^{-1}t^{n}\otimes u\otimes
v-(\lambda +1)z^{-1}t^{n}\otimes u\otimes v.\end{eqnarray*}
Then $\theta L_{\lambda}=L_{\lambda +1}$ is clear.
For $a \in V, u \in M^{1}, v \in M^{2}$ we have:
\begin{eqnarray*}& &Y_{t}(a,x_{1})(Y_{t}(L(-1)u,x_{2})\otimes
v-z^{-1}{\partial \over
\partial x_{2}}Y_{t}(u,x_{2})\otimes v-\lambda
(zx_{2})^{-1}Y_{t}(u,x_{2})\otimes v)\\
&=&Y_{t}(L(-1)u,x_{2})\otimes Y(a,x_{1})v-z^{-1}{\partial \over
\partial x_{2}}Y_{t}(u,x_{2})\otimes Y(a,x_{1})v\\
& &-\lambda
(zx_{2})^{-1}Y_{t}(u,x_{2})\otimes Y(a,x_{1})v\\
& &+{\rm Res}_{x_{0}}(zx_{2})^{-1}\delta
\left(\frac{x_{1}-x_{0}}{zx_{2}}\right)Y_{t}(Y(a,x_{0})L(-1)u,x_{2})\otimes
v\\
& &-x^{-1}{\partial \over
\partial x_{2}}{\rm Res}_{x_{0}}(zx_{2})^{-1}\delta \left(\frac{x_{1}-x_{0}}
{zx_{2}}\right)Y_{t}(Y(a,x_{0})L(-1)u,x_{2})\otimes v\\
& &-{\rm Res}_{x_{0}}\lambda
(zx_{2})^{-2}\delta\left(\frac{x_{1}-x_{0}}{zx_{2}}\right)
Y_{t}(Y(a,x_{0})u,x_{2})\otimes v\\
&=&{\rm Res}_{x_{0}}(zx_{2})^{-1}\delta\left(\frac{x_{1}-x_{0}}{zx_{2}}\right)
Y_{t}(L(-1)Y(a,x_{0})u,x_{2})\otimes v\\
& &-{\rm Res}_{x_{0}}(zx_{2})^{-1}\delta\left(\frac{x_{1}-x_{0}}{zx_{2}}
\right){\partial \over
\partial x_{0}}Y_{t}(Y(a,x_{0})u,x_{2})\otimes v\\
& &-{\rm Res}_{x_{0}}x^{-1}(zx_{2})^{-1}\delta \left(\frac{x_{1}-x_{0}}
{zx_{2}}\right){\partial \over
\partial x_{2}}Y_{t}(Y(a,x_{0})u,x_{2})\otimes v\\
& &-{\rm Res}_{x_{0}}x^{-1}({\partial \over
\partial x_{2}}(zx_{2})^{-1}\delta\left(\frac{x_{1}-x_{0}}{zx_{2}}\right))
Y_{t}(Y(a,x_{0})u,x_{2})\otimes v\\
& &-{\rm Res}_{x_{0}}\lambda x^{-1}_{2}
(zx_{2})^{-1}\delta\left(\frac{x_{1}-x_{0}}{zx_{2}}\right)
Y_{t}(Y(a,x_{0})u,x_{2})\otimes
v \;\;\;\mbox{mod }L_{\lambda}\\
&=&-{\rm Res}_{x_{0}}(zx_{2})^{-1}\delta\left(\frac{x_{1}-x_{0}}{zx_{2}}
\right){\partial \over
\partial x_{0}}Y_{t}(Y(a,x_{0})u,x_{2})\otimes v\\
& &-{\rm Res}_{x_{0}}x^{-1}({\partial \over
\partial x_{2}}(zx_{2})^{-1}\delta\left(\frac{x_{1}-x_{0}}{zx_{2}}\right))
Y_{t}(Y(a,x_{0})u,x_{2})\otimes
v \;\;\;\mbox{mod }L_{\lambda}\\
&=&{\rm Res}_{x_{0}}(zx_{2})^{-1}({\partial \over
\partial x_{0}}\delta\left(\frac{x_{1}-x_{0}}{zx_{2}}\right))
Y_{t}(Y(a,x_{0})u,x_{2})\otimes v\\
& &-{\rm Res}_{x_{0}}x^{-1}{\partial \over \partial x_{2}}({1 \over
x_{1}-x_{0}}\delta\left(\frac{x_{1}-x_{0}}{zx_{2}}\right)
Y_{t}(Y(a,x_{0})u,x_{2})\otimes v\\
&=&-{\rm Res}_{x_{0}}(zx_{2})^{-2}\delta' \left(\frac{x_{1}-x_{0}}{zx_{2}}
\right)Y_{t}(Y(a,x_{0})u,x_{2})\otimes v\\
& &+{\rm Res}_{x_{0}}(zx_{2})^{-2}\delta' \left(\frac{x_{1}-x_{0}}{zx_{2}}
\right)Y_{t}(Y(a,x_{0})u,x_{2})\otimes v\\
&=&0.\; \;\;\;\;\Box
\end{eqnarray*}

{\bf Lemma 5.2.5}. {\it If} $u \in M^{1}(m), v \in M^{2}(n), k \in {\bf
Z}$, {\it then}
\begin{eqnarray}
L(0)(t^{k}\otimes u\otimes v)\equiv (m+n-1-k+\lambda
+h_{1}+h_{2})(t^{k}\otimes u \otimes v) \;\;\;\mbox{mod
}L_{\lambda}.\end{eqnarray}

{\bf Proof}. By definition, we have
\begin{eqnarray*}& &Y_{t}(\omega,x_{1})(Y_{t}(u,x_{2})\otimes v)\\
&=&Y_{t}(u,x_{2})\otimes
Y(\omega,x_{1})v+{\rm Res}_{x_{0}}(zx_{2})^{-1}e^{-x_{0}{\partial \over
\partial x_{1}}}\delta\left({x_{1}\over zx_{2}}\right)Y_{t}(u,x_{2})\otimes v\\
&=&Y_{t}(u,x_{2})\otimes
Y(\omega,x_{1})v+\sum_{i=0}^{\infty}\frac{(-1)^{i}}{i!}\left({\partial
\over \partial x_{1}}\right)^{i}\delta\left({x_{1}\over
zx_{2}}\right)(zx_{2})^{-1}Y_{t}(L(i-1)u,x_{2})\otimes v.\end{eqnarray*}
Then
\begin{eqnarray}
& &L(0)(Y_{t}(u,x_{2})\otimes v)\nonumber\\
&=&Y_{t}(u,x_{2})\otimes L(0)v+zx_{2}Y_{t}(L(-1)u,x_{2})\otimes
v+Y_{t}(L(0)u,x_{2})\otimes v \\
&\equiv& Y_{t}(u,x_{2})\otimes L(0)v+Y_{t}(L(0)u,x_{2})\otimes v\nonumber\\
& &+\lambda Y_{t}(u,x_{2})\otimes v+x_{2}{\partial \over \partial
x_{2}}Y_{t}(u,x_{2})\otimes v\;\;\;\mbox{mod }L_{\lambda}.\end{eqnarray}
Thus
$$L(0)(t^{k}\otimes u\otimes v)\equiv (m+n-k-1+\lambda
+h_{1}+h_{2})(t^{k}\otimes u\otimes v)\;\;\;\mbox{mod}\:L_{\lambda}.
\;\;\;\;\Box$$

Next we set
\begin{eqnarray}
F_{1}(M^{1},M^{2})=F_{0}(M^{1},M^{2})/J.
\end{eqnarray}

{\bf Remark 5.2.6}. $F_{1}(M^{1},M^{2})$ is a lower-truncated ${\bf
Z}$-graded  $g(V)$-module of level one, so that
the vacuum property, the $L(-1)$-derivative property and the
commutator formula (2.1.3) automatically hold. Furthermore, for any $a \in
V,w \in F_{1}(M^{1},M^{2})$, $Y_{t}(a,x)w$ involves only
finitely many negative powers of $x$.

{\bf Remark 5.2.7}. Notice that the action (5.2.6) of $g(V)$ on
$F_{0}(M^{1},M^{2})$ only
reflects the commutator formula (2.1.3), which is weaker than the Jacobi
identity, unlike the situation in the classical Lie
algebra theory. In the next step,  we consider the whole Jacobi identity
relation for an intertwining operator. This step in our approach
corresponds to the ``compatibility condition'' in Huang and Lepowsky's
approach [HL0].

Let $\pi_{1}$ be the quotient map from $F_{0}(M^{1},M^{2})$ onto
$F_{1}(M^{1},M^{2})$ and
let $J_{1}$ be the subspace of $F_{1}(M^{1},M^{2})$,
linearly spanned by all coefficients of monomials
$x_{0}^{m}x_{1}^{n}x_{2}^{k}$ in the following expressions:
\begin{eqnarray}
& &x_{0}^{-1}\delta\left(\frac{x_{1}-zx_{2}}{x_{0}}\right)Y_{t}(a,x_{1})\pi
_{1} (Y_{t}(u,x_{2})\otimes v)\nonumber\\
& &-x_{0}^{-1}\delta\left(\frac{zx_{2}-x_{1}}{-x_{0}}\right)\pi _{1}
(Y_{t}(u,x_{2})\otimes Y(a,x_{1})v)\nonumber\\
&-&(zx_{2})^{-1}\delta\left(\frac{x_{1}-x_{0}}{zx_{2}}\right)\pi _{1}
(Y_{t}(Y(a,x_{0})u,x_{2}))\otimes v  \end{eqnarray}
for $a \in V, u \in M^{1}, v \in M^{2}$.

{\bf Proposition 5.2.8.} {\it The  subspace $J_{1}$ is
a graded $g(V)$-submodule of $F_{1}(M^{1},M^{2})$.}

{\bf Proof}. For $a,b\in V,u\in M^{1},v\in M^{2}$, we have
\begin{eqnarray}
& &x_{0}^{-1}\delta\left(\frac{x_{1}-zx_{2}}{x_{0}}\right)Y_{t}(b,x_{3})
Y_{t}(a,x_{1})\pi _{1}(Y_{t}(u,x_{2})\otimes v)\nonumber \\
&=&x_{0}^{-1}\delta\left(\frac{x_{1}-zx_{2}}{x_{0}}\right) Y_{t}(a,x_{1})
 Y_{t}(b,x_{3})\pi _{1}(Y_{t}(u,x_{2})\otimes v)\nonumber \\
& &+{\rm Res}_{x_{4}}x_{0}^{-1}\delta\left(\frac{x_{1}-zx_{2}}{x_{0}}\right)
x_{1}^{-1}\delta\left(\frac{x_{3}-x_{4}}{x_{1}}\right)Y_{t}(Y(b,x_{4})a,x_{1})
\pi _{1}(Y_{t}(u,x_{2})\otimes v)\nonumber \\
&=&x_{0}^{-1}\delta\left(\frac{x_{1}-zx_{2}}{x_{0}}\right)Y_{t}(a,x_{1})
\pi _{1}(Y_{t}(u,x_{2})\otimes Y(b,x_{3})v)\nonumber \\
& &+{\rm Res}_{x_{4}}x_{0}^{-1}\delta\left(\frac{x_{1}-zx_{2}}{x_{0}}\right)
(zx_{2})^{-1}\delta\left(\frac{x_{3}-x_{4}}{zx_{2}}\right)Y_{t}(a,x_{1})
\pi_{1}(Y_{t}(Y(b,x_{4})u,x_{2})\otimes v)\nonumber \\
& &+{\rm Res}_{x_{4}}x_{0}^{-1}\delta\left(\frac{x_{1}-zx_{2}}{x_{0}}\right)
x_{1}^{-1}\delta\left(\frac{x_{3}-x_{4}}{x_{1}}\right)Y_{t}(Y(b,x_{4})a,x_{1})
\pi_{1}(Y_{t}(u,x_{2})\otimes v); \nonumber\\
&&\mbox{}
\end{eqnarray}
\begin{eqnarray}
& &Y_{t}(b,x_{3})\pi_{1}(Y_{t}(u,x_{2})\otimes Y(a,x_{1})v)\nonumber \\
&=&\pi_{1}(Y_{t}(u,x_{2})\otimes Y(b,x_{3})Y(a,x_{1})v)\nonumber\\
& &+{\rm Res}_{x_{4}}(zx_{2})^{-1}\delta\left(\frac{x_{3}-x_{4}}{zx_{2}}
\right)\pi_{1}
(Y_{t}(Y(b,x_{4})u,x_{2})\otimes Y(a,x_{1})v)\nonumber\\
&=&\pi_{1}(Y_{t}(u,x_{2})\otimes Y(a,x_{1})Y(b,x_{3})v)\nonumber\\
& &+{\rm Res}_{x_{4}}x_{1}^{-1}\delta\left(\frac{x_{3}-x_{4}}{x_{1}}\right)
\pi_{1}
(Y_{t}(u,x_{2})\otimes Y(Y(b,x_{4})a,x_{1})v)\nonumber \\
& &+{\rm Res}_{x_{4}}(zx_{2})^{-1}\delta\left(\frac{x_{3}-x_{4}}{zx_{2}}
\right)\pi_{1}
(Y_{t}(Y(b,x_{4})u,x_{2})\otimes Y(a,x_{1})v);
\end{eqnarray}
and
\begin{eqnarray}
& &(zx_{2})^{-1}\delta\left(\frac{x_{1}-x_{0}}{zx_{2}}\right)Y_{t}(b,x_{3})
\pi_{1}
(Y_{t}(Y(a,x_{0})u,x_{2})\otimes v)\nonumber\\
&=&(zx_{2})^{-1}\delta\left(\frac{x_{1}-x_{0}}{zx_{2}}\right)
\pi_{1}(Y_{t}(Y(a,x_{0})u,x_{2})
\otimes Y(b,x_{4})v)\nonumber\\
& &+{\rm Res}_{x_{4}}(zx_{2})^{-1}\delta\left(\frac{x_{1}-x_{0}}{zx_{2}}\right)
(zx_{2})^{-1}\delta\left(\frac{x_{3}-x_{4}}{zx_{2}}\right)\pi_{1}(Y_{t}
(Y(b,x_{4})Y(a,x_{0})u,x_{2})\otimes v)\nonumber \\
&=&(zx_{2})^{-1}\delta\left(\frac{x_{1}-x_{0}}{zx_{2}}\right)Y_{t}(b,x_{3})
\pi_{1}
(Y_{t}(Y(a,x_{0})u,x_{2})\otimes v)\nonumber\\
& &+{\rm Res}_{x_{4}}(zx_{2})^{-1}\delta\left(\frac{x_{1}-x_{0}}{zx_{2}}\right)
(zx_{2})^{-1}\delta\left(\frac{x_{3}-x_{4}}{zx_{2}}\right)\pi_{1}(Y_{t}
(Y(a,x_{0})Y(b,x_{4})u,x_{2})\otimes v)\nonumber \\
& &+{\rm Res}_{x_{4}}{\rm Res}_{x_{5}}(zx_{2})^{-1}\delta\left(
\frac{x_{1}-x_{0}}{zx_{2}}\right)
(zx_{2})^{-1}\delta\left(\frac{x_{3}-x_{4}}{zx_{2}}\right) x_{0}^{-1}
\delta\left(\frac{x_{4}-x_{5}}
{x_{0}}\right)\cdot \nonumber \\
& &\cdot \pi_{1}(Y_{t}(Y(Y(b,x_{5})a,x_{0})u,x_{2})\otimes v)\nonumber \\
&=&(zx_{2})^{-1}\delta\left(\frac{x_{1}-x_{0}}{zx_{2}}\right)Y_{t}(b,x_{3})
\pi_{1}
(Y_{t}(Y(a,x_{0})u,x_{2})\otimes v)\nonumber\\
& &+{\rm Res}_{x_{4}}(zx_{2})^{-1}\delta\left(\frac{x_{1}-x_{0}}{zx_{2}}\right)
(zx_{2})^{-1}\delta\left(\frac{x_{3}-x_{4}}{zx_{2}}\right)
(zx_{2})^{-1}\delta\left(\frac{x_{3}-x_{4}}{zx_{2}}\right)\nonumber \\
& &\cdot\pi_{1}(Y_{t}
(Y(a,x_{0})Y(b,x_{4})u,x_{2})\otimes v)\nonumber \\
& &+{\rm Res}_{x_{5}}(zx_{2})^{-1}\delta\left(\frac{x_{1}-x_{0}}{zx_{2}}\right)
(zx_{2})^{-1}\delta\left(\frac{x_{3}-x_{0}-x_{5}}{zx_{2}}\right)\cdot
\nonumber \\
& &\cdot \pi_{1}(Y_{t}(Y(Y(b,x_{5})a,x_{0})u,x_{2})\otimes v)\nonumber \\
&=&(zx_{2})^{-1}\delta\left(\frac{x_{1}-x_{0}}{zx_{2}}\right)Y_{t}(b,x_{3})
\pi_{1}
(Y_{t}(Y(a,x_{0})u,x_{2})\otimes v)\nonumber\\
& &+{\rm Res}_{x_{4}}(zx_{2})^{-1}\delta\left(\frac{x_{1}-x_{0}}{zx_{2}}\right)
(zx_{2})^{-1}\delta\left(\frac{x_{3}-x_{4}}{zx_{2}}\right)
(zx_{2})^{-1}\delta\left(\frac{x_{3}-x_{4}}{zx_{2}}\right)\nonumber \\
& &\cdot \pi_{1}(Y_{t}
(Y(a,x_{0})Y(b,x_{4})u,x_{2})\otimes v)\nonumber \\
& &+{\rm Res}_{x_{4}}(zx_{2})^{-1}\delta\left(\frac{x_{1}-x_{0}}{zx_{2}}
\right)x_{1}^{-1}
\delta\left(\frac{x_{3}-x_{4}}{x_{1}}\right)\pi_{1}
(Y_{t}(Y(Y(b,x_{5})a,x_{0})u,x_{2}) \otimes v).\nonumber\\
& &\mbox{}
\end{eqnarray}
Then it is clear that $J_{1}$ is stable under the action of
$Y_{t}(b,x)$ for any $b\in V$.  $\;\;\;\;\Box$

{\bf Theorem 5.2.9.} {\it The quotient space
$F_{2}(M^{1},M^{2})=F_{1}(M^{1},M^{2})/J_{1}$ is a
lower-truncated ${\bf Z}$-graded weak $V$-module.}

{\bf Proof}. We only need to prove the Jacobi identity. For $a,b\in
V,u\in M^{1},v\in M^{2}$, we have
\begin{eqnarray}
& &x_{0}^{-1}\delta\left(\frac{x_{1}-x_{2}}{x_{0}}\right)Y_{t}(a,x_{1})
Y_{t}(b,x_{2})(Y_{t}(u,x_{3})\otimes v)   \nonumber \\
&=&x_{0}^{-1}\delta\left(\frac{x_{1}-x_{2}}{x_{0}}\right)Y_{t}(a,x_{1})
\left( Y_{t}(u,x_{3})\otimes Y(b,x_{2})v\right) \nonumber \\
& &+{\rm Res}_{x_{4}}x_{0}^{-1}\delta\left(\frac{x_{1}-x_{2}}{x_{0}}\right)
(zx_{3})^{-1}\delta\left(\frac{x_{2}-x_{4}}{zx_{3}}\right)Y_{t}(a,x_{1})\left(
Y_{t}(Y(b,x_{4})u,x_{3})\otimes v\right)\nonumber \\
&=&x_{0}^{-1}\delta\left(\frac{x_{1}-x_{2}}{x_{0}}\right)Y_{t}(u,x_{3})\otimes
Y(a,x_{1})Y(b,x_{2})v \\
& &+ {\rm Res}_{x_{4}}x_{0}^{-1}\delta\left(\frac{x_{1}-x_{2}}{x_{0}}\right)
(zx_{3})^{-1}\delta\left(\frac{x_{1}-x_{4}}{zx_{3}}\right)Y_{t}(Y(a,x_{4})u,
x_{3})\otimes Y(b,x_{2})v \nonumber\\
& &\mbox{}\\
& &+ {\rm Res}_{x_{4}}x_{0}^{-1}\delta\left(\frac{x_{1}-x_{2}}{x_{0}}\right)
(zx_{3})^{-1}\delta\left(\frac{x_{2}-x_{4}}{zx_{3}}\right)
Y_{t}(a,x_{1})(Y_{t}(
Y(b,x_{4})u,x_{3})\otimes v );\nonumber\\
& &\mbox{}
\end{eqnarray}
\begin{eqnarray}
& &x_{0}^{-1}\delta\left(\frac{-x_{2}+x_{1}}{x_{0}}\right)Y_{t}(b,x_{2})
Y_{t}(a,x_{1})(Y_{t}(u,x_{3})\otimes v)   \nonumber \\
&=&x_{0}^{-1}\delta\left(\frac{-x_{2}+x_{1}}{x_{0}}\right)Y_{t}(u,x_{3})\otimes
Y(b,x_{2})Y(a,x_{1})v\\
& &+{\rm Res}_{x_{4}}x_{0}^{-1}\delta\left(\frac{-x_{2}+x_{1}}{x_{0}}\right)
(zx_{3})^{-1}\delta\left(\frac{x_{2}-x_{4}}{zx_{3}}\right)Y_{t}(Y(b,x_{4})u,
x_{3})\otimes Y(a,x_{1})v\nonumber\\
& &\mbox{}\\
& &+{\rm Res}_{x_{4}}x_{0}^{-1}\delta\left(\frac{-x_{2}+x_{1}}{x_{0}}\right)
(zx_{3})^{-1}\delta\left(\frac{x_{1}-x_{4}}{zx_{3}}\right)Y_{t}(b,x_{2})(Y_{t}(
Y(a,x_{4})u,x_{3})\otimes v);\nonumber\\
& &\mbox{}
\end{eqnarray}
and
\begin{eqnarray}
& &x_{2}^{-1}\delta\left(\frac{x_{1}-x_{0}}{x_{2}}\right)Y_{t}(Y(a,x_{0})
b,x_{2})(Y_{t}(u,x_{3})\otimes v)   \nonumber \\
&=&x_{2}^{-1}\delta\left(\frac{x_{1}-x_{0}}{x_{2}}\right)Y_{t}(u,x_{3})\otimes
Y(Y(a,x_{0})b,x_{2})v\\
& &+{\rm Res}_{x_{4}}x_{2}^{-1}\delta\left(\frac{x_{1}-x_{0}}{x_{2}}\right)
(zx_{3})^{-1}\delta\left(\frac{x_{2}-x_{4}}{zx_{3}}\right)Y_{t}(Y(Y(a,x_{0})b,
x_{4})u,x_{3})\otimes v.\nonumber\\
& &\mbox{}
\end{eqnarray}
It follows from the Jacobi identity of $M^{2}$ that
(5.2.24)$-$(5.2.27)=(5.2.30). Since
\begin{eqnarray}
& &x_{0}^{-1}\delta\left(\frac{x_{1}-x_{2}}{x_{0}}\right)(zx_{3})^{-1}\delta
\left(\frac{x_{2}-x_{4}}{zx_{3}}\right)\nonumber \\
&=&x_{0}^{-1}\delta\left(\frac{x_{1}-zx_{3}-x_{4}}{x_{0}}\right)(zx_{3})^{-1}
\delta\left(\frac{x_{2}-x_{4}}{zx_{3}}\right)\nonumber \\
&=&(x_{1}-zx_{3})^{-1}\delta\left(\frac{x_{0}+x_{4}}{x_{1}-zx_{3}}\right)
(zx_{3})^{-1}\delta\left(\frac{x_{2}-x_{4}}{zx_{3}}\right)\nonumber\\
&=&(x_{0}+x_{4})^{-1}\delta\left(\frac{x_{1}-zx_{3}}{x_{0}+x_{4}}\right)
(zx_{3})^{-1}\delta\left(\frac{x_{2}-x_{4}}{zx_{3}}\right);
\end{eqnarray}
\begin{eqnarray}
& &x_{0}^{-1}\delta\left(\frac{-x_{2}+x_{1}}{x_{0}}\right)(zx_{3})^{-1}\delta
\left(\frac{x_{2}-x_{4}}{zx_{3}}\right)\nonumber \\
&=&x_{0}^{-1}\delta\left(\frac{-zx_{3}-x_{4}+x_{1}}{x_{0}}\right)
(zx_{3})^{-1}\delta
\left(\frac{x_{2}-x_{4}}{zx_{3}}\right)\nonumber \\
&=&(x_{0}+x_{4})^{-1}\delta\left(\frac{-zx_{3}+x_{1}}{x_{0}+x_{4}}\right)
(zx_{3})^{-1}\delta\left(\frac{x_{2}-x_{4}}{zx_{3}}\right),
\end{eqnarray}
by the $J_{1}$-defining relation (5.2.20), we have
\begin{eqnarray}
& &(5.2.26)-(5.2.28)\nonumber \\
&=&{\rm Res}_{x_{4}}(zx_{3})^{-1}\delta
\left(\frac{x_{2}-x_{4}}{zx_{3}}\right)(zx_{3})^{-1}\delta\left(\frac{x_{1}
-x_{0}-x_{4}}{zx_{3}}\right)\nonumber \\
& &\cdot Y_{t}(
Y(a,x_{0}+x_{4})Y(b,x_{4})u,x_{3})\otimes v\nonumber\\
&=&{\rm Res}_{x_{4}}{\rm Res}_{x_{5}}x_{5}^{-1}\delta\left(\frac{x_{0}+x_{4}}
{x_{5}}\right)(zx_{3})^{-1}\delta
\left(\frac{x_{2}-x_{4}}{zx_{3}}\right)(zx_{3})^{-1}\delta\left(\frac
{x_{1}-x_{5}}{zx_{3}}\right)\nonumber\\
& &\cdot Y_{t}(Y(a,x_{5})Y(b,x_{4})u,x_{3})\otimes v\nonumber\\
&=&{\rm Res}_{x_{4}}{\rm Res}_{x_{5}}x_{0}^{-1}\delta\left(\frac{x_{5}-x_{4}}
{x_{0}}\right)(zx_{3})^{-1}\delta
\left(\frac{x_{2}-x_{4}}{zx_{3}}\right)(zx_{3})^{-1}\delta\left(\frac
{x_{1}-x_{5}}{zx_{3}}\right)\nonumber\\
& &\cdot Y_{t}(Y(a,x_{5})Y(b,x_{4})u,x_{3})\otimes v.
\end{eqnarray}
Similarly, we obtain
\begin{eqnarray}
& &(5.2.25)-(5.2.29)\nonumber \\
&=&-{\rm Res}_{x_{4}}(zx_{3})^{-1}\delta
\left(\frac{x_{1}-x_{4}}{zx_{3}}\right)(zx_{3})^{-1}\delta\left(\frac
{x_{1}+x_{0}-x_{4}}{zx_{3}}\right)\nonumber\\
& &\cdot Y_{t}(Y(a,-x_{0}+x_{4})Y(b,x_{4})u,x_{3})\otimes v\nonumber\\
&=&-{\rm Res}_{x_{4}}{\rm Res}_{x_{5}}x_{5}^{-1}\delta\left(\frac{-x_{0}+x_{4}}
{x_{5}}\right)(zx_{3})^{-1}\delta
\left(\frac{x_{1}-x_{4}}{zx_{3}}\right)(zx_{3})^{-1}\delta\left(\frac{x_{2}
-x_{5}}{zx_{3}}\right)\nonumber \\
& &\cdot Y_{t}(Y(b,x_{4})Y(a,x_{5})u,x_{3})\otimes v\nonumber\\
&=&-{\rm Res}_{x_{4}}{\rm Res}_{x_{5}}x_{0}^{-1}\delta\left(\frac{-x_{5}+x_{4}}
{x_{0}}\right)(zx_{3})^{-1}\delta
\left(\frac{x_{1}-x_{4}}{zx_{3}}\right)(zx_{3})^{-1}\delta\left(\frac{x_{2}
-x_{5}}{zx_{3}}\right)\nonumber \\
& &\cdot Y_{t}(Y(b,x_{4})Y(a,x_{5})u,x_{3})\otimes v \nonumber\\
&=&-{\rm Res}_{x_{5}}{\rm Res}_{x_{4}}x_{0}^{-1}\delta\left(\frac{-x_{4}+x_{5}}
{x_{0}}\right)(zx_{3})^{-1}\delta
\left(\frac{x_{1}-x_{5}}{zx_{3}}\right)(zx_{3})^{-1}\delta\left(\frac{x_{2}
-x_{4}}{zx_{3}}\right)\nonumber \\
& &\cdot Y_{t}(Y(b,x_{5})Y(a,x_{4})u,x_{3})\otimes v.
\end{eqnarray}
Therefore, we have
\begin{eqnarray}
& &(5.2.25)+(5.2.26)-(5.2.28)-(5.2.29)\nonumber \\
&=&{\rm Res}_{x_{4}}{\rm Res}_{x_{5}}x_{4}^{-1}\delta\left(
\frac{x_{5}-x_{0}}{x_{4}}\right)(zx_{3})^{-1}\delta
\left(\frac{x_{2}-x_{4}}{zx_{3}}\right)
(zx_{3})^{-1}\delta\left(\frac{x_{1}-x_{5}}{zx_{3}}\right)\nonumber \\
& &\cdot Y_{t}(Y(Y(a,x_{0})b,x_{4})u,x_{3})\otimes v\nonumber \\
&=&{\rm Res}_{x_{4}}(zx_{3})^{-1}\delta
\left(\frac{x_{2}-x_{4}}{zx_{3}}\right)x_{2}^{-1}\delta\left(\frac{x_{1}-x_{0}}
{x_{2}}\right)Y_{t}(Y(Y(a,x_{0})b,x_{4})u,x_{3})\otimes v\nonumber\\
&=& (5.2.32).\end{eqnarray}
Here we have used the following fact:
\begin{eqnarray}
& &{\rm Res}_{x_{5}}x_{4}^{-1}\delta\left(
\frac{x_{5}-x_{0}}{x_{4}}\right)(zx_{3})^{-1}\delta
\left(\frac{x_{2}-x_{4}}{zx_{3}}\right)
(zx_{3})^{-1}\delta\left(\frac{x_{1}-x_{5}}{zx_{3}}\right)\nonumber \\
&=&{\rm Res}_{x_{5}}(zx_{3})^{-1}\delta
\left(\frac{x_{2}-x_{4}}{zx_{3}}\right)(zx_{3})^{-1}\delta\left(\frac{x_{1}
-x_{4}-x_{0}}{zx_{3}}\right)x_{5}^{-1}\delta\left(\frac{x_{4}+x_{0}}{x_{5}}
\right) \nonumber \\
&=&(zx_{3})^{-1}\delta
\left(\frac{x_{2}-x_{4}}{zx_{3}}\right)x_{1}^{-1}\delta\left(\frac{x_{2}-x_{4}
+x_{4}+x_{0}}{x_{1}}\right)\nonumber \\
&=&(zx_{3})^{-1}\delta
\left(\frac{x_{2}-x_{4}}{zx_{3}}\right)x_{2}^{-1}\delta\left(\frac{x_{1}
-x_{0}}{x_{2}}\right). \end{eqnarray}
Then we finish the proof. $\;\;\;\;\Box $

Since $F_{2}(M^{1},M^{2})$ is a weak $V$-module, we will freely use
$Y(a,x)$ for $Y_{t}(a,x)$.

{\bf Proposition  5.2.10.} {\it Let $M^{1}$ and $M^{3}$ be two
$V$-modules, let $M$ be a lowest-weight $g(V)$-module
of level-one and let $J$ be the $g(V)$-submodule of $M$
generated from the Jacobi relations (5.2.20), so that $\bar{M}=M/J$
(which may be zero) is a generalized $V$-module.
Let $I(\cdot,x)$ be any intertwining operator of type
$\left(\begin{array}{c}M^{3}\\M^{1},M\end{array}\right)$. Then
$I(\cdot,x)J=0$, so that we obtain an intertwining operator of type
 $\left(\begin{array}{c}M^{3}\\M^{1},\bar{M}\end{array}\right)$.}

{\bf Proof.} This directly follows from the proof of Theorem 3.2.9.
$\;\;\;\;\Box$

Symmetrically, we have:

{\bf Proposition  5.2.11.} {\it Let $M^{2}$ and $M^{3}$ be two
generalized $V$-modules and let $M$, $J$ and $\bar{M}$ be given as in
Proposition 5.2.10.
Let $I(\cdot,x)$ be any intertwining operator of type
$\left(\begin{array}{c}M^{3}\\M,M^{2}\end{array}\right)$. Then
$I(J,x)=0$, so that we obtain an intertwining operator of type
 $\left(\begin{array}{c}M^{3}\\\bar{M},M^{2}\end{array}\right)$.}

{\bf Proof.} The proof of this proposition does not directly follow
from the proof of Theorem 5.2.9, but it follows from Proposition
5.2.10 and the notion of transpose intertwining operator. Since
$I^{t}(\cdot,x)$ is an intertwining operator of type
$\left(\begin{array}{c}M^{3}\\M^{2},M\end{array}\right)$, by
Proposition 5.2.10 we have: $I^{t}(\cdot,x)J=0$. Thus $I(J,x)=0$.
$\;\;\;\;\Box$

{\bf Remark 5.2.12.} Let $\tilde{g}$ be an affine Lie algebra and let
$\ell$ be a positive integer. It has been proved ([DL], [FZ], [L2])
that $L(\ell,0)$ is a rational vertex operator algebra with the set of
equivalence classes of irreducible modules being the set of
equivalence classes of standard $\tilde{g}$-modules of level $\ell$.
Thus any lowest-weight $\tilde{g}$-module of level $\ell$ is a
$L(\ell,0)$-module if and only if it is an (irreducible) standard module.
Suppose $M^{2}$ and $M^{3}$ are standard modules of level $\ell$ (equivalently,
$L(\ell,0)$-modules). Let $M^{1}$ be a generalized Verma $\tilde{g}$-module
of level $\ell$ and let $I(\cdot,x)$ be an intertwining operator of
type $\left(\begin{array}{c}M^{3}\\M^{1},M^{2}\end{array}\right)$.
Then by Proposition 5.2.11 $I(\cdot,x)$ gives an intertwining operator among
three $L(\ell,0)$-modules (equivalently, three standard
$\tilde{g}$-modules). In this sense Proposition 5.2.11
could be considered as a generalization of
Tsuchiya and Kanie's ``nuclear democracy theorem'' [TK] (we will discuss
this in Chapter 7).

It follows from Lemmas 5.2.4 and 5.2.5 that $\theta$ induces a surjective
$V$-endomorphism $\bar{\theta}$ of $F_{1}(M^{1},M^{2})$.
For any complex number $\lambda$, we define a quotient module
\begin{eqnarray}
F_{2}(M^{1},M^{2})^{(\lambda)}=F_{1}(M^{1},M^{2})/(J_{1}+
L_{\lambda}). \end{eqnarray}
Let $\sigma _{\lambda}$ be the quotient map from
$F_{2}(M^{1},M^{2})$ onto
$F_{2}(M^{1},M^{2})^{(\lambda)}$. It follows from Lemma 5.2.4 that
$\bar{\theta}$ induces a surjective $V$-homomorphism
$\bar{\theta}_{\lambda}$ from $F_{2}(M^{1},M^{2})^{(\lambda)}$ onto
$F_{2}(M^{1},M^{2})^{(\lambda+1)}$. Then we get a sequence of
generalized $V$-modules:
\begin{eqnarray}
\cdots \rightarrow F_{2}(M^{1},M^{2})^{(\lambda -1)}
\stackrel{\bar{\theta}_{\lambda -1}}{\longrightarrow}
F_{2}(M^{1},M^{2})^{(\lambda)}
\stackrel{\bar{\theta}_{\lambda}}{\longrightarrow}
F_{2}(M^{1},M^{2})^{(\lambda +1)} \longrightarrow \cdots \end{eqnarray}

{\bf Lemma 5.2.13.} {\it If $\lambda +h_{1}+h_{2}$ is not the lowest weight of
a generalized $V$-module, then $\bar{\theta}_{\lambda}$ is an isomorphism}.

{\bf Proof}. It suffices to prove that the linear map $\theta ^{-1}$ induces a
$V$-homomorphism from $F_{2}(M^{1},M^{2})^{(\lambda +1)}$ to
$F_{2}(M^{1},M^{2})^{(\lambda)}$. Since degree-zero subspace of
$F_{2}(M^{1},M^{2})^{(\lambda)}$ is the lowest-weight subspace of weight
$\lambda +h_{1}+h_{2}$ (from Lemma 5.2.5), then the degree-zero subspace
must be zero. By considering the following sequence:
\begin{eqnarray}
F_{2}(M^{1},M^{2})\stackrel{\theta^{-1}}{\longrightarrow}
F_{2}(M^{1},M^{2})\stackrel{\pi}{\longrightarrow}
F_{2}(M^{1},M^{2})\stackrel{\sigma_{\lambda}}{\longrightarrow}
F_{2}(M^{1},M^{2})^{(\lambda)},
\end{eqnarray}
we obtain
\begin{eqnarray}
\sigma _{\lambda}\pi \theta ^{-1}(t^{m+n+k+1}\otimes
M^{1}(m)\otimes M^{2}(n))\subseteq \sigma _{\lambda}\pi
(t^{m+n+k}\otimes M^{1}(m)\otimes M^{2}(n))\;\left(=0\right)\nonumber\\
 & &\mbox{}
\end{eqnarray}
for all $m,n,k \in {\bf N}$ and
\begin{eqnarray}
\sigma _{\lambda}\pi \theta ^{-1}(t^{m+n}\otimes
M^{1}(m)\otimes M^{2}(n))\subseteq \sigma _{\lambda}\pi
(t^{m+n-1}\otimes M^{1}(m)\otimes M^{2}(n))\;\;\left(=0 \right)\nonumber\\
& &\mbox{}
\end{eqnarray}
for all $m,n \in {\bf N}.$
By (5.2.41), (5.2.42) and (5.2.16), $\theta ^{-1}$ induces a
$V$-homomorphism from $F_{2}(M^{1},M^{2})^{(\lambda +1)}$ to
$F_{2}(M^{1},M^{2})^{(\lambda)}.\;\;\;\;\Box$

{\bf Remark 5.2.14.} In general, $L(0)$ may not act
semisimply on $F_{2}(M^{1},M^{2})$, but from Lemma 5.2.5,
$L(0)$ acts semisimply on
$F(M^{1},M^{2})^{(\lambda)}$, and it has lowest weight $\lambda
+h_{1}+h_{2}$.

Let $W$ be any $V$-module with lowest weight $h$ such that
$W=\oplus_{n=0}^{\infty}W_{(n+h)}$ and let $I(\cdot,x)$ be an
intertwining operator of type
$\left(\begin{array}{c}W\\M^{1},M^{2}\end{array}\right)$.
Let $I^{o}(\cdot,x)=x^{h_{1}+h_{2}-h}I(\cdot,x)$ be the normalization.
Then we define
\begin{eqnarray}
\psi_{I}: F_{2}(M^{1},M^{2})\rightarrow W,\;t^{n}\otimes
u \otimes v \mapsto I_{u}(n)vz^{-n-1} \end{eqnarray}
for $u \in M^{1},v \in M^{2},n \in {\bf Z}.$
In terms of generating elements, $\psi_{I}$ can be written as:
\begin{eqnarray}
\psi_{I}(Y_{t}(u,x)\otimes v)=
I^{o}(u,zx)v\;\;\;\mbox{for }u \in M^{1},v \in M^{2}.
\end{eqnarray}

{\bf Lemma 5.2.15.} {\it The linear map $\psi_{I}$ is a ${\bf
g}(V)$-homomorphism. In other words},
\begin{eqnarray}
\psi_{I}(Y_{t}(a,x)w)=Y(a,x)\psi_{I}(w)\;\;\;\mbox{{\it for} } a \in
V, w \in F_{2}(M^{1},M^{2}).
\end{eqnarray}

{\bf Proof.} For $a \in V, u \in M^{1},v \in M^{2}$, we have
\begin{eqnarray*}& &\psi_{I}^{z}(Y_{t}(a,x_{1})(Y_{t}(u,x_{2})\otimes v))\\
&=&\psi_{I}^{z}(Y_{t}(u,x_{2})\otimes Y(a,x_{1})v)
+{\rm
Res}_{x_{0}}(zx_{2})^{-1}\delta\left(\frac{x_{1}-x_{0}}{zx_{2}}\right)\psi_{I}^{z}(Y_{t}(Y(a,x_{0})u,x_{2})\otimes
v)\\
&=&I^{o}(u,zx_{2})Y(a,x_{1})v+{\rm
Res}_{x_{0}}(zx_{2})^{-1}\delta\left(\frac{x_{1}-x_{0}}{zx_{2}}\right)I^{o}(Y(a,x_{0})u,zx_{2})v\\
&=&Y(a,x_{1})I^{o}(u,zx_{2})v \\
&=&Y(a,x_{1})\psi_{I}(Y_{t}(u,x_{2})\otimes v).\;\;\;\;\Box
\end{eqnarray*}

{\bf Corollary 5.2.16.} {\it The linear map $\psi_{I}$ induces a
$V$-module homomorphism $\bar{\psi}_{I}$ from
$T(M^{1},M^{2})$ to $W$ such that $\bar{\psi}_{I}$
preserves the
${\bf Z}$-gradings and $\bar{\psi}_{I}=\pi \psi_{I}$, where
$\pi$ is the quotient map from $F_{0}(M^{1},M^{2})$ to
$T(M^{1},M^{2})$. Furthermore, $\psi _{I}$ induces a
$V$-homomorphism
$\psi ^{\lambda}_{I}$ from
$F_{2}(M^{1},M^{2})^{(\lambda)}$ to $W$ where
$\lambda=h-h_{1}-h_{2}$}.

{\bf Proof}. It follows from Proposition 4.1.2 that $J \subseteq
{\rm ker}\;\psi_{I}$.
By the Jacobi identities for a $V$-module and for an intertwining
operator we get: $J+J_{1} \subseteq {\rm ker}\;\psi_{I}$. Then we
 have an
induced linear map $\bar{\psi} _{I}$ from
$F_{2}(M^{1},M^{2})$ to $W$. By Lemma 5.2.10
$\bar{\psi} _{I}$ is a $V$-homomorphism. Furthermore, by (4.1.12),
the $L(-1)$-derivative formula for $I(\cdot,x)$ is equivalent to the
following formula:
\begin{eqnarray}
I^{o}(L(-1)u^{1},xz)u^{2}-x^{-1}{d\over
dz}I^{o}(u^{1},xz)u^{2}-\lambda (xz)^{-1}I^{o}(u^{1},xz)u^{2}=0.
\end{eqnarray}
Thus
$$\psi_{I}\left(Y_{t}(L(-1)u,z)\otimes v-x^{-1}{d\over dz}Y_{t}(u,z)\otimes
v-\lambda (xz)^{-1}Y_{t}(u,z)\otimes v\right)=0.$$
So $L_{\lambda}\subseteq {\rm ker}\;\bar{\psi}_{I}$. Therefore we obtain a
$V$-homomorphism $\bar{\psi}_{I}(\lambda)$ from
$F_{2}(M^{1},M^{2})^{(\lambda)}$ to $W.\;\;\;\;\Box$

As before, let $W$ be a $V$-module in $S$ with lowest weight $h$. Then
we define the following linear map:
\begin{eqnarray}
\bar{\psi}(\lambda):& &I\left(\begin{array}{c}W\\M^{1},M^{2}\end{array}\right)
 \rightarrow
\mbox{Hom}_{V}(F_{2}(M^{1},M^{2})^{(\lambda)},W)\nonumber\\
& &I(\cdot,x)\mapsto
\bar{\psi}_{I}(\lambda),
\end{eqnarray}
 where $\lambda =h-h_{1}-h_{2}$.

{\bf Theorem 5.2.17.} {\it The defined linear map}
$\bar{\psi}(\lambda)$ {\it is a linear isomorphism.}

{\sl Proof}. Since $\bar{\psi}(\lambda)$ is clearly injective by
(5.2.43), we only need to prove the surjectivity.
Let $f$ be a V-homomorphism from $F(M^{1},M^{2})^{(\lambda)}$ to
$W$. Then we define an operator $I(\cdot,x)$ from
$M^{1}\otimes M^{2}$ to $W\{x\}$ as follows:
\begin{eqnarray}
I(u,x)v=f\pi (Y_{t}(u,x)\otimes
v)x^{h-h_{1}-h_{2}} \mbox{   for }u \in M^{1},v \in M^{2}.
\end{eqnarray}
Then $I(\cdot,x)$ is an intertwining operator because of
the defining relations (5.2.11), (5.2.12) and (5.2.14).$\;\;\;\;\Box$

Let $\mbox{Hom}^{0}_{V}(F_{2}(M^{1},M^{2}),W)$ be the space of all
$V$-homomorphisms from $F_{2}(M^{1},M^{2})$ to $W$ which
preserve the ${\bf Z}$-gradings.

{\bf Theorem 5.2.18}. {\it The map} $\bar{\psi} :
I\left(\begin{array}{c}W\\M^{1},M^{2}\end{array}\right)\rightarrow
\mbox{Hom}^{0}_{V}(F_{2}(M^{1},M^{2}),W); I \mapsto
\bar{\psi}_{I}$ {\it is a linear isomorphism}.

{\bf Proof}. By Theorem 5.2.17, it is enough to prove that for any
 $V$-homomorphism $f$ from $F_{2}(M^{1},M^{2})$ to $W$
preserving the grading, $f$ maps the image of
$L_{h-h_{1}-h_{2}}$ to zero.
For $k \in {\bf Z},m,n \in {\bf N}; u \in M^{1}(m), v \in M^{2}$,
we have:
 $$\deg\:(t^{k}\otimes u \otimes v)=m+n-k-1.$$
Therefore
\begin{eqnarray}
L(0)f\pi (t^{k}\otimes u\otimes v)=(h+m+n-k-1)f\pi
(t^{k}\otimes u\otimes v). \end{eqnarray}
By definition, we have:
\begin{eqnarray}
& &L(0)(t^{k}\otimes u \otimes v)\nonumber \\
&=&t^{k}\otimes u \otimes
L(0)v+xt^{k+1}\otimes u \otimes v+t^{k}\otimes L(0)u \otimes
v\nonumber \\
&=&xt^{k+1}\otimes L(-1)u \otimes v+(h_{1}+h_{2}+m+n)t^{k}\otimes
u\otimes v. \end{eqnarray}
Therefore
$$f\pi \left(xt^{k+1}\otimes L(-1)u\otimes
v+(h_{1}+h_{2}-h+k+1)t^{k}\otimes u\otimes v\right)=0.$$
Observing (5.2.14) we have $f\pi (L_{h-h_{1}-h_{2}})=0$.$\;\;\;\;\Box$

{\bf Remark 5.2.19}. From Theorem 5.2.18, $F_{2}(M^{1},M^{2})$
almost satisfies
the requirement for being a tensor product for the pair $(M^{1},M^{2})$. In
general, not any $V$-module homomorphism from
$F_{2}(M^{1},M^{2})$ to $W$ preserves the ${\bf
Z}$-gradings. But any
$V$-module homomorphism maps any lowest-weight vector to a
lowest-weight vector. If $W$ is irreducible, then any $V$-module
homomorphism from $F_{2}(M^{1},M^{2})$ to $W$ maps any lowest-weight vector of
non-zero degree to zero. This fact leads us to consider the
quotient module of $F_{2}(M^{1},M^{2})$.

For any lower-truncated ${\bf Z}$-graded weak $V$-module
$M=\oplus_{n\in {\bf
N}}M(n)$, we define
the {\it radical} of $M$ to be the maximal graded
submodule $R(M)$
such that $R(M)\cap M(0)=0$. There is another definition of radical
[Zhu] as follows:
Let $M'=\oplus_{n\in N}M(n)^{*}$.
Define $R^{1}(M)$ to be the subspace of $M$ consisting of all elements
$u\in M$ such that:
\begin{eqnarray}
\langle u',Y(a,x)u\rangle =0\mbox{   for all }a\in V,u'\in
M(0)^{*}.\end{eqnarray}
By using the associativity of vertex operator algebras one can easily see that
$R^{1}(M)$ is a graded submodule. Furthermore $R(M)=R^{1}(M)$.
Set $S(M)=M/R(M)$. If $M$ is
completely reducible, then $S(M)$ is the submodule generated by the
lowest-degree subspace $M(0)$.

{\bf Definition 5.2.20}. Define $T(M^{1},M^{2})$ to be the
quotient module of $F_{2}(M^{1},M^{2})$ divided by the radical of
$F_{2}(M^{1},M^{2})$.

Suppose that $V$ is rational in the sense of [Zhu]. Then
$F_{2}(M^{1},M^{2})$ is completely reducible and $T(M^{1},
M^{2})$ can be considered as the submodule generated by degree-zero
subspace $F_{2}(M^{1},M^{2})(0)$ of $F_{2}(M^{1},M^{2})$. For any
complex number $h$, let $W_{h}$ be the direct sum of all irreducible
submodules of $T(M^{1},M^{2})$ with lowest weight $h$.
Then $T(M^{1},M^{2})$ is the direct sum of all $W_{h}$
for $h\in {\bf C}$. Let $P_{h}$ be the projection map of $T(M^{1},M^{2})$
onto $W_{h}$. Then we define:
\begin{eqnarray}
F(\cdot,x) :& &M^{1} \rightarrow \left({\rm Hom}_{{\bf
C}}(M^{2},T(M^{1},M^{2})) \right)\{x\};\nonumber\\
& &u^{1}\mapsto F(u^{1},x)\;\;\;\;\mbox{  for }u^{1}\in M^{1}
\end{eqnarray}
where $F(u^{1},x)u^{2}=\sum_{h\in {\bf C}}x^{h-h_{1}-h_{2}}P_{h}\pi
(Y_{t}(u^{1},x)\otimes u^{2})$ for $u^{1}\in m^{1}, u^{2}\in M^{2}$.

{\bf Proposition 5.2.21.} {\it Suppose that $V$ is rational in the sense of
[Zhu]. Then $F(\cdot,x)$ is an intertwining operator of type
$\left(\begin{array}{c}T(M^{1},M^{2})\\M^{1},M^{2}
\end{array}\right)$.}

{\bf Proof.} Let $F_{h}(u^{1},x)u^{2}=x^{h-h_{1}-h_{2}}P_{h}\pi
(Y_{t}(u^{1},x)\otimes u^{2})$. Then it is enough to prove that each
$F_{h}(\cdot,x)$ is an intertwining operator of type
$\left(\begin{array}{c}W_{h}\\M^{1},M^{2}\end{array}\right)$.
This easily follows from Theorem 5.2.18.$\;\;\;\;\Box$

{\bf Theorem 5.2.22}. {\it If $V$ is rational in the sense of [Zhu] and
$M^{i}$ (i=1,2,3) are irreducible $V$-modules, then
$(T(M^{1},M^{2}), F(\cdot,x))$ satisfies the universal property
in Definition 5.1.1, i.e., it is a tensor product for the ordered pair
$(M^{1},M^{2})$. }

{\bf Proof}. Since any $V$-module is completely reducible, by Remark
5.1.9, it is sufficient to check the universal property for each
irreducible $V$-module $W$.
Let $W$ be an irreducible $V$-module and let $I(\cdot,x)$ be any
intertwining operator of type
$\left(\begin{array}{c}W\\M^{1},M^{2}\end{array}\right)$. Then from
Theorem 5.2.18, we get a $V$-homomorphism $\bar{\psi} _{I}$ from
$T(M^{1},M^{2})$ to $W$ preserving the gradings. By
Remark 5.2.19, $\bar{\psi}_{I}$
maps the radical $R(T(M^{1},M^{2}))$ to zero. Then we get a
$V$-homomorphism $g$ from
$T(M^{1},M^{2})$ to $W$ satisfying the condition:
\begin{eqnarray}
g\pi (Y_{t}(u^{1},x)\otimes u^{2})=I^{o}(u^{1},x)u^{2}\;\;\;\;\mbox{for }
u^{1}\in M^{1}, u^{2}\in M^{2}.
\end{eqnarray}
Thus $g\circ F(u^{1},x)u^{2}=I(u^{1},x)u^{2}$ for $u^{1}\in M^{1},
u^{2}\in M^{2}$.

Conversely, let $f$ be a $V$-homomorphism from $T(M^{1}
,M^{2})$ to $W$. Since $T(M^{1},M^{2})$ is completely
reducible and generated from degree-zero subspace (the lowest-weight
subspace), then any lowest-weight vector of $T(M^{1},M^{2})$
must be of degree-zero. So $f$ preserves the ${\bf Z}$-grading. Therefore the
pull-back of $f$ is a grading preserving $V$-homomorphism from
$F_{2}(M^{1},M^{2})$ to $W$. By Theorem 5.2.18, we get an
intertwining
operator $I(\cdot,x)$. It follows from the construction that $\psi
_{I}=f.\;\;\;\;\Box$

{}From Theorem 5.2.17, we find that $F_{2}(M^{1},M^{2})^{(\lambda)}$ posses
a similar property as the desired tensor product. If we put all of
them together, what we get is too big. If we just take one from the
sequence (5.2.39) for each $\lambda +{\bf Z}\in {\bf C/Z}$, it is not
enough. We have to consider a sort of ``cover'' or the direct limit
for the whole sequence. If the left
direct limit exists for any $\lambda +{\bf Z} \in {\bf C/Z}$, we
denote $F_{2}(M^{1},M^{2})^{\bar{\lambda}}$ the direct limit of sequence
(5.2.39). This ``if'' corresponds to that ``if there is a maximal
submodule in $M^{1}\Box M^{2}$'' in Huang and Lepowsky's approach [HL0]-[HL2].

{\bf Theorem 5.2.23}. {\it If $V$ is a rational vertex operator algebra, then}
\begin{eqnarray}
T(M^{1},M^{2})=\oplus _{\bar{\lambda} \in {\bf
C/Z}}F_{2}(M^{1},M^{2})^{\bar{\lambda}}.\end{eqnarray}

{\bf Proof}. It suffices to check the universal property for the right
hand side of (5.2.54). The proof is exactly the same as that of Theorem
5.2.17 except making a shift when defining $I$ and $I(\cdot,x).\;\;\;\;\Box$

{\bf Definition 5.2.24.} Let $M^{1}=\oplus_{\alpha \in
A}M^{\alpha}, M^{2}=\oplus _{\beta \in B}M^{\beta}$
where $M^{\alpha}, M^{\beta}$ are $V$-modules from $S$. Then we define
\begin{eqnarray}
T(M^{1},M^{2})=\oplus_{\alpha \in A,\beta \in
B}T(M^{\alpha},M^{\beta}).
\end{eqnarray}
Then it is easy to see that $(T(M^{1},M^{2}),F(\cdot,x))$ is a tensor
product for the pair $(M^{1},M^{2})$ where $F(\cdot,x)$ is the sum of
all $F_{\alpha,\beta}(\cdot,x)$.

\newpage

\chapter{Unital property and commutativity for tensor products}

In this chapter we shall prove that the adjoint module satisfies the
unital property and the defined tensor product satisfies  a certain
commutative property, so that the fusion algebra is
a commutative algebra with identity. We also prove that for
irreducible $V$-modules $M^{1}$ and $M^{2}$, Frenkel and Zhu's
$A(V)$-module $A(M^{1})\otimes _{A(V)}M^{2}(0)$ (recalled in Chapter 4) is
isomorphic to the
degree-zero (the lowest-degree) subspace of $T(M^{1},M^{2})$ defined
in Chapter 5.

Throughout this chapter we shall assume that
$V$ is a vertex operator algebra.

\section {The unital property of the adjoint module}

In this section we shall prove that in the defined tensor category of
$V$-modules, $V$ satisfies the unital property.
The main result of this section may be considered as an interpretation
of Lemma 4.1.6  in terms of tensor product language.

Let $M$ be any $V$-module and let $u$ be a vector of $M$. Denote by
$V\cdot u$ the linear span of all elements $a_{n}u$ for $a\in V,n\in
{\bf Z}$.

{\bf Lemma 6.1.1.} {\it Let $u$ be any element of a
$V$-module $M$. Then $V\cdot u$ is a $V$-submodule of $M$.}

{\bf Proof.} Let $a$ and $b$ be any two elements of $V$. Let
$k$ be a positive integer such that $a_{m}u=0$ for $m\ge k$. Then
taking ${\rm Res}_{x_{1}}x_{1}^{k}$ of the Jacobi identity
for $(a,b,u)$, we obtain
\begin{eqnarray}
(x_{0}+x_{2})^{k}Y(a,x_{0}+x_{2})Y(b,x_{2})u=
(x_{2}+x_{0})^{k}Y(Y(a,x_{0})b,x_{2})u.
\end{eqnarray}
Then
$(x_{0}+x_{2})^{k}
Y(a,x_{0}+x_{2})Y(b,x_{2})u\in (V\cdot u)((x_{0}))((x_{2}))$.
Thus $Y(a,x)(V\cdot u)\subseteq (V\cdot u)((x))$ for any $a\in V$. This
proves that $(V\cdot u)$ is a $V$-submodule of $M$. $\;\;\;\;\Box$

Let $M^{1}$ and $M^{2}$ be $V$-modules
and let $\pi_{1}$, $\pi _{2}$ and $\pi$ be the quotient maps from
$F_{0}(M^{1},M^{2})$ to $F_{1}(M^{1},M^{2})$,
from $F_{0}(M^{1},M^{2})$ to
$F_{2}(M^{1},M^{2})$, and from $F_{0}(M^{1},M^{2})$ to
$T(M^{1},M^{2})$, respectively.

{\bf Proposition 6.1.2}. {\it Let $M$ be a $V$-module in $S$ with
lowest weight $h$ such that for any nonzero submodule $W$, $W_{(h)}\ne
0$ and that $M_{(h)}$ generates $M$. Then} $T(V,M)\simeq M.$

{\bf Proof}. Let $k$ $(k\le 0)$ be the lowest weight of
$V$ and define a linear map:
\begin{eqnarray}
f:& & F_{0}(V,M)={\bf C}[t,t^{-1}]\otimes V\otimes M
\rightarrow M;\nonumber \\
& & t^{n}\otimes a\otimes u\mapsto z^{-n-1}a_{n+k}u\;\;\;\mbox{for }
a\in V, u\in M, n\in {\bf Z}.
\end{eqnarray}
Equivalently, in terms of generating functions, we have
\begin{eqnarray}
f(Y_{t}(a,x)\otimes u)=(zx)^{k}Y(a,zx)u.
\end{eqnarray}
For $ a, b\in V, u\in M$, by definition we have:
\begin{eqnarray}
& &f\left(Y(b,x_{1})(Y_{t}(a,x_{2})\otimes u)\right)\nonumber\\
&=&(zx_{2})^{k}Y(a,zx_{2})Y(b,x_{1})u+{\rm
Res}_{x_{0}}(zx_{2})^{-1}\delta\left(\frac{x_{1}-x_{0}}{zx_{2}}\right)
(zx_{2})^{k}Y(Y(b,x_{0})a,zx_{2})u\nonumber\\
&=&(zx_{2})^{k}Y(b,x_{1})Y(a,zx_{2})u\nonumber\\
&=&Y(b,x_{1})f(Y_{t}(a,x_{2})\otimes u).
\end{eqnarray}
It is clear that ${\it f}$ maps the generating elements of $J$
to zero, so that
${\it f}$ induces a $V$-homomorphism $\bar{f}$ from $F_{2}(V,M)$ to $M$.
Then from the condition on $M$, $\bar{f}$ gives rise to a
$V$-homomorphism from $T(V,M)$ to $M$.

On the other hand, we construct an inverse map as follows: Define
\begin{eqnarray}
g:& & M \rightarrow T(V,M);\nonumber\\
& &u \mapsto \pi \left(t^{-1-k}\otimes {\bf 1}\otimes u\right)
 \;\;\;\mbox{for } u\in M.
\end{eqnarray}
Equivalently, $g(u)={\rm Res}_{x}x^{-1-k}\psi(Y_{t}({\bf 1},x)\otimes u)$.
For $a\in V, u\in M$, by definition we have:
\begin{eqnarray}
& &Y(a,x_{1})g(u)\nonumber\\
&=&{\rm Res}_{x_{2}}Y(a,x_{1})x^{-1-k}_{2}\pi (Y_{t}({\bf
1},x_{2})\otimes u)\nonumber\\
&=&{\rm Res}_{x_{2}}x^{-1-k}_{2}\pi(Y_{t}({\bf 1},x_{2})\otimes
Y(a,x_{1})u)\nonumber\\
& &+{\rm Res}_{x_{0}}{\rm
Res}_{x_{2}}z^{-1}x^{-2-k}_{2}\delta\left(\frac{x_{1}-x_{0}}{zx_{2}}\right)
\pi (Y_{t}(Y(a,x_{0}){\bf 1},x_{2})\otimes u)\nonumber\\
&=&\pi(t^{-1-k}\otimes {\bf 1}\otimes
Y(a,x_{1})u)\nonumber\\
&=&g(Y(a,x_{1})u).
\end{eqnarray}
Then $g$ is a $V$-homomorphism from $M$ to $T(V,M)$. Since
\begin{eqnarray}
\bar{f}g(u)=\bar{f}\pi(t^{-1-k}\otimes {\bf 1}\otimes u)={\bf 1}_{-1}u=u
\;\;\;\mbox{for all}\;u\in M,
\end{eqnarray}
$g$ is injective. The remain of the proof is to prove the surjective
property of $g$. First, for any $m\in {\bf Z}, u\in M_{(h)}$, we have:
\begin{eqnarray}
a_{n}\pi(t^{m}\otimes {\bf 1}\otimes u)=t^{m}\otimes {\bf 1}\otimes a_{n}u
+\sum_{i=0}^{\infty}\left(\begin{array}{c}n\\i\end{array}\right)
\pi(t^{m+n-i}\otimes a_{i}{\bf 1}\otimes u)=0
\end{eqnarray}
for any $a\in V, n\in {\bf Z}$ such that $\deg a_{n}<0$. By the
definition of $T(V,M)$, we have:
\begin{eqnarray}
\pi(t^{m}\otimes {\bf 1}\otimes u)=0 \;\;\;\mbox{ for }m\ne -k-1.
\end{eqnarray}
Thus $\pi(t^{m}\otimes {\bf 1}\otimes u)\in g(M)$ for any $m\in
{\bf Z}, u\in M_{(h)}$.

Let $m,n\in {\bf Z}, u\in M_{(h)}, a\in V$ such that $\deg a_{n}>0$. Then
\begin{eqnarray}
& &\pi(t^{m}\otimes {\bf 1}\otimes a_{n}u)\nonumber\\
&=&a_{n}\pi(t^{m}\otimes {\bf 1}\otimes u)
-\sum_{i=0}^{\infty}\left(\begin{array}{c}n\\i\end{array}\right)
\pi(t^{m+n-i}\otimes a_{i}{\bf 1}\otimes u)\nonumber\\
&=&a_{n}\pi(t^{m}\otimes {\bf 1}\otimes u)\in g(M).
\end{eqnarray}
It follows from Proposition 6.1.1 that $\pi(t^{m}\otimes {\bf
1}\otimes u)\in g(M)$ for any $m\in {\bf Z}, u\in M$.

Let $a\in V, m\in {\bf Z}, u\in M$. Then
\begin{eqnarray}
& &\pi(t^{m}\otimes a\otimes u)\nonumber\\
&=&{\rm Res}_{x_{0}}{\rm Res}_{x_{2}}x_{0}^{-1}x_{2}^{m}
\pi(Y_{t}(Y(a,x_{0}){\bf 1},x_{2})\otimes u)\nonumber\\
&=&{\rm Res}_{x_{1}}{\rm Res}_{x_{2}}x_{2}^{m}(x_{1}-x_{2})^{-1}
Y(a,x_{1})\pi(Y_{t}({\bf 1},x_{2})\otimes u)\nonumber\\
& &-{\rm Res}_{x_{1}}{\rm Res}_{x_{2}}x_{2}^{m}(-x_{2}+x_{1})^{-1}
\pi(Y_{t}({\bf 1},x_{2})\otimes Y(a,x_{1})u).
\end{eqnarray}
Then $\pi(t^{m}\otimes a\otimes u)\in g(M)$ for any $m\in {\bf Z},
a\in V, u\in M$. Thus $g(M)=T(V,M)$ so that $g$ is surjective.
Therefore $M$ is isomorphic to $T(V,M)$.$\;\;\;\;\Box$

\section{A commutativity for tensor products}
In this section we shall prove that the defined tensor products
satisfy a certain commutative property.

Let $M^{1}$ and $M^{2}$ be $V$-modules from $S$. Let $z$ be the same
complex number we used in Section 5.2 and define a linear map $\tau $
as follows:
\begin{eqnarray}
\tau :& & F_{0}(M^{1},M^{2}) \rightarrow F_{2}(M^{2},M^{1}),\nonumber \\
& &Y_{t}(u,x)\otimes v \mapsto e^{zxL(-1)}\pi_{2} (Y_{t}(v,-x)\otimes
u)\;\;\;\mbox{for } u\in M^{1}, v\in M^{2}.
\end{eqnarray}

{\bf Proposition 6.2.1}. {\it The linear map $\tau$ is a ${\bf
g}(V)$-homomorphism. In other words}:
\begin{eqnarray}
\tau (Y_{t}(a,x_{1})(Y_{t}(u,x_{2})\otimes v))=Y_{t}(a,x_{1})\tau
(Y_{t}(u,x_{2})\otimes v)
\end{eqnarray}
for $ a\in V, u\in M^{1}, v\in M^{2}$.

{\bf Proof}. For $a\in V, u\in M^{1}, v\in M^{2}$, from the defining
relation $J_{1}$ (the Jacobi identity relation) we have:
\begin{eqnarray}
& &x^{-1}_{1}\delta\left(\frac{x_{0}+zx_{2}}{x_{1}}\right)Y(a,x_{0})\pi_{2}
(Y_{t}(v,-x_{2})\otimes u)\nonumber\\
& &-x^{-1}_{1}\delta\left(\frac{zx_{2}+x_{0}}{x_{1}}\right)\pi_{2}
(Y_{t}(v,-x_{2})\otimes Y(a,x_{0})u)\nonumber\\
&=&(-zx_{2})^{-1}\delta\left(\frac{x_{0}-x_{1}}{-zx_{2}}\right)\pi_{2}
(Y_{t}(Y(a,x_{1})v,-x_{2})\otimes u).
\end{eqnarray}
By using the properties of delta-functions we obtain:
\begin{eqnarray}
& &x_{0}^{-1}\delta\left(\frac{x_{1}-zx_{2}}{x_{0}}\right)Y(a,x_{0})\pi_{2}
(Y_{t}(v,-x_{2})\otimes u)\nonumber\\
& &-(zx_{2})^{-1}\delta\left(\frac{x_{1}-x_{0}}{zx_{2}}\right)\pi_{2}
(Y_{t}(v,-x_{2})\otimes Y(a,x_{0})u)\nonumber\\
&=&x_{0}^{-1}\delta\left(\frac{-zx_{2}+x_{1}}{x_{0}}\right)\pi_{2}
(Y_{t}(Y(a,x_{1})v,-x_{2})\otimes u).
\end{eqnarray}
That is,
\begin{eqnarray}
& &x_{0}^{-1}\delta\left(\frac{x_{1}-zx_{2}}{x_{0}}\right)Y(a,x_{0})\pi_{2}
(Y_{t}(v,-x_{2})\otimes u)\nonumber\\
& &-x_{0}^{-1}\delta\left(\frac{-zx_{2}+x_{1}}{x_{0}}\right)\pi_{2}
(Y_{t}(Y(a,x_{1})v,-x_{2})\otimes u)\nonumber\\
&=&(zx_{2})^{-1}\delta\left(\frac{x_{1}-x_{0}}{zx_{2}}\right)\pi_{2}
(Y_{t}(v,-x_{2})\otimes Y(a,x_{0})u).
\end{eqnarray}
Multiplying (6.2.5) by $e^{zx_{2}L(-1)}$ from the left side, we obtain:
\begin{eqnarray}
& &x_{0}^{-1}\delta\left(\frac{x_{1}-zx_{2}}{x_{0}}\right)Y(a,x_{1})
e^{zx_{2}L(-1)}\pi_{2} (Y_{t}(v,-x_{2})\otimes u)\nonumber\\
& &-x_{0}^{-1}\delta\left(\frac{-zx_{2}+x_{1}}{x_{0}}\right)e^{zx_{2}L(-1)}
\pi_{2}(Y_{t}(Y(a,x_{1})v,-x_{2})\otimes u)\nonumber\\
&=&(zx_{2})^{-1}\delta\left(\frac{x_{1}-x_{0}}{zx_{2}}\right)e^{zx_{2}L(-1)}
\pi_{2}(Y_{t}(v,-x_{2})\otimes Y(a,x_{0})u).
\end{eqnarray}
Thus
\begin{eqnarray}
& &x^{-1}_{0}\delta\left(\frac{x_{1}-zx_{2}}{x_{0}}\right)Y(a,x_{1})\tau
(Y_{t}(u,x_{2})\otimes v)\nonumber\\
& &-x^{-1}_{0}\delta\left(\frac{-zx_{2}+x_{1}}{x_{0}}\right)\tau
(Y_{t}(u,x_{2})\otimes Y(a,x_{1})v)\nonumber\\
&=&(zx_{2})^{-1}\delta\left(\frac{x_{1}-x_{0}}{zx_{2}}\right)\tau
(Y_{t}(Y(a,x_{0})u,x_{2})\otimes v).
\end{eqnarray}
Taking ${\rm Res}_{x_{0}}$, we obtain
\begin{eqnarray}
& &Y(a,x_{1})\tau (Y_{t}(u,x_{2})\otimes v)\nonumber\\
&=&\tau (Y_{t}(u,x_{2})\otimes
Y(a,x_{1})v) +{\rm
Res}_{x_{0}}x^{-1}_{2}\delta\left(\frac{x_{1}-x_{0}}{x_{2}}\right)\tau
(Y_{t}(Y(a,x_{0})u,x_{2})\otimes v)\nonumber\\
&=&\tau (Y(a,x_{1})(Y_{t}(u,x_{2})\otimes v)).\; \;\;\;\Box
\end{eqnarray}

{\bf Corollary 6.2.2}. {\it $\tau $ induces a $V$-isomorphism from
$F_{2}(M^{1},M^{2})${\it to} $F_{2}(M^{2},M^{1})$}.

{\bf Proof}. Writing (6.2.1) into components, we obtain
\begin{eqnarray}
\tau (t^{r}\otimes u\otimes v)=\sum
^{\infty}_{k=0}\frac{(-1)^{k+r+1}L(-1)^{k}}{k!}\pi_{2} (t^{k+r}\otimes
v\otimes u)\;\;\;\mbox{for } u\in M^{1}, v\in M^{2}.
\end{eqnarray}
Then $\tau (J)=0$ by the definition of $J$ and Proposition 6.2.1.
By (6.2.2) and Proposition 6.2.1, we can easily find that $\tau$ induces a
$V$-homomorphism from $F_{2}(M^{1},M^{2})$ to $F_{2}(M^{2},M^{1})$.
Since
\begin{eqnarray}
\tau ^{2}(Y_{t}(u,x)\otimes v)&=&\tau
(e^{zxL(-1)}\pi_{2}(Y_{t}(v,-x)\otimes u))\nonumber\\
&=&e^{zxL(-1)}e^{-zxL(-1)}\pi_{2}(Y_{t}(u,x)\otimes v)\nonumber\\
&=&\pi_{2}(Y_{t}(u,x)\otimes v),
\end{eqnarray}
$\tau$ is an isomorphism. $\;\;\;\;\Box$

The following theorem directly follows from the definition of tensor
product and Corollary 6.2.2.

{\bf Theorem 6.2.3}. {\it The linear map $\tau$ induces a $V$-isomorphism from
$T(M^{1},M^{2})$ onto $T(M^{2},M^{1})$.}

{\bf Proposition 6.2.4}. {\it For any} $\lambda\in {\bf C}$, $\tau $ {\it
induces an isomorphism from} $F_{2}(M^{1},M^{2})^{(\lambda)}$ {\it onto}
$F_{2}(M^{2},M^{1})^{(\lambda)}$.

{\bf Proof.} Let $\tau _{\lambda}=\sigma _{\lambda} \tau$ from
$F_{2}(M^{1},M^{2})$ to $F_{2}(M^{2},M^{1})^{(\lambda)}$. If $\tau
_{\lambda}(L_{\lambda})=0$, then $\tau _{\lambda}$ induces a
$V$-homomorphism from $F_{2}(M^{1},M^{2})^{(\lambda)}$ to
$F_{2}(M^{2},M^{1})^{(\lambda)}$.
For $u\in M^{1}, v\in M^{2}$, we have:
\begin{eqnarray}
& &\tau _{\lambda}(Y_{t}(u,x)\otimes v-{d\over
dx}Y_{t}(u,x)\otimes
v-\lambda x^{-1}Y_{t}(u,x)\otimes v)\nonumber\\
&=&e^{zxL(-1)}\sigma _{\lambda}\pi_{2} (Y_{t}(v,-x)\otimes u)-{d \over
dx}(e^{zxL(-1)}\sigma _{\lambda}\pi_{2} (Y_{t}(v,-x)\otimes u))\nonumber\\
& &-\lambda x^{-1}e^{zxL(-1)\sigma _{\lambda}\pi_{2}}(Y_{t}(v,-x)\otimes
u)\nonumber \\
&=&e^{zxL(-1)}\sigma _{\lambda}\pi_{2}(Y_{t}(v,-x)\otimes u-L(-1)\sigma
_{\lambda}\pi_{2}(Y_{t}(v,-x)\otimes u))\nonumber\\
& &-e^{zxL(-1)}\sigma _{\lambda}\pi_{2} \left({d\over dx}Y_{t}(v,-x)\otimes u-
\lambda x^{-1}Y_{t}(v,-x)\otimes u\right)\nonumber\\
&=&-e^{zxL(-1)}\sigma _{\lambda}\pi_{2} ((Y_{t}(L(-1)v,-x)+{d\over
dx}Y_{t}(v,-x)-\lambda
x^{-1}Y_{t}(v,-x))\otimes u).\nonumber\\
& &\mbox{}
\end{eqnarray}
By the definition of $L_{\lambda}$ (the formula (5.2.15))
we have:
\begin{eqnarray}
(Y_{t}(L(-1)v,-x)+{d\over dx}Y_{t}(v,-x)-\lambda
x^{-1}Y_{t}(v,-x))\otimes u
\end{eqnarray}
is
\begin{eqnarray}
(-1)^{n+1}(t^{n}\otimes L(-1)v\otimes v+nt^{n-1}\otimes v\otimes
u-\lambda t^{n-1}\otimes v\otimes u)\subseteq L_{\lambda}[[x,x^{-1}]].
\end{eqnarray}
Then $\tau _{\lambda}(L_{\lambda})=0$.

Similarly to the proof of Corollary 6.2.2, we can prove that $\tau
_{\lambda}$ is an
isomorphism. $\;\;\;\;\Box$

\section { Identifying $T(M^{1},M^{2})(0)$ with
$A(M^{1})\otimes_{A(V)}M^{2}(0)$ }
It is clear that our approach to the tensor product theory is a formal
variable approach while Huang and Lepowsky's approach is analytic.
It is also easy to see that Frenkel and Zhu's approach is also a formal
variable approach, i.e., an intertwining vertex operator is studied
as a formal series of operators.
In this section we shall prove that as an $A(V)$-module the
lowest-degree subspace of
$T(M^{1},M^{2})$ is isomorphic to $A(M^{1})\otimes _{A(V)}M^{2}(0)$.

For any $V$-module $M$ and any nonzero complex number $z$, slightly
generalizing [FZ], we define $A(V)$-bimodules $A_{z}(M)$ continuously
depending on $z$. Let $O_{z}(M)$ be the subspace of $M$ linearly spanned by
\begin{eqnarray}
{\rm Res}_{x}Y(a,x){\frac{(z+x)^{{\rm wt}\:a}}{x^{2}}}u
\end{eqnarray}
for all $u\in M$ and all homogeneous $a\in V$.

Define a left action and a right action for $V$ on $M$ as follows:
For each homogeneous $a\in V$, $u\in M$, we define
\begin{eqnarray}
a*_{z}u&=&{\rm Res}_{x}Y(a,x){\frac{(z+x)^{{\rm wt}\:a}}{x}}u;\\
u*_{z}a&=&{\rm Res}_{x}zY(a,x){\frac{(z+x)^{{\rm
wt}\:a-1}}{x}}u.\end{eqnarray}
Then extend the definition linearly to any element of $V$. By
slightly modifying Frenkel-Zhu's proof we can easily get:

{\bf Proposition 6.3.1} {\it Under the definitions (6.3.2) and
(6.3.3), we have:}
\begin{eqnarray}
& &V*_{z}O_{z}(M)\subseteq O_{z}(M),\;O_{z}(M)*_{z}V\subseteq O_{z}(M);\\
& &O_{1}(V)*_{z}O_{z}(M)=0,\;O_{z}(M)*_{z}O_{1}(V)=0.
 \end{eqnarray}

Set $A_{z}(M)=M/O_{z}(M)$. Then we have:

{\bf Proposition 6.3.2.} {\it For any nonzero complex number $z$, $A_{z}(M)$
is an $A(V)$-bimodule. In other words,}
\begin{eqnarray}
& &(a*_{1}b)*_{z}u=a*_{z}(b*_{z}u)\;\;\;\;{\rm mod}\:O_{z}(M);\\
& &u*_{z}(a*_{1}b)=(u*_{z}a)*_{z}b\;\;\;\;{\rm mod}\:O_{z}(M);\\
& &a*_{z}(u*_{z}b)=(a*_{z}u)*_{z}b\;\;\;\;{\rm mod}\:O_{z}(M).
\end{eqnarray}

{\bf Theorem 6.3.3.} {\it If $V$ is rational and $M^{i}$ (i=1,2,3) are
irreducible
$V$-modules, then ${\rm
Hom}_{A(V)}(A_{z}(M^{1})\otimes _{A(V)}M^{2}(0),M^{3}(0))$ is linearly
isomorphic to
$I\left(\begin{array}{c}M^{3}\\M^{1},M^{2}\end{array}\right)$.}

Let $T(M^{1},M^{2})(0)$ be the degree-zero subspace of
$T(M^{1},M^{2})(0)$. Then $T(M^{1},M^{2})(0)$ is linearly spanned by
\begin{eqnarray}
\pi_{2} (t^{m+n-1}\otimes M^{1}(m)\otimes M^{2}(n))\;\;\;\mbox{for
all }m,n\in {\bf N}.
\end{eqnarray}

{\bf Lemma 6.3.4.} {\it If $M^{2}(0)$ generates $M^{2}$, then
$T(M^{1},M^{2})(0)$ is linearly spanned by}
\begin{eqnarray}
\pi_{2} (t^{m-1}\otimes M^{1}(m)\otimes M^{2}(0))\;\;\;\mbox{for all }m
\in {\bf N}.
\end{eqnarray}

{\bf Proof.} For any homogeneous elements $a\in V,\;u^{1}\in
M^{1},\;u^{2}\in M^{2}(0)$, by (5.2.6) we have:
\begin{eqnarray}
& &\pi (Y_{t}(u^{1},x_{2})\otimes Y(a,x_{1})u^{2})\nonumber\\
&=&Y(a,x_{1})\pi(Y_{t}(u^{1},x_{2})\otimes u^{2})+{\rm
Res}_{x_{0}}(zx_{2})^{-1}\delta\left(\frac{x_{1}-x_{0}}{zx_{2}}\right)\pi
(Y_{t}(Y(a,x_{0})u^{1},x_{2})\otimes u^{2}).\nonumber\\
& &\mbox{}
\end{eqnarray}
Then by considering the degree-zero components and using Lemma 6.1.1,
we see that $T(M^{1},M^{2})(0)$ is linearly spanned by (6.3.10). $\;\;\;\;\Box$

We may naturally identify $\displaystyle{\oplus _{m=0}^{\infty}t^{m-1}\otimes
M^{1}(m)}$ with $M^{1}$.

{\bf Lemma 6.3.5.} $\pi (O_{z}(M^{1})\otimes M^{2}(0))=0.$

{\bf Proof.} For $a \in V_{(m)}, u \in M^{1}(n), v \in M^{2}(0)$, we have:
\begin{eqnarray}
& &\pi \left({\rm Res}_{x}\left(Y(a,x)\frac{(z+x)^{{\rm
wt}a}}{x^{2}}u\right)\otimes v\right)\nonumber\\
&=&\sum _{i=0}^{\infty}\left(\begin{array}{c}{\rm wt}a\\ i\end{array}\right)
x^{{\rm wt}\:a-i}\pi (a_{i-2}u\otimes v)\nonumber\\
&\equiv&\sum_{i=0}^{\infty}\left(\begin{array}{c}{\rm wt}a\\i\end{array}\right)
z^{{\rm wt}a-i}\pi (t^{{\rm wt}a+{\rm deg}u-i}\otimes a_{i-2}u\otimes
v)\nonumber \\
&=&{\rm Res}_{x_{2}}{\rm Res}_{x_{0}}\sum
_{i=0}^{\infty}\left(\begin{array}{c}{\rm wt}a\nonumber\\
i\end{array}\right)x_{0}^{i-2}z^{{\rm wt}a-i}x_{2}^{{\rm
wt}a+{\rm deg}u-i}\pi (Y_{t}(Y(a,x_{0})u,x_{2})\otimes v)\nonumber\\
&=&{\rm Res}_{x_{2}}{\rm Res}_{x_{0}}\frac{(xz+x_{0})^{{\rm
wt}a}x_{2}^{{\rm deg}u}}{x_{0}^{2}}\pi (Y_{t}(Y(a,x_{0})u,x_{2})\otimes
v)\nonumber \\
&=&{\rm Res}_{x_{2}}{\rm Res}_{x_{0}}{\rm Res}_{x_{1}}\frac{x_{1}^{{\rm
wt}a}x_{2}^{{\rm deg}u}}{x_{0}^{2}}x_{1}^{-1}\delta\left(\frac{zx_{2}+x_{0}}
{x_{1}}\right)\pi (Y_{t}(Y(a,x_{0})u,x_{2})\otimes v)\nonumber \\
&=&{\rm Res}_{x}{\rm Res}_{x_{1}}x_{1}^{{\rm wt}a}x_{2}^{{\rm deg}\:u}
(x_{1}-zx_{2})^{-2}Y(a,x_{1})\pi (Y_{t}(u,x_{2})\otimes v)\nonumber\\
& &-{\rm Res}_{x_{2}}{\rm Res}_{x_{1}}x_{1}^{{\rm wt}a}x_{2}^{{\rm deg}\:u}
(-zx_{2}+x_{1})^{-2}\pi (Y_{t}(u,x_{2})\otimes Y(a,x_{1})v)\nonumber\\
&=&0.\;\;\;\;\Box\end{eqnarray}

The following lemma can be easily proved by using the same proof as
that Frenkel and Zhu used in [FZ].

{\bf Lemma 6.3.6.} {\it For} $a\in V,u\in A_{z}(M),v\in M^{2}(0)$, {\it
we have}
\begin{eqnarray}
\pi (u*_{z}a\otimes v)=\pi (u\otimes a*_{z}v).
\end{eqnarray}

By Lemmas 6.3.4, 6.3.5 and 6.3.6 we obtain an $A(V)$-module
homomorphism $\varphi$ from
$A_{z}(M^{1})\otimes _{A(V)}M^{2}(0)$ onto $T(M^{1},M^{2})(0)$.
Combining Theorem 5.2.16 with Theorem 6.3.3, we have:

{\bf Corollary 6.3.7.} {\it $T(M^{1},M^{2})(0)$ and
$A_{z}(M^{1})\otimes _{A(V)}M^{2}(0)$ are isomorphic left $A(V)$-modules}.

If we could prove Corollary 6.3.7 directly, FZ's Theorem 6.3.3 would be
proved. But the defining relations $J$ is too complicated to be
made explicit.

\newpage

\chapter{An analogue of the ``Hom''-functor}

In the classical Lie theory the ''Hom''-functor is a very
useful functor which is also related to the $\otimes$-functor (tensor
product). This chapter is toward to give an analogue of the ``${\rm
Hom}$''-functor for vertex operator algebra theory and to find certain
relations with the tensor product functor $T$ we defined in Chapter 5.

For any two modules $M^{1}$ and $M^{2}$ for a vertex operator algebra $V$,
the obvious candidate $\left({\rm Hom}_{{\bf C}}(M^{1},M^{2})\right)\{x\}$
is too big to be a natural $V$-module. The right candidate turns out
to be $G(M^{1},M^{2})$, the subspace consisting of what we call ``generalized
intertwining operators'' from $M^{1}$ to $M^{2}$. The definition
 of generalized intertwining operators from $M^{1}$ to $M^{2}$
(Definition 7.1.1) exactly reflects the main features of $I(u,x)$ for
$u\in M$, where $M$ is a $V$-module and $I(\cdot,x)$ is an
intertwining operator of type
$\left(\begin{array}{c}M^{2}\\M,M^{1}\end{array}\right)$. As desired,
$G(M^{1},M^{2})$ is proved to be a (generalized) $V$-module (Theorem
7.1.6), which satisfies a certain universal property in terms of
the space of intertwining operators of a certain type (Theorem 7.2.1).
If the vertex operator algebra $V$ satisfies certain finiteness
and semisimplicity conditions, it is proved that there exists a unique
maximal submodule
$\Delta(M^{1},M^{2})$ of $G(M^{1},M^{2})$ and that the contragredient
module of $\Delta(M^{1},M^{2})$ is proved to be a tensor product for
the ordered pair $(M^{1}, (M^{2})')$ (Theorem 7.2.6). Using
Theorem 7.2.6 we derive a generalized version of
the well-known
Tsuchiya and Kanie's nuclear democracy theorem for WZW model [TK]
(Proposition 7.3.3).

Throughout this chapter, $V$ will be a fixed vertex operator algebra.

\section{Generalized intertwining operators }
In this section we shall first define generalized intertwining operators
from one $V$-module to another and then
we prove that the space of all generalized intertwining
operators becomes a generalized $V$-module under a certain action of
$V$. We show that this generalized module gives us another
construction of the corresponding tensor product module under a
certain condition.

{\bf Definition 7.1.1.} Let $M^{1}$ and $M^{2}$ be two $V$-modules.  A
{\it generalized intertwining operator} from $M^{1}$
to $M^{2}$ is an element
$\phi (x)=\displaystyle{\sum_{\alpha\in {\bf C}}\phi_{\alpha}x^{-\alpha-1}}
\in \left({\rm Hom}(M^{1},M^{2})\right)\{x\}$ satisfying the following
conditions:

(G1)\hspace{0.25cm}  For any $\alpha\in {\bf C}, u^{1}\in M^{1}$,
$\phi_{\alpha+n}u^{1}=0$ for $n\in {\bf Z}$ sufficiently large;

(G2)\hspace{0.25cm} $[L(-1),\phi (x)]=\phi'(x)
 \left(=\displaystyle{{d\over dx}}\phi(x)\right);$

(G3)\hspace{0.25cm} For any $a\in V$, there exists a positive integer
$k$ such that
$$(x_{1}-x_{2})^{k}Y_{M^{2}}(a,x_{1})\phi(x_{2})
=(x_{1}-x_{2})^{k}\phi(x_{2})Y_{M^{1}}(a,x_{1}).$$

A generalized intertwining  operator $\phi(x)$ is said to be
{\it homogeneous of weight $h$} if it satisfies the following condition:

(G4)\hspace{0.25cm} $[L(0),\phi(x)]=\left(h+x\displaystyle{{d\over
dx}}\right)\phi(x)$.

Denote by ${\rm G}(M^{1},M^{2})_{(h)}$ the space of all weight-$h$
homogeneous generalized intertwining operators from $M^{1}$ to
$M^{2}$ and set
\begin{eqnarray}
{\rm G}(M^{1},M^{2})=\oplus_{h\in {\bf C}}{\rm G}(M^{1},M^{2})_{(h)}.
\end{eqnarray}

Let $W(M^{1},M^{2})$ be the space consisting of each element
$\phi(x)\in \left({\rm Hom}_{{\bf C}}(M^{1},M^{2})\right)$
which satisfies the conditions (G1) and (G2). For any $a\in V$, we
define the left and the right actions of $\hat{V}$ on  $W(M^{1},M^{2})$
as follows:
\begin{eqnarray}
Y_{t}(a,x_{0})*\phi(x_{2}):&=&{\rm Res}_{x_{1}}x_{0}^{-1}\delta\left(
\frac{x_{1}-x_{2}}{x_{0}}\right)Y_{M^{2}}(a,x_{1})\phi(x_{2})\\
&=&Y_{M^{2}}(a,x_{0}+x_{2})\phi(x_{2}).\\
\phi(x_{2})* Y_{t}(a,x_{0}):&=&{\rm Res}_{x_{1}}x_{0}^{-1}\delta\left(
\frac{-x_{2}+x_{1}}{x_{0}}\right)\phi(x_{2})Y_{M^{1}}(a,x_{1})\\
&=&\phi(x_{2})(Y_{M^{1}}(a,x_{0}+x_{2})-Y_{M^{1}}(a,x_{2}+x_{0})).
\end{eqnarray}

{\bf Proposition 7.1.2.} {\it a) $W(M^{1},M^{2})$ is a left ${\bf
g}(V)$-module of level one under the defined left action.}

{\it b) $W(M^{1},M^{2})$ is a right ${\bf g}(V)$-module of level zero under
the defined right action.}

{\bf Proof.} a) First we check that $W(M^{1},M^{2})$ is closed under
the left action. For any $a\in V, m\in {\bf Z}, \phi(x)\in
W(M^{1},M^{2}), u\in M^{1}$, by definition we have:
\begin{eqnarray}
\left((t^{m}\otimes a)*\phi(x)\right)u&=&{\rm Res}_{x_{0}}x_{0}^{m}
Y_{M^{2}}(a,x_{0}+x)\phi(x_{2})u\nonumber\\
&=&\sum_{i=0}^{\infty}
\left(\begin{array}{c}-m+i-1\\i\end{array}\right)x^{i}a_{m-i}\phi(x)u.
\end{eqnarray}
Then it is clear that $(t^{m}\otimes a)*\phi(x_{2})$ satisfies (G1). Since
\begin{eqnarray}
& &[L(-1), Y_{t}(a,x_{0})*\phi(x_{2})]\nonumber\\
&=&[L(-1),Y_{M^{2}}(a,x_{0}+x_{2})\phi(x_{2})]\nonumber\\
&=&[L(-1),Y_{M^{2}}(a,x_{0}+x_{2})]\phi(x_{2})+Y_{M^{2}}(a,x_{0}+x_{2})
[L(-1),\phi(x_{2})]\nonumber\\
&=&{\partial\over \partial x_{2}}\left(Y_{M^{2}}(a,x_{0}+x_{2})\phi(x_{2})
\right)\nonumber\\
&=&{\partial\over \partial x_{2}}Y_{t}(a,x_{0})*\phi(x_{2}),
\end{eqnarray}
$(t^{m}\otimes a)*\phi(x_{2})$ satisfies (G2).

Next, we check the defining relations for ${\bf g}(V)$. By
definition we have
\begin{eqnarray}
Y_{t}({\bf 1},x_{0})*\phi(x_{2})=Y_{M^{2}}({\bf 1},x_{0}+x_{2})\phi(x_{2})
=\phi(x_{2})
\end{eqnarray}
and
\begin{eqnarray}
Y_{t}(L(-1)a,x_{0})*\phi(x_{2})&=&Y_{M^{2}}(L(-1)a,x_{0}+x_{2})\phi(x_{2})
\nonumber\\
&=&{\partial\over \partial x_{0}}Y(a,x_{0}+x_{2})\phi(x_{2})
\nonumber\\
&=&{\partial\over \partial x_{0}}Y_{t}(a,x_{0})*\phi(x_{2}).
\end{eqnarray}
Furthermore, for any $a,b\in V$, we have
\begin{eqnarray}
& &Y_{t}(a,x_{1})*Y_{t}(b,x_{2})*\phi(x_{3})\nonumber\\
&=&Y_{t}(a,x_{1})*\left(Y_{M^{2}}(b,x_{2}+x_{3})\phi(x_{3})\right)
\nonumber\\
&=&Y_{M^{2}}(a,x_{1}+x_{3})Y_{M{2}}(b,x_{2}+x_{3})\phi(x_{3}).
\end{eqnarray}
Similarly, we have
\begin{eqnarray}
Y_{t}(b,x_{2})*Y_{t}(a,x_{1})*\phi(x_{3})=Y_{M{2}}(b,x_{2}+x_{3})
Y_{M^{2}}(a,x_{1}+x_{3})\phi(x_{3}).
\end{eqnarray}
Therefore
\begin{eqnarray}
& &Y_{t}(a,x_{1})*Y_{t}(b,x_{2})*\phi(x_{3})-Y_{t}(b,x_{2})*Y_{t}(a,x_{1})*
\phi(x_{3})\nonumber\\
&=&{\rm Res}_{x_{0}}(x_{2}+x_{3})^{-1}\delta\left(\frac{x_{1}+x_{3}-x_{0}}
{x_{2}+x_{3}}\right)Y_{M^{2}}(Y(a,x_{0})b,x_{2}+x_{3})\phi(x_{3})
\nonumber\\
&=&{\rm Res}_{x_{0}}x_{1}^{-1}\delta\left(\frac{x_{2}+x_{0}}{x_{1}}\right)
Y_{M^{2}}(Y(a,x_{0})b,x_{2}+x_{3})\phi(x_{3})\nonumber\\
&=&{\rm Res}_{x_{0}}x_{2}^{-1}\delta\left(\frac{x_{1}-x_{0}}{x_{2}}\right)
Y_{t}(Y(a,x_{0})b,x_{2})*\phi(x_{3}).
\end{eqnarray}
Then a) is  proved.

The proof of b) is similar to the proof of a), but for completeness, we also
write the details. For any $a\in V, \phi(x)\in W(M^{1},M^{2})$, by
definition we have
\begin{eqnarray}
& &{\partial\over \partial x_{2}}\left(\phi(x_{2})*Y_{t}(a,x_{0})\right)
\nonumber\\
&=&{\partial\over \partial x_{2}}\left(\phi(x_{2})(Y_{M^{1}}(a,x_{0}+x_{2})
-Y_{M^{1}}(a,x_{2}+x_{0}))\right)\nonumber\\
&=&\phi'(x_{2})(Y_{M^{1}}(a,x_{0}+x_{2})
-Y_{M^{1}}(a,x_{2}+x_{0}))\nonumber\\
& &+\phi(x_{2})(Y_{M^{1}}(L(-1)a,x_{0}+x_{2})
-Y_{M^{1}}(L(-1)a,x_{2}+x_{0}))\nonumber\\
&=&[L(-1),\phi(x_{2})*Y_{t}(a,x_{0})],
\end{eqnarray}
and
\begin{eqnarray}
& &\phi(x_{2})*Y_{t}(L(-1)a,x_{0})\nonumber\\
&=&\phi(x_{2})(Y_{M^{1}}(L(-1)a,x_{0}+x_{2})-Y_{M^{1}}(L(-1)a,x_{2}+x_{0}))
\nonumber\\
&=&{\partial\over\partial x_{0}}\left(\phi(x_{2})(Y_{M^{1}}(a,x_{0}+x_{2})
-Y_{M^{1}}(a,x_{2}+x_{0})\right)\nonumber\\
&=&{\partial\over\partial x_{0}}\phi(x_{2})*Y_{t}(a,x_{0}).
\end{eqnarray}
For any $a,b\in V$, we have
\begin{eqnarray}
& &\phi(x_{3})*Y_{t}(a,x_{1})*Y_{t}(b,x_{2})\nonumber\\
&=&{\rm Res}_{x_{4}}x_{1}^{-1}\delta\left(\frac{-x_{3}+x_{4}}{x_{1}}\right)
\left(\phi(x_{3})Y_{M^{1}}(a,x_{4})\right)*Y_{t}(b,x_{2})\nonumber\\
&=&{\rm Res}_{x_{4}}{\rm Res}_{x_{5}}x_{1}^{-1}\delta\left(\frac{-x_{3}+x_{4}}
{x_{1}}\right)x_{2}^{-1}\delta\left(\frac{-x_{3}+x_{5}}
{x_{2}}\right)\phi(x_{3})Y_{M^{1}}(a,x_{4})Y_{M^{1}}(b,x_{5}).\nonumber\\
& &\mbox{}
\end{eqnarray}
Similarly, we have
\begin{eqnarray}
& &\phi(x_{3})*Y_{t}(b,x_{2})*Y_{t}(a,x_{1})\nonumber\\
&=&{\rm Res}_{x_{4}}{\rm Res}_{x_{5}}x_{1}^{-1}\delta\left(\frac{-x_{3}+x_{4}}
{x_{1}}\right)x_{2}^{-1}\delta\left(\frac{-x_{3}+x_{5}}
{x_{2}}\right)\phi(x_{3})Y_{M^{1}}(b,x_{5})Y_{M^{1}}(a,x_{4}).\nonumber\\
& &\mbox{}
\end{eqnarray}
Thus
\begin{eqnarray}
& &\phi(x_{3})*Y_{t}(a,x_{1})*Y_{t}(b,x_{2})-\phi(x_{3})*Y_{t}(b,x_{2})*
Y_{t}(a,x_{1})\nonumber\\
&=&{\rm Res}_{x_{4}}{\rm Res}_{x_{5}}{\rm Res}_{x_{0}}x_{1}^{-1}\delta\left(
\frac{-x_{3}+x_{4}}{x_{1}}\right)x_{2}^{-1}\delta\left(\frac{-x_{3}+x_{5}}
{x_{2}}\right)x_{5}^{-1}\delta\left(\frac{x_{4}-x_{0}}{x_{5}}\right)\nonumber\\
& &\cdot \phi(x_{3})Y_{M^{1}}(Y(a,x_{0})b,x_{5})\nonumber\\
&=&{\rm Res}_{x_{5}}{\rm Res}_{x_{0}}x_{1}^{-1}\delta\left(
\frac{-x_{3}+x_{5}+x_{0}}{x_{1}}\right)x_{2}^{-1}\delta\left(
\frac{-x_{3}+x_{5}}{x_{2}}\right)\phi(x_{3})Y_{M^{1}}(Y(a,x_{0})b,x_{5})
\nonumber\\
&=&{\rm Res}_{x_{5}}{\rm Res}_{x_{0}}x_{1}^{-1}\delta\left(
\frac{x_{2}+x_{0}}{x_{1}}\right)x_{2}^{-1}\delta\left(
\frac{-x_{3}+x_{5}}{x_{2}}\right)\phi(x_{3})Y_{M^{1}}(Y(a,x_{0})b,x_{5})
\nonumber\\
&=&\phi(x_{3})*{\rm Res}_{x_{0}}x_{2}^{-1}\delta\left(\frac{x_{1}-x_{0}}
{x_{2}}\right)Y_{t}(Y(a,x_{0})b,x_{2}).
\end{eqnarray}
Then the proof is complete.$\;\;\;\;\Box$

For any $a\in V, \phi(x)\in W(M^{1},M^{2})$, we define
\begin{eqnarray}
Y_{t}(a,x_{0})\circ \phi(x_{2})=Y_{t}(a,x_{0})*\phi(x_{2})-\phi(x_{2})*
Y_{t}(a,x_{0}).
\end{eqnarray}
{}From the classical Lie algebra theory, we have:

{\bf Corollary 7.1.3.} {\it Under the defined action $''\circ''$,
$W(M^{1},M^{2})$ becomes a ${\bf g}(V)$-module of level one.}

{\bf Lemma 7.1.4.} {\it Let $\phi (x)\in W(M^{1},M^{2})$
satisfying (G4) for some complex number $h$ and let $a$ be any
homogeneous element of $V$. Then}
\begin{eqnarray}
[L(0), Y_{t}(a,x_{0})\circ \phi (x_{2})]
=\left({\rm wt}a+h+x_{0}{\partial\over\partial
x_{0}}+x_{2}{\partial\over\partial x_{2}}\right)Y_{t}(a,x_{0})\circ
\phi (x_{2}).
\end{eqnarray}

{\bf Proof.} By definition we have:
\begin{eqnarray}
& &[L(0),Y_{t}(a,x_{0})\circ \phi (x_{2})]\nonumber\\
&=&{\rm Res}_{x_{1}}x_{0}^{-1}\delta\left(\frac{x_{1}-x_{2}}{x_{0}}\right)
[L(0),Y(a,x_{1})\phi (x_{2})]\nonumber\\
& &-{\rm Res}_{x_{1}}x_{0}^{-1}\delta\left(\frac{-x_{2}+x_{1}}{x_{0}}\right)
[L(0),\phi (x_{2})Y(a,x_{1})]\nonumber\\
&=&{\rm Res}_{x_{1}}x_{0}^{-1}\delta\left(\frac{x_{1}-x_{2}}{x_{0}}\right)
\left({\rm wt}a+x_{1}{\partial\over\partial
x_{1}}+h+x_{2}{\partial\over\partial x_{2}}\right)Y(a,x_{1})\phi (x_{2})
\nonumber\\
& &-{\rm Res}_{x_{1}}x_{0}^{-1}\delta\left(\frac{x_{2}-x_{1}}{-x_{0}}\right)
\left({\rm wt}a+x_{1}{\partial\over\partial
x_{1}}+h+x_{2}{\partial\over\partial x_{2}}\right)\phi (x_{2})Y(a,x_{1})
\nonumber\\
&=&({\rm wt}a+h)Y_{t}(a,x_{0})\circ \phi (x_{2})\nonumber\\
& &-{\rm Res}_{x_{1}}\left({\partial\over\partial x_{1}}x_{1}x_{0}^{-1}
\delta\left(\frac{x_{1}-x_{2}}{x_{0}}\right)\right)Y(a,x_{1})\phi (x_{2})
\nonumber\\
& &+{\rm Res}_{x_{1}}x_{0}^{-1}\delta\left(\frac{x_{1}-x_{2}}{x_{0}}\right)
x_{2}{\partial\over\partial x_{2}}Y(a,x_{1})\phi (x_{2})\nonumber\\
& &-{\rm Res}_{x_{1}}x_{0}^{-1}\delta\left(\frac{x_{2}-x_{1}}{-x_{0}}\right)
x_{2}{\partial\over\partial x_{2}}\phi (x_{2})Y(a,x_{1})\nonumber\\
& &+{\rm Res}_{x_{1}}\left({\partial\over\partial x_{1}}x_{1}x_{0}^{-1}\delta
\left(\frac{x_{2}-x_{1}}{-x_{0}}\right)\right)\phi (x_{2})Y(a,x_{1}).
\end{eqnarray}
Since
\begin{eqnarray}
{\partial\over\partial x_{0}}x_{0}^{-1}\delta\left(
\frac{x_{1}-x_{2}}{x_{0}}\right)
={\partial\over\partial x_{2}}x_{0}^{-1}\delta\left(
\frac{x_{1}-x_{2}}{x_{0}}\right)
=-{\partial\over\partial x_{1}}x_{0}^{-1}\delta\left(
\frac{x_{1}-x_{2}}{x_{0}}\right),
\end{eqnarray}
we have
\begin{eqnarray}
& &{\partial\over\partial x_{1}}\left(x_{1}x_{0}^{-1}\delta\left(
\frac{x_{1}-x_{2}}{x_{0}}\right)\right)\nonumber\\
&=&{\partial\over\partial x_{1}}\left((x_{0}+x_{2})x_{0}^{-1}\delta\left(
\frac{x_{1}-x_{2}}{x_{0}}\right)\right)\nonumber\\
&=&x_{0}{\partial\over\partial x_{1}}x_{0}^{-1}\delta\left(
\frac{x_{1}-x_{2}}{x_{0}}\right)+x_{2}{\partial\over\partial x_{1}}x_{0}^{-1}
\delta\left(\frac{x_{1}-x_{2}}{x_{0}}\right)\nonumber\\
&=&-x_{0}{\partial\over\partial x_{0}}x_{0}^{-1}\delta\left(
\frac{x_{1}-x_{2}}{x_{0}}\right)
-x_{2}{\partial\over\partial x_{2}}x_{0}^{-1}\delta\left(
\frac{x_{1}-x_{2}}{x_{0}}\right).
\end{eqnarray}
Similarly, we have
\begin{eqnarray}
{\partial\over\partial x_{1}}\left(x_{1}x_{0}^{-1}\delta\left(
\frac{x_{2}-x_{1}}{-x_{0}}\right)\right)
=-\left(x_{0}{\partial\over\partial x_{0}}+x_{2}{\partial\over\partial x_{2}}
\right)x_{1}x_{0}^{-1}\delta\left(\frac{x_{2}-x_{1}}{-x_{0}}\right).
\end{eqnarray}
Therefore, we obtain
\begin{eqnarray}
& &[L(0),Y_{t}(a,x_{0})\circ \phi(x_{2})]\nonumber\\
&=&\left({\rm wt}a+h+x_{0}{\partial\over\partial x_{0}}+
x_{2}{\partial\over\partial x_{2}}\right)Y_{t}(a,x_{0})\circ \phi(x_{2}).
\;\;\;\;\Box \end{eqnarray}

{\bf Proposition 7.1.5.} {\it ${\rm G}(M^{1},M^{2})$ is a
${\bf g}(V)$-submodule of $W(M^{1},M^{2})$ and ${\rm
G}(M^{1},M^{2})$ itself is a ${\bf C}$-graded ${\bf g}(V)$-module.}

{\bf Proof.} It is enough to prove that for any $a\in V, n\in {\bf Z},
\phi(x)\in {\rm G}(M^{1},M^{2})$, $(t^{n}\otimes a)\circ \phi(x)$
satisfies (G3). Let $b$ be any element of $V$ and
let $r$ be a positive integer such that $r+n>0$ and that the
following identities hold:
\begin{eqnarray}
& &(x_{1}-x_{2})^{r}Y_{M^{i}}(a,x_{1})Y_{M^{i}}(b,x_{2})
=(x_{1}-x_{2})^{r}Y_{M^{i}}(b,x_{2})Y_{M^{i}}(a,x_{1});\\
& &(x_{1}-x_{2})^{r}Y_{M^{2}}(a,x_{1})\phi(x_{2})=(x_{1}-x_{2})^{r}\phi(x_{2})
Y_{M^{1}}(a,x_{1});\\
& &(x_{1}-x_{2})^{r}Y_{M^{2}}(b,x_{1})\phi(x_{2})=(x_{1}-x_{2})^{r}\phi(x_{2})
Y_{M^{1}}(b,x_{1}).
\end{eqnarray}
By definition, we have
\begin{eqnarray}
& &(t^{n}\otimes a)\circ \phi(x_{2})\nonumber\\
&=&{\rm Res}_{x_{1}}
\left((x_{1}-x_{2})^{n}Y_{M^{2}}(a,x_{1})\phi(x_{2})
-(-x_{2}+x_{1})^{n}\phi(x_{2})Y_{M^{1}}(a,x_{1})\right).
\end{eqnarray}
Since
\begin{eqnarray}
& &(x_{2}-x_{3})^{4r}\left((x_{1}-x_{2})^{n}Y_{M^{2}}(a,x_{1})\phi(x_{2})
-(-x_{2}+x_{1})^{n}\phi(x_{2})Y_{M^{2}}(a,x_{1})\right)Y_{M^{1}}(b,x_{3})
\nonumber\\
&=&\sum_{s=0}^{3r}\left(\begin{array}{c}3r\\s\end{array}\right)
(x_{2}-x_{1})^{3r-s}(x_{1}-x_{3})^{s}(x_{2}-x_{3})^{r}\nonumber\\
& &\cdot \left((x_{1}-x_{2})^{n}Y_{M^{2}}(a,x_{1})\phi(x_{2})Y_{M^{1}}(b,x_{3})
-(-x_{2}+x_{1})^{n}\phi(x_{2})Y_{M^{1}}(a,x_{1})Y_{M^{1}}(b,x_{3})\right)
\nonumber\\
&=&\sum_{s=r+1}^{3r}\left(\begin{array}{c}3r\\s\end{array}\right)
(x_{2}-x_{1})^{3r-s}(x_{1}-x_{3})^{s}(x_{2}-x_{3})^{r}\nonumber\\
& &\cdot \left((x_{1}-x_{2})^{n}Y_{M^{2}}(a,x_{1})\phi(x_{2})Y_{M^{1}}(b,x_{3})
-(-x_{2}+x_{1})^{n}\phi(x_{2})Y_{M^{1}}(a,x_{1})Y_{M^{1}}(b,x_{3})\right)
\nonumber\\
&=&\sum_{s=r+1}^{3r}\left(\begin{array}{c}3r\\s\end{array}\right)
(x_{2}-x_{1})^{3r-s}(x_{1}-x_{3})^{s}(x_{2}-x_{3})^{r}\nonumber\\
& &\cdot \left((x_{1}-x_{2})^{n}Y_{M^{2}}(b,x_{3})Y_{M^{2}}(a,x_{1})\phi(x_{2})
-(-x_{2}+x_{1})^{n}Y_{M^{2}}(b,x_{3})\phi(x_{2})Y_{M^{1}}(a,x_{1})\right)
\nonumber\\
&=&(x_{2}-x_{3})^{4r}Y_{M^{2}}(b,x_{3})
\left((x_{1}-x_{2})^{n}Y_{M^{2}}(a,x_{1})\phi(x_{2})
-(-x_{2}+x_{1})^{n}\phi(x_{2})Y_{M^{1}}(a,x_{1})\right),\nonumber\\
& &\mbox{}
\end{eqnarray}
we have
\begin{eqnarray}
(x_{2}-x_{3})^{4r}\left((t^{n}\otimes
a)\circ\phi(x_{2})\right)Y_{M^{1}}(b,x_{3})=(x_{2}-x_{3})^{4r}
Y_{M^{2}}(b,x_{3})
\left((t^{n}\otimes a)\circ \phi(x_{2})\right).\nonumber\\
& &\mbox{}
\end{eqnarray}
Then $(t^{n}\otimes a)\circ \phi (x)\in G(M^{1},M^{2})$. Thus
$G(M^{1},M^{2})$ is a submodule of $W(M^{1},M^{2})$.$\;\;\;\;\Box$

{\bf Note:} The essential idea of the proof of Proposition 7.1.5 (also
see [L2]) belongs to Professor Chongying Dong.

{\bf Theorem 7.1.6.} {\it The $g(V)$-module $G(M^{1},M^{2})$ is
a generalized $V$-module.}

{\bf Proof.} For any $h\in {\bf C}$, set
\begin{eqnarray}
G(M^{1},M^{2})^{h}=\oplus_{n\in {\bf Z}}G(M^{1},M^{2})_{(n+h)}.
\end{eqnarray}
Then $G(M^{1},M^{2})^{h}$ is a $g(V)$-submodule of $G(M^{1},M^{2})$ and
\begin{eqnarray}
G(M^{1},M^{2})=\sum_{h\in {\bf C}}G(M^{1},M^{2})^{h}.
\end{eqnarray}
Therefore it is enough to prove that each $G(M^{1},M^{2})^{h}$ is a
generalized $V$-module. This follows from Proposition 5.2.17. $\;\;\;\;\Box$

\section{A relation between $G(M^{1},M^{2})$ and $T(M^{1}, (M^{2})')$}
In this section we shall prove that just as in the classical case,
$G(M^{1},M^{2})$ is closely related to $T(M^{1}, (M^{2})')$
 under certain finiteness and semisimplicity
conditions.

Let $M$ be another $V$-module and let $\phi\in {\rm Hom}_{V}(M,
G(M^{1},M^{2}))$. Then we define a linear map $I_{\phi}(\cdot,x)$ by:
\begin{eqnarray}
\phi : & &M\rightarrow ({\rm Hom}_{{\bf
C}}(M^{1},M^{2}))\{x\}\nonumber\\
& &u\mapsto I_{\phi}(u,x)=\phi(u)(x)\;\;\;\mbox{ for }u\in M.
\end{eqnarray}
By definition, we have
\begin{eqnarray}
& &I_{\phi}(L(-1)u,x)=\phi(L(-1)u)(x)=L(-1)\circ (\phi(u)(x))={d\over
dx}\phi(u)(x)={d\over dx}I_{\phi}(u,x).\nonumber\\
& &\mbox{}
\end{eqnarray}
Furthermore, for $a\in V, u\in M$, by (7.1.18) we have
\begin{eqnarray}
& &I_{\phi}(Y(a,x_{0})u,x_{2})\nonumber\\
&=&{\rm Res}_{x_{1}}\left(x_{0}^{-1}\delta\left(\frac{x_{1}-x_{2}}{x_{0}}
\right)Y(a,x_{1})I_{\phi}(u,x_{2})-x_{0}^{-1}\delta\left(\frac{-x_{2}+x_{1}}
{x_{0}}\right)I_{\phi}(u,x_{2})Y(a,x_{1})\right).\nonumber\\
& &\mbox{}
\end{eqnarray}
It is well-known (see for example [FHL]) that this
iterate formula
implies the associativity. Furthermore, (G3) gives the commutativity
(without involving matrix-coefficients) for
$I(\cdot,x)$. Therefore it follows from [FHL] that
$I_{\phi}(\cdot,x)$ satisfies the Jacobi
identity, so that $I_{\phi}(\cdot,x)$ is an intertwining operator of
type $\left(\begin{array}{c}M^{2}\\M,M^{1}\end{array}\right)$. If
$I_{\phi}(\cdot,x)=0$, then $\phi(u)(x)=0$ for all $u\in M$. Thus
$\phi=0$. Therefore, we obtain an injective linear map
\begin{eqnarray}
\pi : & &{\rm Hom}_{V}(M,G(M^{1},M^{2}))\rightarrow I\left(
\begin{array}{c}M^{2}\\M,M^{1}\end{array}\right)\nonumber\\
& &\phi\mapsto I_{\phi}(\cdot,x).
\end{eqnarray}

On the other hand, for any intertwining operator $I(\cdot,x)$ of type
 $\left(\begin{array}{c}M^{2}\\M,M^{1}\end{array}\right)$, it is clear
that $I(u,x)\in G(M^{1},M^{2})$ for any $u\in M$. Then we
obtain a linear map $f_{I}$ from $M$ to $G(M^{1},M^{2})$
defined by $f_{I}(u)=I(u,x)$. Tracing back the argument above, we see
that $f_{I}$ is a $V$-homomorphism such that $I_{f_{I}}=I(\cdot,x)$.
Therefore we have proved the following universal property:

{\bf Theorem 7.2.1}. {\it Let $M^{1}$ and $M^{2}$ be $V$-modules.
Then for any $V$-module $M$ and any intertwining operator $I(\cdot,x)$
of type $\left(\begin{array}{c}M^{2}\\M,M^{1}\end{array}\right)$,
there exists a unique $V$-homomorphism $\psi$ from $M$ to
$G(M^{1},M^{2})$ such that $I(u,x)=\psi(u)$ for $u\in M$.}

{\bf Corollary 7.2.2.} {\it The linear space ${\rm Hom}_{V}(M,
G(M^{1},M^{2}))$ is linearly isomorphic to
$I\left(\begin{array}{c}M^{2}\\M,M^{1}
\end{array}\right)$ for any $V$-module $M$.}

The universal property in Theorem 7.2.1 looks very
much like the universal property for a tensor product in Definition
5.1.1 and also in [HL1]. In Lie algebra theory, $(M^{1})^{*}\otimes
M^{2}$ can be
naturally embedded into
${\rm Hom}_{{\bf C}}(M^{1},M^{2})$ as a Lie algebra module.
In vertex operator algebra theory,
$G(M^{1},M^{2})$ is closely related to the contragredient module
of tensor product of $M^{1}$ and $(M^{2})'$. First we consider a
special case with $M^{1}=V$.

{\bf Proposition 7.2.3}. {\it Let $M$ be a $V$-module.
Then $G(V,M)\simeq M$.}

{\bf Proof}. For any $u\in M$, we define
\begin{eqnarray}
\phi _{u}(x)&:&V\rightarrow M[[x,x^{-1}]];\nonumber\\
& &\phi _{u}(x)a=e^{xL(-1)}Y(a,-x)u\;\;\;\;\mbox{for any }a\in V.
\end{eqnarray}
By definition, we get
\begin{eqnarray}
[L(-1),\phi _{u}(x)]a
&=&L(-1)e^{xL(-1)}Y(a,-x)u-e^{xL(-1)}Y(L(-1)a,-x)u\nonumber\\
&=&\left({d\over dx}e^{xL(-1)}\right)Y(a,-x)u+e^{xL(-1)}\left({d\over
dx}Y(a,-x)u \right)\nonumber\\
&=&{d\over dx}\phi _{u}(x)a.
\end{eqnarray}
For any $b\in V$, there is a positive integer $k$ such that
\begin{eqnarray}
(x_{0}+x_{2})^{k}Y(b,x_{0}+x_{2})Y(a,x_{2})u=(x_{0}+x_{2})^{k}Y(Y(b,x_{0})
a,x_{2})u\end{eqnarray}
for any $a\in V$, where $k$ is independent of $a$. By definition, we have:
\begin{eqnarray}
(x_{0}+x_{2})^{k}Y(b,x_{0}+x_{2})e^{x_{2}L(-1)}\phi _{u}(-x_{2})=
(x_{0}+x_{2})^{k}e^{x_{2}L(-1)}\phi _{u}(-x_{2})Y(b,x_{0})a.\end{eqnarray}
By the conjugation formula ([FLM], [FHL]) we get
\begin{eqnarray}
(x_{0}+x_{2})^{k}Y(b,x_{0})\phi _{u}(-x_{2})=(x_{0}+x_{2})^{k}\phi _{u}
(-x_{2})Y(b,x_{0}).\end{eqnarray}
Thus $\phi _{u}(x)\in G(V,M)$ for any $u\in M$. Similarly to
the proof of Theorem 7.2.1, one
can prove that the linear map $\phi$ from $M$ to $G(V,M)$ is a
$V$-homomorphism.

Conversely, we prove that any generalized intertwining operator from
$V$ to $M$ is equal to a certain $\phi_{u}$ for some $u\in M$.
Let $\psi (x)$ be any generalized intertwining
operator from $V$ to $M$. Then
\begin{eqnarray}
\psi (x)=\sum_{\alpha\in {\bf C}}\psi _{\alpha}x^{-\alpha -1}
=\sum_{\alpha\in {\bf C},0\le {\rm Re}\alpha<1}
\sum_{n\in {\bf Z}}\psi_{\alpha +n}x^{-\alpha-n-1}.
\end{eqnarray}
It is easy to see that for any $\alpha\in {\bf C}$,
$\sum_{n\in {\bf Z}}\psi_{\alpha +n}x^{-\alpha-n-1}$ is a generalized
intertwining operator
again. Then it suffices for us to prove that $\sum_{n\in {\bf
Z}}\psi_{\alpha +n}x^{-\alpha-n-1}=0$ if $\alpha$ is not an integer
and $\sum_{n\in {\bf
Z}}\psi_{n}x^{-n-1}=\phi_{u}(x)$ for some $u\in M$.
Let $h$ be a complex number and let $\phi(x)=\sum_{n\in
h+{\bf Z}}\phi_{n}x^{-n-1}$ is a generalized intertwining operator
from $V$ to $M$. From condition (G2), we
have:
\begin{eqnarray}
[L(-1),\phi_{n}]=-n\phi _{n-1}\;\;\mbox{for any }n\in h+{\bf Z}.
\end{eqnarray}
If $\phi _{n}{\bf 1}=0$ for some nonzero $n$, then
\begin{eqnarray}
\phi_{n-1}{\bf 1}=-{1\over n}(L(-1)\phi _{n}-\phi_{n}L(-1)){\bf 1}=0.
\end{eqnarray}
If $h$ is not an integer, it follows from (G1) that $\phi _{n}{\bf
1}=0$ for all $n\in h+{\bf Z}$. Let $U=\{a\in V|\phi (x)a=0\}$. Then
$U$ is an ideal of $V$, which contains the vacuum vector. Thus $U=V$.
Therefore,  $\phi (x)=0$ if $h$ is not an integer.
For the rest of the proof, we assume $h=0$.
By condition (G1), it follows that $\phi _{n}{\bf 1}=0$ for any
nonnegative $n$. Therefore $\phi (x){\bf 1}$ involves only
nonnegative powers of $x$. Let $\phi _{-1}{\bf 1}=u\in M$. By using
(G2), one can easily get $\phi (x){\bf 1}=e^{xL(-1)}u$.
For any
$a\in V$, let $k$ be a positive integer such that
\begin{eqnarray}
& &(x-x_{1})^{k}\phi (x)Y(a,x_{1})=(x-x_{1})^{k}Y(a,x_{1})\phi (x),\\
& &x^{k}Y(a,x)u\in M[[x]].
\end{eqnarray}
Then
\begin{eqnarray}
x^{k}\psi (x)a
&=&\lim _{x_{1}\rightarrow 0}(x-x_{1})^{k}\phi (x)Y(a,x_{1}){\bf 1}\nonumber\\
&=&\lim _{x_{1}\rightarrow 0}(x-x_{1})^{k}Y(a,x_{1})\phi (x){\bf 1}\nonumber\\
&=&\lim _{x_{1}\rightarrow 0}(x-x_{1})^{k}Y(a,x_{1})e^{xL(-1)}u\nonumber\\
&=&\lim _{x_{1}\rightarrow 0}(x-x_{1})^{k}e^{xL(-1)}Y(a,x_{1}-x)u\nonumber\\
&=&x^{k}Y(a,-x)u.\end{eqnarray}
Thus $\phi (x)a=e^{xL(-1)}Y(a,-x)u$. That is, $\phi(x)=\phi_{u}(x)$.
Therefore, $G(V,M)$ is isomorphic to $M$ as a $V$-module.$\;\;\;\;\Box$

{\bf Remark 7.2.4}. If $M=V$, then $V=G(V,V)$. That is, any
generalized intertwining operator is a vertex operator. For this
special case, Proposition 7.2.3 was proved by Goddard [Go] where
he assumed $\phi (x){\bf 1}=e^{xL(-1)}a$ for some $a\in
V$ instead of proving this.

For any two $V$-modules $M^{1}$ and $M^{2}$, let $\Delta
(M^{1},M^{2})$ be the sum of all $V$-modules inside the
generalized $V$-module $G(M^{1},M^{2})$.

{\bf Proposition 7.2.5.} {\it Let $V$ be a vertex operator
algebra satisfying the following conditions: (1) There are finitely
many inequivalent irreducible $V$-modules. (2) Any $V$-module is
completely reducible. (3)  Any fusion rule for three modules is
finite.  Then for any $V$-modules $M^{1}$ and $M^{2}$,
$\Delta(M^{1},M^{2})$ is the unique maximal $V$-module inside
the generalized module $G(M^{1},M^{2})$. }

{\bf Proof.} It follows from the condition (2) that
$\Delta(M^{1},M^{2})$ is a direct sum of irreducible $V$-modules. It
follows from Corollary 7.2.2 and the condition (3) that the multiplicity
of each irreducible $V$-module in  $\Delta(M^{1},M^{2})$ is finite.
Therefore $\Delta(M^{1},M^{2})$ is a direct sum of finitely many
irreducible $V$-modules. That is, $\Delta(M^{1},M^{2})$ is a
$V$-module. By the definition of $\Delta(M^{1},M^{2})$, it is clear
that $\Delta(M^{1},M^{2})$ is the unique
maximal $V$-module inside the generalized $V$-module ${\rm G}(M^{1},M^{2})$.
$\;\;\;\;\Box$

Let $V$ be a vertex operator algebra satisfying the conditions (1)-(3)
of Proposition 7.2.5  and let $M^{1}$ and $M^{2}$ be any two
$V$-modules. Let $F(\cdot,x)$ be
the natural intertwining operator of type
$\left(\begin{array}{c}M^{2}\\
\Delta(M^{1},M^{2}),M^{1}\end{array}\right)$ defined by:
\begin{eqnarray}
F(u,x)=u(x)\;\;\;\mbox{ for any }u\in \Delta (M^{1},M^{2}).
\end{eqnarray}
Then by Proposition 4.1.4,  the transpose operator $F^{t}(\cdot,x)$ of
$F(\cdot,x)$ is an
intertwining operator of type $\left(\begin{array}{c}M^{2}\\
M^{1},\Delta(M^{1},M^{2})\end{array}\right)$. Furthermore, it follows
from Proposition 4.1.4 that $(F^{t})'(\cdot,x)$ is an intertwining
operator of type $\left(\begin{array}{c}(\Delta(M^{1},M^{2}))'\\ M^{1},
(M^{2})'\end{array}\right)$.

{\bf Theorem 7.2.6.} {\it If $V$ satisfies the conditions (1)-(3) of
Proposition 7.2.5, then the pair
$((\Delta (M^{1},M^{2})', (F^{t})'(\cdot,x))$ is a tensor product
for the ordered pair $(M^{1},(M^{2})')$.}

{\bf Proof.} Let $M$ be any $V$-module and let $I(\cdot,x)$ be any
intertwining operator of type $\left(
\begin{array}{c}M\\M^{1},(M^{2})'\end{array}\right)$. It follows from
Proposition 4.1.4 that $(I')^{t}(\cdot,x)$ is an intertwining operator
of type $\left(
\begin{array}{c}M^{2}\\M',M^{1}\end{array}\right)$. From Theorem 7.2.1,
there exists a (unique) $V$-homomorphism $\psi$ from $M'$ to $
G(M^{1},M^{2})$ such that $(I')^{t}(u,x)=\psi (u)(x)$ for any $u\in M'$.
It follows from the definition of $\Delta(M^{1},M^{2})$ that $\psi$ is
a $V$-homomorphism from $M'$ to $\Delta(M^{1},M^{2})$. Therefore,
we obtain a $V$-homomorphism $\psi '$ from $(\Delta(M^{1},M^{2}))'$ to $M$.
It is clear that $\psi '$ satisfies the condition of the universal
property in Definition 5.1.1. $\;\;\;\;\Box$

\section{A generalized nuclear democracy theorem}
In this section we shall use Theorem 7.1.6 to give a generalized
version of Tsuchiya
and Kanie's nuclear democracy theorem [TK]. By applying this result to
 vertex operator algebras associated to an affine Lie algebra we  give
an alternate proof for Tsuchiya
and Kanie's nuclear democracy theorem [TK].

{\bf Theorem 7.3.1.} {\it Let $M^{1}$ and $M^{2}$ be two $V$-modules.
Let $U$ be a $g(V)_{0}$-module and let $I_{0}(\cdot,x)$
be a linear injective map
from $U$ to $\left({\rm Hom}_{{\bf C}}(M^{1},M^{2})\right)\{x\}$
such that for any $u\in U$, $I_{0}(u,x)$ satisfies the truncation
condition $(G1)$, the $L(-1)$-derivative property $(G2)$ and the
following condition:}
\begin{eqnarray}
& &(x_{1}-x_{2})^{{\rm wt}a-1}Y_{M^{2}}(a,x_{1})I_{0}(u,x_{2})-
(-x_{2}+x_{1})^{{\rm wt}a-1}I_{0}(u,x_{2})Y_{M^{1}}(a,x_{1})\nonumber\\
&=&x_{1}^{-1}\delta\left({x_{2}\over x_{1}}\right)I_{0}(\bar{a}u,x_{2})
\end{eqnarray}
{\it for any $a\in V, u\in U$. Then $U$ is an $A(V)$-module and there is a
unique intertwining operator $I(\cdot,x)$ of type
$\left(\begin{array}{c}M^{2}\\ \bar{M}(U),M^{1}\end{array}\right)$
extending $I_{0}(\cdot,x)$.}

{\bf Proof.} Since
$\displaystyle{(x_{1}-x_{2})\delta\left({x_{2}\over x_{1}}\right)=0}$,
we have:
\begin{eqnarray}
(x_{1}-x_{2})^{{\rm wt}a+i}Y_{M^{2}}(a,x_{1})I_{0}(u,x_{2})
=(-x_{2}+x_{1})^{{\rm wt}a+i}I_{0}(u,x_{2})Y_{M^{1}}(a,x_{1})
\end{eqnarray}
for $a\in V, u\in U,i\in {\bf Z}_{+}$. Then by definition
$I_{0}(u,x)\in G(M^{1},M^{2})$ for any $u\in U$ and
\begin{eqnarray}
& &a_{m}\circ I_{0}(u,x)=0\;\;\;\mbox{for }m\ge {\rm wt }a,\\
& &a_{{\rm wt}a-1}\circ I_{0}(u,x)=I_{0}(\bar{a}u,x).
\end{eqnarray}
Then $\bar{U}:=\{I(u,x)|u\in U\}\subseteq \Omega (G(M^{1},M^{2}))$ and
$\bar{U}$ is a $g(V)_{0}$-submodule of
$G(M^{1},M^{2})$. Thus $\bar{U}$ is an $A(V)$-module. Since $\bar{U}$
as a $g(V)_{0}$-module is
isomorphic to $U$, $U$ is an $A(V)$-module.
Let $W=U(g(V))\bar{U}$ be the $V$
or $g(V)$-submodule of $G(M^{1},M^{2})$. Then $W=U(g_{+})\bar{U}$ is a
lower-truncated ${\bf Z}$-graded weak $V$-module. Then we have a
natural intertwining operator of type
$\left(\begin{array}{c}M^{2}\\ W,M^{1}\end{array}\right)$, so that we
have a natural intertwining operator $I(\cdot,x)$ of type
$\left(\begin{array}{c}M^{2}\\ \bar{M}(U),M^{1}\end{array}\right)$ by
using the natural $V$-homomorphism from $M(\bar{U})$ to $W$. The
uniqueness is clear.$\;\;\;\;\Box$.

{\bf Corollary 7.3.2.} {\it Under the assumption of Theorem 7.3.1, if
any $V$-module is completely reducible, then there is a unique
intertwining operator $I(\cdot,x)$ of type
$\left(\begin{array}{c}M^{2}\\ L(U),M^{1}\end{array}\right)$
extending $I_{0}(\cdot,x)$.}

One of the well-known and important classes of vertex operator algebras
is the one associated to highest-weight modules for an affine or the
Virasoro Lie algebra. Next we shall apply our general results to this
class of vertex operator algebras.

Let ${\bf g}$ be a finite-dimensional simple Lie algebra with a fixed
Cartan subalgebra ${\bf h}$. Let $\Delta$ be the set of all roots of
$({\bf g},{\bf h})$ and let $\Delta_{+}$ be the set of positive roots
with respect to a fixed Weyl chamber. Let $\theta$ be the longest root
of $\Delta$. Let $\langle\cdot,\cdot\rangle$ be the normalized Killing
form on ${\bf g}$ such that $\langle \theta,\theta\rangle =2$. For any
linear functional $\lambda\in {\bf h}^{*}$, we denote by $L(\lambda)$
(resp. $L^{*}(\lambda)$) the
irreducible highest (resp. lowest) weight  ${\bf g}$-module with
highest (resp. lowest) weight $\lambda$.

Let $\tilde{{\bf g}}={\bf C}[t,t^{-1}]\otimes {\bf g}\oplus {\bf C}c$
be the corresponding affine Lie algebra and let $\hat{{\bf
g}}=\tilde{{\bf g}}\oplus {\bf C}d$ be the extended affine algebra.
For any $\ell\in {\bf C},
\lambda\in {\bf h}^{*}$, denote by $M(\ell,\lambda)$ the generalized
Verma $\tilde{{\bf g}}$-module (or Weyl module) of level $\ell$ with
lowest-weight subspace
$L(\lambda)$ and denote by $L(\ell,\lambda)$ the irreducible quotient
module of $M(\ell,\lambda)$.
For any ${\bf g}$-module $U$, let $\hat{U}$ be the loop $\tilde{{\bf
g}}$-module ${\bf C}[t,t^{-1}]\otimes U$ of level $0$.

It has been proved ([FZ], [L2]) that each $M(\ell,0)$ has a
vertex operator algebra structure except when $\ell$ is the negative dual
Coxeter number, and that each $M(\ell,\lambda)$ is a $M(\ell,0)$-module
with lowest weight $h=\frac{\langle \lambda,
\lambda+\rho\rangle}{2(\ell+h^{\lor})}$ where $\rho$ is the half-sum
of positive roots and $h^{\lor}$ is the dual Coxeter number.
For a fixed $\ell$, let $V=L(\ell,0)$ be the simple vertex operator algebra.
Suppose $L(\ell,\lambda_{i})$ $(i=1,2,3)$ are $V$-modules. Let
$I(\cdot,x)$ be an intertwining operator of type
$\left(\begin{array}{c}L(\ell,\lambda_{3})\\L(\ell,\lambda_{1}),
L(\ell,\lambda_{2})\end{array}\right)$. As before, set
\begin{eqnarray}
I^{o}(u^{1},x)=x^{h_{1}+h_{2}-h_{3}}I(u^{1},x)=\sum_{n\in {\bf
Z}}u^{1}(n)x^{-n-1}\;\;\;\mbox{ for any }u^{1}\in L(\ell,\lambda_{1}).
\end{eqnarray}
Set
\begin{eqnarray}
Y_{t}(u^{1},x)=\sum_{n\in {\bf Z}}(t^{n}\otimes
u^{1})x^{-n-1}\;\;\;\mbox{for }u^{1}\in L(\ell,\lambda_{1}).
\end{eqnarray}
Then $I^{o}(\cdot,x)$ may be considered as a linear map from ${\bf
C}[t,t^{-1}]\otimes L(\ell,\lambda_{1})\otimes L(\ell,\lambda_{2})$ to
$L(\ell,\lambda_{3})$. Let $\phi_{I}$ be the restriction of
$I^{o}(\cdot,x)$ on ${\bf
C}[t,t^{-1}]\otimes L(\lambda_{1})\otimes L(\ell,\lambda_{2})$. By the
commutator formula, we have:
\begin{eqnarray}
& &[a(m), I(u,x)]=x^{m}I(au,x);\\
& &[L(m), I(u,x)]=\left((m+1)h_{1}+ x{d\over dx}\right)I(u,x)
\end{eqnarray}
for any $a\in {\bf g}\subseteq L(\ell,0), u\in L(\lambda_{1}), m\in
{\bf Z}$, where
$h_{1}$ is the lowest weight of $L(\ell,\lambda_{1})$.

It is easy to
see that (7.3.7) is equivalent to that
the map $\phi_{I}$ is a  $\tilde{{\bf g}}$-homomorphism from
$\hat{L}(\lambda_{1})\otimes L(\ell,\lambda_{2})$ to
$L(\ell,\lambda_{3})$. If we
consider $\hat{L}(\lambda_{1})$ as a $\hat{{\bf g}}$-module with
$d={d\over dt}\otimes 1$ and consider $L(\ell,\lambda)$ as a
$\hat{{\bf g}}$-module with $d=L(0)-h$ where $h$ is the lowest weight,
then it follows from (7.3.7) for $m=0$ that $\phi_{I}$ is a $\hat{{\bf
g}}$-homomorphism. Then we obtain a linear map:
\begin{eqnarray}
\phi :& &I\left(\begin{array}{c}L(\ell,\lambda_{3})\\L(\ell,\lambda_{1}),
L(\ell,\lambda_{2})\end{array}\right)\rightarrow {\rm
Hom}_{\hat{{\bf g}}}(\hat{L}(\lambda_{1})\otimes
L(\ell,\lambda_{2}), L(\ell,\lambda_{3}));\nonumber\\
& &I(\cdot,x)\mapsto \phi_{I}.
\end{eqnarray}

In [TK] or in many physics references, an intertwining operator
$I(\cdot,x)$ of vertex
$\left(\begin{array}{c}L(\ell,\lambda_{3})\\L(\ell,\lambda_{1}),
L(\ell,\lambda_{2})\end{array}\right)$ is defined to be a linear map:
\begin{eqnarray}
I(\cdot,x):& &L(\lambda_{1})\rightarrow ({\rm
Hom}(L(\ell,\lambda_{2}),L(\ell,\lambda_{3})))\{x\};\nonumber\\
& &u\mapsto I(u,x)
\end{eqnarray}
satisfying the conditions (7.3.7) and (7.3.8).
It is clear that this definition is weaker than Definition 4.1.1. But,
if $\ell$ is a positive integer, Tsuchiya and
Kanie's nuclear democracy theorem [TK] says that $I(\cdot,x)$ can be
uniquely extended to $L(\ell,\lambda_{1})$. (To be precise, this was
proved only for ${\bf g}=sl_{2}$ in [TK].) Next By using Theorem 7.1.6
or Theorem 7.3.1 we
give an alternative proof.

{\bf Proposition 7.3.3.} {\it The linear map $\phi$ is an isomorphism if
$\ell$ is a positive integer.}

{\bf Proof.} It follows from the Jacobi identity that $\phi_{I}=0$
implies $I^{o}(\cdot,x)=0$. Therefore, $\phi$ is injective.

On the
other hand, let $\psi$ be a $\tilde{{\bf g}}$-homomorphism from
$\hat{L}(\lambda_{1})\otimes L(\ell,\lambda_{2})$ to
$L(\ell,\lambda_{3})$.
Then we may naturally consider $\psi$ as a
linear map from $\hat{L}(\lambda_{1})$ to
${\rm Hom}(L(\ell,\lambda_{2}),L(\ell,\lambda_{3}))$. Define
\begin{eqnarray}
\bar{\psi}:& & L(\lambda_{1})\rightarrow
({\rm Hom}(L(\ell,\lambda_{2}),L(\ell,\lambda_{3})))\{x\};\nonumber\\
& &u^{1}\mapsto x^{h_{3}-h_{1}-h_{2}}\psi(Y_{t}(u^{1},x))
\end{eqnarray}
for $u^{1}\in L(\lambda_{1})$. Since $L(\lambda_{1})$ is an
irreducible ${\bf g}$-module, it is clear that $\bar{\psi}$ is
injective. By writing (7.3.7) in terms of generating
functions, we get
\begin{eqnarray}
& &[Y(a,x_{1}),I(u^{1},x_{2})]=x_{2}^{-1}\delta\left({x_{1}\over x_{2}}\right)
I(au^{1},x_{2}).
\end{eqnarray}
Since $\displaystyle{(x_{1}-x_{2})\delta\left({x_{1}\over
x_{2}}\right)=0}$, we have:
\begin{eqnarray}
(x_{1}-x_{2})[Y(a,x_{1}),I(u^{1},x_{2})]=0
\end{eqnarray}
for any $a\in {\bf g}, u^{1}\in L(\lambda_{1})$. Since ${\bf g}$
generates $L(\ell,0)$ as a vertex operator algebra, similarly to
Proposition 7.1.4 we can prove that $I(u^{1},x)$ satisfies (G3) for
any $a\in L(\ell,0)$. Furthermore, (7.3.7) implies (G2).
Therefore
$L(\lambda_{1})$ is embedded into
${\rm G}(L(\ell,\lambda_{2}),L(\ell,\lambda_{2}))$ through
$\bar{\psi}$. Let $W$ be the $V$-submodule of  ${\rm
G}(L(\ell,\lambda_{2}),L(\ell,\lambda_{2}))$ generated by
$L(\lambda_{1})$. From (7.3.13), we have:
\begin{eqnarray}
a(m)\circ I(u^{1},x_{2})=0\;\;\;\;\mbox{for any }a\in {\bf g}, m>0,
u^{1}\in L(\lambda_{1}).
\end{eqnarray}
Then $W$ is a certain quotient module of $M(\ell,\lambda_{1})$. If
$\ell$ is a positive integer, the vertex operator algebra $L(\ell,0)$
is rational ([FZ], [DL], [L2]). Thus
$W=L(\ell,\lambda_{1})$. Therefore, we obtain an intertwining vertex
operator of type $\left(\begin{array}{c}L(\ell,\lambda_{3})\\L(\ell,
\lambda_{1}),L(\ell,\lambda_{2})\end{array}\right)$. Thus $\phi$ is an
isomorphism. $\;\;\;\;\Box$

Let $L(c,h)$ be the irreducible module of the Virasoro algebra Vir with
central charge ${\it c}$ and lowest weight ${\it h}$. It has been
proved ([FZ], [H2], [L2]) that
$L(c,0)$ is a vertex operator algebra. (But not all $L(c,h)$ are
$L(c,0)$-modules in general.)

{\bf Remark 7.3.4.} Suppose that $L(c,h_{1})$ and $L(c,h_{2})$ are two
modules for the vertex operator algebra $L(c,0)$. Let $\phi(x)\in
\left({\rm Hom}_{{\bf C}}(L(c,h_{1}),L(c,h_{2}))\right)\{x\}$ such that
\begin{eqnarray}
[L(m),\phi(x)]=\left((m+1)h+x{d\over dx}\right)\phi(x)
\end{eqnarray}
for some complex number $h$. Then $\phi(x)$ generates a lowest-weight
module $M$ of lowest weight $h$ for the Virasoro algebra inside ${\rm
G}(L(c,h_{1}),L(c,h_{2}))$. If $c$ is among the minimal series, then the
vertex operator algebra $L(c,0)$ is rational [W]. Therefore
$M=L(c,h)$. Then we obtain an intertwining vertex operator of type
$\left(\begin{array}{c}L(c,h_{2})\\L(c,h),L(c,h_{2})\end{array}\right)$.

{\bf Proposition 7.3.5.} {\it Each fusion rule is either 0 or 1 among
irreducible $L(c,0)$-modules}.

{\bf Proof.}  Let $L(c,h_{i})$ (i=1,2,3) be irreducible
$L(c,0)$-modules and let $h_{i},\;u^{i}$ be the lowest weight and
lowest weight vector of $L(c,h_{i})$ (i=1,2,3), respectively. Let
$I(\cdot,x)$ be any
intertwining operator of type
$\left(\begin{array}{c}L(c,h_{3})\\L(c,h_{1}),L(c,h_{2})\end{array}\right)$.
Then
\begin{eqnarray}
I(u,x)=\sum _{n\in {\bf
Z}}I_{n}(u)x^{-n-1}x^{h_{3}-h_{1}-h_{2}}\;\;\;{\rm for}\; u\in L(c,h_{1}).
\end{eqnarray}
{}From [FHL] we have the following formula:
\begin{eqnarray}
{\rm wt}(I_{n}(u^{1})u^{2})=h_{3}-n-1\;\;\;{\rm for}\;n \in {\bf
Z}.\end{eqnarray}

Claim: $I(\cdot,x)$ is uniquely determined by $I_{-1}(u^{1})u^{2}$.
Equivalently, $I_{-1}(u^{1})u^{2}=0$ implies $I(\cdot,x)=0$.

If $I(\cdot,x)$ is a nontrivial intertwining operator, then there is
an integer $k$ such that
\begin{eqnarray}I_{k}(u^{1})u^{2}\ne 0\;\;{\rm and}\;\;
I_{n}(u^{1})u^{2}=0\;\;{\rm if}\;n > k.\end{eqnarray}
By the commutator formula (2.1.3), we get
\begin{eqnarray}& &[Y(\omega,x_{1}), I(u^{1},x_{2})]\nonumber\\&=&\sum
_{i=0}^{\infty}\frac{(-1)^{i}}{i!}x_{2}^{-1}\left({\partial \over \partial
x_{1}}\right)^{i}\delta\left({x_{1}\over
x_{2}}\right)I(L(i-1)u^{1},x_{2})\nonumber \\
&=&x_{2}^{-1}\delta \left({x_{1}\over
x_{2}}\right)I(L(-1)u^{1},x_{2})-x_{2}^{-1}{\partial\over \partial
x_{1}}\delta \left({x_{1}\over x_{2}}\right)I(L(0)u^{1},x_{2}).\end{eqnarray}
Equivalently,
\begin{eqnarray}
[L(n),I_{m}(u^{1})]=(-m-n-1+h_{3}-h_{1}-h_{2}+(n+1)h_{1})I_{m+n}(u^{1}).
\end{eqnarray}
Therefore
\begin{eqnarray}L(n)I_{k}(u^{1})u^{2}=0 \;\;\;\mbox {for all }n\ge 1.
\end{eqnarray}
That is, $I_{k}(u^{1})u^{2}$ is a lowest weight vector of
$L(c,h_{3})$. Since the lowest weight vector is unique up to a
constant multiple, by (7.3.17) we get $k=-1$. So $I_{-1}(u^{1})u^{2}\ne
0.\;\;\;\;\Box$

Similarly, we have:

{\bf Proposition 7.3.6.} {\it Let $V=L(\ell,0)$ be a vertex operator
algebra associated
to a finite-dimensional simple Lie algebra ${\bf g}$.
Let $M^{i}$ (i=1,2,3)
be three irreducible $V$-modules. Then}
\begin{eqnarray}
\dim I\left(\begin{array}{c}M^{3}\\M^{1},M^{2}\end{array}\right)\le
 \dim {\rm Hom}_{{\bf
g}}(M^{1}(0)\otimes M^{2}(0),M^{3}(0)).\end{eqnarray}

{\bf Proof}. Similarly to the proof of Proposition 7.3.5, we can prove
that any intertwining operator $I(\cdot,x)$ of type
$\left(\begin{array}{c}M^{3}\\M^{1},M^{2}\end{array}\right)$ is
uniquely determined by the following linear map
\begin{eqnarray}f_{I} :& & M^{1}(0)\otimes M^{2}(0)\rightarrow
M^{3}(0)\nonumber \\
& &u\otimes v\mapsto I_{-1}(u)v\;\;\;\mbox{for }u\in M^{1}(0),v\in
M^{2}(0). \end{eqnarray}
For any $a\in {\bf g}=V_{(1)}$, by (2.1.3) we have
\begin{eqnarray}
a_{0}I_{-1}(u)v=I_{-1}(a_{0}u)v+I_{-1}(u)a_{0}v.\end{eqnarray}
Then $f_{I}$ is a ${\bf g}$-homomorphism.$\;\;\;\;\Box$

\newpage

\makeatletter
\@addtoreset{equation}{chapter}
\def\theequation{\thechapter.\arabic{equation}}
\makeatother

\appendix

\chapter{A counterexample}

In this appendix, we shall give an example to show that the conditions
on $M^{2}$ and $M^{3}$ of Theorem 4.2.4 are necessary.

For any complex numbers $c$ and $h$, let $M(c,h)$ be the Verma module
for the Virasoro Lie algebra $Vir$ with lowest weight $h$ of central
charge $c$. Let $L(c,h)$ be the corresponding irreducible quotient
module. Let ${\bf 1}$ be a lowest-weight vector of weight-zero for
$M(c,0)$. Then $L(-1){\bf 1}$ is a lowest-weight vector of weight-one.
Let $M_{c}$ be the quotient module of $M(c,0)$ divided by the submodule
generated from $L(-1){\bf 1}$. It is well-known ([FZ], [H2], [L2])
that $M_{c}$ has a
natural vertex operator algebra structure and that any Verma module
$M(c,h)$ is a module for this vertex operator algebra $M_{c}$.

The following proposition easily follows from [FZ].

{\bf Proposition A.1.} {\it For any complex number $c$,
$A(M_{c})$ is isomorphic to the polynomial algebra ${\bf C}[t]$ in one
indeterminant $t$. For any complex number $h$, $A(M(c,h))$ is
isomorphic to the vector space ${\bf C}[t_{1},t_{2}]$ where $t_{1}$
and $t_{2}$ are two independent indeterminants and the left and the right
actions are given by}
\begin{eqnarray}
 t^{n}\cdot f(t_{1},t_{2})&=&t_{1}^{n}f(t_{1},t_{2}),\\
f(t_{1},t_{2})\cdot t^{n}&=&(t_{1}-t_{2})^{n}f(t_{1},t_{2})
\end{eqnarray}
{\it for any $n\in {\bf N}, f(t_{1},t_{2})\in {\bf C}[t_{1},t_{2}]$.}

For any coprime positive integers $p$ and $q$ ($p,q\ge 2$), set
\begin{eqnarray}
c_{p,p}=1-{\frac{6(p-q)^{2}}{pq}}.
\end{eqnarray}
It follows from Kac's determinant formula that $M_{c}$ is irreducible
if $c\ne c_{p,q}$ for any coprime numbers $p$ and $q$. In this case,
$M_{c}=L(c,0)$.

{\bf Proposition A.2 [W].} {\it The vertex operator algebra $L(c,0)$ is
rational if and only if $c=c_{p,q}$ for some coprime positive integers
$p$ and $q$.}

Suppose $c$ is not among the minimal series (A.3). Then $A(M_{c})={\bf
C}[t]$. Let $h$ be a complex number such that $M(c,h)$ is irreducible.
Then
\begin{eqnarray}
& &A(M_{c})\otimes _{A(M_{c})}M(c,h)(0)=M(c,h)(0),\\
& &A(M(c,h))\otimes _{A(M_{c})}M_{c}(0)\simeq {\bf C}[t],
\end{eqnarray}
where ${\bf C}[t]$ is the adjoint $A(M_{c})$-module.
Let $h_{1}$ and $h_{2}$ be two different complex numbers such that
$M(c,h_{1})$ and $M(c,h_{2})$ are irreducible. Then
$I\left(\begin{array}{c}M(c,h_{2})\\M_{c},M(c,h_{1})\end{array}\right)=0$.
By Corollary 4.1.5, we have:
$I\left(\begin{array}{c}M(c,h_{2})\\M(c,h_{1}),M_{c}\end{array}\right)=0$.
If Frenkel and Zhu's fusion
formula were true, then
\begin{eqnarray}
\dim I\left(\begin{array}{c}M(c,h_{2})\\M(c,h_{1}),M_{c}\end{array}\right)
&=&\dim {\rm Hom}_{A(V)}(A(M(c,h_{1}))\otimes _{A(V)}M_{c}(0), M(c,h_{2})(0))
\nonumber\\
&=&\dim {\rm Hom}_{A(V)}(A(M_{c}), M(c,h_{2})(0))\nonumber\\
&=&1.
\end{eqnarray}
This is a contradiction. In this example, the condition
$\bar{M}(M^{2}(0))=M^{2}$ of Theorem 4.2.4 is violated since
$\bar{M}(M_{c}(0))=M(c,0)\ne M_{c}$.

\chapter{An example of the tensor product module being zero}
In the classical Lie algebra case, the tensor product module for any
two nonzero modules is not zero. But the tensor product module of two
nonzero modules for a vertex operator algebra may be zero.
Since in the classical associative
algebra (or ring) case, the tensor product module of a nonzero bimodule with a
nonzero left module may be a zero module,
from Frenkel and Zhu's theory, this is not surprising.
In this appendix we shall give an example such that the tensor product
module of two nonzero modules is zero for a vertex operator algebra.

Let ${\bf g}$ be a finite-dimensional simple Lie algebra with a fixed
Cartan subalgebra ${\bf h}$ and let $\tilde{{\bf g}}$ be the
corresponding affine Lie algebra. Let $\ell$ be a positive
integer and let $L(\ell,\lambda)$ be the irreducible highest-weight
$\tilde{{\bf g}}$-module of level $\ell$ with highest-weight $\lambda$.
Let $M(\ell,{\bf C})$ be the generalized Verma $\tilde{{\bf
g}}$-module of level $\ell$. Then $M(\ell,{\bf C})$ is a vertex
operator algebra and any highest-weight $\tilde{{\bf g}}$-module of
level $\ell$ is a module for this vertex operator algebra ([FZ], [L2]).
Let $V=M(\ell,{\bf C})$ and let $\bar{V}=L(\ell,0)$ be the irreducible
quotient module of  $M(\ell,{\bf C})$. Then $\bar{V}$ is a vertex
operator algebra and the
standard $\tilde{{\bf g}}$-modules of level $\ell$ form the set of
equivalence classes of irreducible $\bar{V}$-modules ([DL], [FZ], [L2]).

{\bf Proposition B.1.}
{\it Suppose that $\lambda$ is not an integral dominant weight. Then
 the tensor product module for $V$-modules
$\bar{V}$ and $L(\ell,\lambda)$ is zero.}

{\bf Proof.} Since $\lambda$ is not an integral dominant weight,
$L(\ell,\lambda)$ is a $V$-module, but it is not a $\bar{V}$-module.
To prove that the
zero module is a tensor product for the ordered pair $(\bar{V},
L(\ell,\lambda))$, we need to check the universal property of
Definition 5.1.1. Let $M$ be any $V$-module and let $I(\cdot,x)$ be
any intertwining operator of type $\left(\begin{array}{c}M\\ \bar{V},
L(\ell,\lambda)\end{array}\right)$. Then we need to prove that $I(\cdot,x)=0$.
Since $a_{i}{\bf 1}=0$ for any $a\in V, i\in {\bf Z}_{+}$,
$I({\bf 1},x_{2})$ commutes with every vertex operator $Y(a,x_{1})$.
Since $\displaystyle{{d\over dx}}I({\bf 1},x)=I(L(-1){\bf 1},x)=0$,
$I({\bf 1},x)$ is a constant. Thus $I({\bf 1},x)$ is a $V$-homomorphism
from $L(\ell,\lambda)$ to $M$. Let $W$ be the image of $I({\bf 1},x)$.
Then it follows from the Jacobi identity that
$I(a,x)L(\ell,\lambda)\subseteq W\{x\}$ for any $a\in V$.  Suppose
$I(\cdot,x)\ne 0$.
Since $L(\ell,\lambda)$ is irreducible, $W=L(\ell,\lambda)$. Thus
$I({\bf 1},x)=\alpha {\rm id}$ for some nonzero number $\alpha$. Then
we obtain an intertwining operator $F(\cdot,x)=\alpha ^{-1}I(\cdot,x)$ of type
$\left(\begin{array}{c}L(\ell,\lambda)\\
\bar{V},L(\ell,\lambda)\end{array}\right)$ such that $F({\bf
1},x)={\rm id}$. Therefore $L(\ell,\lambda)$ is a $\bar{V}$-module.
This is a contradiction. Thus $I(\cdot,x)=0$. Then the zero module is a
tensor product module for the ordered pair $(\bar{V},
L(\ell,\lambda))$. $\;\;\;\;\Box$

{\bf Remark B.2.} Suppose that the vertex operator algebra $V$ is
selfdual (a necessary and sufficient condition was given in [L1]) and
that the tensor product functor $T$ or Huang and Lepowsky's box tensor
satisfies a certain  associativity (this has been proved by Huang and Lepowsky
under a certain condition). Let $M$ and $W$ be any two irreducible
$V$-modules. Since $I\left(\begin{array}{c}M\\V,M\end{array}\right)\ne
0$, it follows from Corollary 4.1.5 that
$I\left(\begin{array}{c}V'\\M',M\end{array}\right)\ne 0$. Thus
$I\left(\begin{array}{c}V\\M',M\end{array}\right)\ne 0$. It follows
from Proposition 5.1.5 that there is
a nonzero $V$-homomorphism $\psi$ from $T(M',M)$ to $V$. Then
\begin{eqnarray}
T(M',T(M,W)) \simeq T(T(M',M),W)\stackrel{T(\psi,W)}\longrightarrow
T(V,W) \simeq W.
\end{eqnarray}
Therefore $T(M,W)\ne 0$.

\bibliography{thesis}
\bibliographystyle{plain}

\end{document}